\documentclass[letterpaper,12pt]{article}
\pdfoutput=1
\usepackage{jheppub}
\usepackage{epsfig}
\usepackage{amsmath}
\usepackage{graphicx}
\usepackage{capt-of}
\usepackage{tikz-cd}
\usepackage[missing=]{gitinfo2}
\usepackage{float}
\usepackage{verbatim}
\usepackage{hyperref}
\restylefloat{table}

\DeclareSymbolFont{AMSa}{U}{msa}{m}{n}
\DeclareSymbolFont{AMSb}{U}{msb}{m}{n}
\let\Box\relax
\DeclareMathSymbol{\Box}{\mathord}{AMSa}{"03}

\allowdisplaybreaks

\definecolor{mymaroon}{rgb}{0.8, 0.25, 0.33}

\newcommand{\ti}[1]{\textit{#1}}

\newcommand{\cH}{{\mathcal H}}

\newcommand{\cN}{{\mathcal N}}

\newcommand{\cF}{{\mathcal F}}

\newcommand{\fg}{{\mathfrak g}}

\newcommand{\eps}{{\epsilon}}

\newcommand{\C}{{\mathbb C}}
\newcommand{\PP}{{\mathbb P}}
\newcommand{\bbZ}{{\mathbb Z}}

\newcommand{\bbR}{{\mathbb R}}
\newcommand{\Z}{{\bbZ}}
\newcommand{\R}{{\bbR}}

\newcommand{\fro}{{\overline{\underline\Omega}}}

\newcommand{\abs}[1]{\lvert#1\rvert}

\newcommand{\de}{{\mathrm d}}
\newcommand{\e}{{\mathrm e}}
\newcommand{\HOMFLY}{{\mathrm {HOMFLY}}}
\newcommand{\I}{{\mathrm i}}

\newcommand{\tM}{\widetilde M}
\newcommand{\tL}{\widetilde L}

\newcommand{\tC}{\widetilde C}

\newcommand{\IP}[1]{\langle#1\rangle}

\newcommand{\fapprox}{F_{\mathrm{approx}}}

\DeclareMathOperator{\Tr}{Tr}
\DeclareMathOperator{\Sk}{Sk}

\DeclareMathOperator{\im}{Im}

\DeclareMathOperator{\fgl}{{\mathfrak{gl}}}
\DeclareMathOperator{\fsl}{{\mathfrak{sl}}}

\newcommand{\insfigsvg}[3]{

\medskip
\noindent
\begin{minipage}{\linewidth}

\makebox[\linewidth]{\includegraphics[keepaspectratio=true,scale=#2]{figures/#1.pdf}}

\captionof{figure}{#3}

\label{fig:#1}
\end{minipage}
\medskip

}

\newcommand{\insfigpng}[3]{

\medskip
\noindent
\begin{minipage}{\linewidth}

\makebox[\linewidth]{\includegraphics[keepaspectratio=true,scale=#2]{figures/#1.png}}

\captionof{figure}{#3}

\label{fig:#1}
\end{minipage}
\medskip

}

\title{The quantum UV-IR map for line defects in $\mathfrak{gl}(3)$-type class $S$ theories}

\author[1]{Andrew Neitzke}
\author[2]{and Fei Yan}
\affiliation[1]{Department of Mathematics, Yale University}
\affiliation[2]{NHETC and Department of Physics and Astronomy, Rutgers University}
\emailAdd{andrew.neitzke@yale.edu}
\emailAdd{fyan.hepth@gmail.com}

\abstract{We consider the quantum UV-IR map for line defects in class $S$ theories of $\mathfrak{gl}(3)$-type. 
This map computes the protected spin character which counts framed BPS states with spin for the bulk-defect system. 
We give a geometric method of computing this map motivated by the 
physics of five-dimensional $\cN=2$ supersymmetric
Yang-Mills theory, and compute it explicitly in various examples.
As a spin-off we propose a new way of computing a certain specialization of the 
HOMFLY polynomial for links in $\R^3$, as a sum over BPS webs attached to the link.}

\begin{document}

\noindent{{\tiny \color{gray} \tt \gitAuthorIsoDate \gitAbbrevHash}}

\maketitle
\flushbottom

\section{Introduction}\label{sec:introduction}

This paper is a continuation of \cite{Neitzke:2020jik}.
In that paper we considered the {\it quantum UV-IR map} (or $q$-nonabelianization map).
In general, this map
takes links $L$ in a 3-manifold $M$ to $\Z[q,q^{-1}]$-linear
combinations of links $\tL$ in a branched $N$-fold cover $\tM$. 
It has two important specializations:

\begin{itemize}
	\item When $M = C \times \R$ for a compact surface $C$, 
the quantum UV-IR map gives the IR effective description for supersymmetric line 
defects in a class $S$ theory of type $\fgl(N)$ at a point in its Coulomb branch.
\item 
When $M = \R^3$, the quantum UV-IR 
map gives a new way of computing a certain 1-parameter specialization of the
HOMFLY polynomial of a link.
\end{itemize}

In \cite{Neitzke:2020jik} we gave a description of the quantum UV-IR
map, motivated from the 
physics of five-dimensional $\cN=2$ $U(N)$ super Yang-Mills theory.
We also gave an explicit construction of the quantum UV-IR map in the case of $N=2$ 
and $M=C\times \R$ or $M = \R^3$. 
The extension from $N=2$ to $N>2$ brings in many new features and complications.
In this paper we give a construction of the quantum UV-IR map 
for $N=3$ (in cases where no multi-loop BPS
webs appear), and compute it in various cases.

The constructions in this paper and in \cite{Neitzke:2020jik}
are inspired by
various previous works in the physics literature, 
especially \cite{Gaiotto:2012rg,Gaiotto:2011nm,Galakhov:2014xba,Gabella:2016zxu},
and in the math literature, particularly \cite{Fock-Goncharov,alex2019quantum}, 
\cite{Bonahon2010} for the $N=2$
case, and \cite{douglas2021quantum} for $N=3$. 
We discuss the comparison to those works in more detail in
\autoref{sec:results-outlook} below.

In the rest of this introduction we review the motivation for our constructions;
in \autoref{sec:results-outlook} we give a summary of results and outlook.

\subsection{Motivation}\label{sec:motiv}

Line operators are important tools in quantum field theory. Wilson and 't Hooft lines are well-known and important observables in the study of phases of 4d gauge theories \cite{PhysRevD.10.2445,THOOFT19781}. More recently the study of line operators 
has led to insights on the global structure of quantum field theories. Additionally, studying the response of a quantum field theory to insertion of line defects modeling infinitely massive particles can also inform us about the spectrum in the original theory. For a sampling of recent developments in the study of lines in 4d theories see e.g. \cite{Kapustin_2006,Gomis:2006sb,Pestun:2007rz,DHoker:2007mci,Drukker:2009id,Drukker:2009tz,Gaiotto:2010be,Drukker:2010jp,Ito:2011ea,Aharony_2013,Cordova:2013bza,Lewkowycz:2013laa,Gaiotto:2014kfa,Fiol:2015spa,Moore:2015szp,Cordova:2016uwk,Bianchi:2018zpb,Giombi:2018hsx,Brennan:2018rcn,Brennan:2018yuj,Cirafici:2019otj,Ang_2020,Agmon:2020pde,Rudelius:2020orz,Bhardwaj:2021pfz,Costello:2021zcl,Apruzzi:2021phx,Bhardwaj:2021zrt,Cuomo:2021rkm}. 

In this paper, we study supersymmetric line defects in a large class of 
four-dimensional $\cN=2$ theories called class
$S$ \cite{Gaiotto:2009we,Gaiotto:2009hg}. Examples of such line defects include not only familiar supersymmetric Wilson-'t Hooft lines in supersymmetric gauge theories, but also generalizations in non-Lagrangian theories. 

A special case occurs in abelian gauge theories, where it is easier to describe supersymmetric line defects explicitly. An abelian gauge theory has an electromagnetic charge lattice $\Gamma$. Supersymmetric line defects in the theory are then labeled by charges $\gamma\in\Gamma$, and can be understood as representing the insertion of an infinitely massive BPS dyon with charge $\gamma$. As a simple example, we consider 4d $\cN=2$ $U(1)$ gauge theory, and take $\gamma$ to be purely electric; then a supersymmetric Wilson line extended in the time direction can be explicitly written as follows, 
\begin{equation}\label{eqn:abelianline}
\mathbb{L}(\gamma,\zeta)= \text{exp}\left[\text{i}\gamma\int \left(A+\frac{1}{2}(\zeta^{-1}\phi+\zeta\bar{\phi})\right)\right],
\end{equation}
where the integral is along the time direction, and $A$ and $\phi$ are the $U(1)$ gauge field and complex scalar in the $\cN=2$ vector multiplet. 
This line defect preserves half of the supercharges of the $\cN=2$ theory;
which half of the supercharges are preserved is determined by the parameter $\zeta\in \C^\times$.

As it is often simpler to study line defects in abelian gauge theories, one strategy to study line defects in a 4d $\cN=2$ theory is to deform to a generic point $u$ in the Coulomb branch \cite{Gaiotto:2010be}, where the low energy effective theory is $U(1)^r$ gauge theory \cite{Seiberg:19941,Seiberg:19942}. Following a supersymmetric line defect $\mathbb{L}$ into 
the infrared, its IR limit can be expanded as 
a superposition of IR dyonic line defects  $\mathbb{L}_\gamma^{\text{IR}}$ in the abelian theory,
with integer coefficients $\fro(\mathbb{L},u,\gamma)$ counting the ground states of the bulk-line system in the IR charge sector labeled by $\gamma$ \cite{Gaiotto:2010be}. We will give a proper definition of $\fro(\mathbb{L},u,\gamma)$ in \autoref{sec:PSC}.\footnote{One could imagine a richer notion of UV-IR map which takes account of the expectation that line defects form a category rather than just a set or an algebra. In \cite{Neitzke:2020jik} and this paper, we do not attempt to go in this direction.}

\subsection{The quantum UV-IR map and protected spin characters}\label{sec:PSC}

It was argued in \cite{Gaiotto:2010be} that the coefficients 
$\fro(\mathbb{L},u,\gamma)$
in the IR expansion of a $1/2$-BPS
line defect $\mathbb{L}$ can equivalently be understood as
an index counting framed BPS states of $\mathbb{L}$, as follows. We consider the Hilbert space of the 4d N=2 theory on $\R^3$, where the insertion of line defect $\mathbb{L}$ modifies the Hilbert space at the origin of $\R^3$. Moreover the vacuum at infinity is fixed corresponding to a point $u$ in the Coulomb branch. This defect Hilbert space $\cH_{\mathbb{L},u}$ then admits a decomposition into different superselection sectors labeled by the IR electromagnetic (and flavor) charge $\gamma$:
\begin{equation}
\cH_{\mathbb{L},u}=\bigoplus_{\gamma}\cH_{\mathbb{L},u,\gamma}
\end{equation}
Within each charge sector, one can define an index $\fro(\mathbb{L},u,\gamma)$ counting the lowest-energy states. As the $1/2$-BPS line defect $\mathbb{L}$ also preserves the $SU(2)_P$ spatial rotation around the line and the $SU(2)_R$ R-symmetry, we can refine the counting of lowest-energy states by the spin data. This gives the {\it protected spin character} \footnote{In this paper and in \cite{Neitzke:2020jik} we have adopted a different convention from \cite{Gaiotto:2010be}. In particular, the framed no-exotics conjecture, which states that framed BPS states form trivial representation under $SU(2)_R$, implies that $\fro(\mathbb{L},u,\gamma,q)$ are actually characters (in $-q$) of the $SU(2)_P$ rotation.} \cite{Gaiotto:2010be}:
\begin{equation}\label{eqn:PSC}
\fro(\mathbb{L},u,\gamma,q):=\text{Tr}_{\cH_{\mathbb{L},u,\gamma}}(-q)^{2J}q^{2R},
\end{equation}
where $J$ and $R$ are Cartan generators of $SU(2)_P$ rotation and $SU(2)_R$ R-symmetry respectively. Generically $\fro(\mathbb{L},u,\gamma,q)$ takes values in $\Z[q,q^{-1}]$. Taking $q= -1$ then recovers the integer-valued indices $\fro(\mathbb{L},u,\gamma)$ introduced in \autoref{sec:motiv}. 

The data of $\fro(\mathbb{L},u,\gamma,q)$ could be collected into the following generating function for protected spin characters
\begin{equation}\label{eqn:qUVIR}
f(\mathbb{L},u):=\bigoplus_\gamma \fro(\mathbb{L},u,\gamma,q) X_\gamma
\end{equation}
We denote $f$ as the {\it quantum UV-IR map}, which maps a line defect $\mathbb{L}$ to the generating function for its protected spin characters at at a point $u$ in the Coulomb branch. Here $X_\gamma$ are formal variables corresponding to IR dyonic line defects with charge $\gamma$. Determining the quantum UV-IR map $f$ at the point $u$ is equivalent to solving the problem of 
computing the spin-refined framed BPS spectra of all line defects in the theory at $u$.

\subsection{The case of class \texorpdfstring{$S$}{S} theories}\label{sec:introclassS}

In this paper we consider line defects in specific 
$\cN=2$ theories known as class $S$ theories \cite{Gaiotto:2009we,Gaiotto:2009hg}, obtained by partially-twisted compactification of 6d $(2,0)$ theory on a Riemann surface $C$. 
For these theories,
as we describe in \autoref{sec:skeinalgebras} below:
\begin{itemize}
\item There is a class of $1/2$-BPS line defects $\mathbb{L}$
corresponding to simple closed curves\footnote{As we will describe in \autoref{sec:skeinalgebras}, in general $\mathbb{L}$ can also correspond to laminations or contain junctions. We defer systematic study of these cases to future work.} $\ell$ on the Riemann surface
$C$ (up to isotopy). This correspondence comes from the 6d construction. Consider a two-dimensional surface defect in 6d, wrapping $\ell \subset C$ and extending along the time direction. After the compactification on $C$, this produces a line defect $\mathbb{L}$ extending along the time direction.

\item The IR line defects likewise correspond to simple closed curves
on an $N$-fold cover $\tC$ of $C$ (up to homology), i.e. the Seiberg-Witten curve for the class S theory at the point $u$ in the Coulomb branch.

\item The operator products of line defects correspond to 
natural skein algebra structures in the space of formal linear combinations of
closed curves, both in the UV and IR.
\end{itemize}
Due to supersymmetry, computing the line defects OPE  commutes with the RG flow. It then follows that
the quantum UV-IR map $f$ must induce a homomorphism of skein algebras,
\begin{equation}\label{eqn:Fhom-intro}
F: \Sk(M,\fgl(N)) \to \Sk(\tM,\fgl(1))
\end{equation}
where $M = C \times \R_h$ and $\tM = \tC \times \R_h$.
This property, together with the fact that line defects only depend on simple closed curves up to isotopy, give very strong constraints on the quantum UV-IR map.
We exploit these constraints heavily in computing it. 

\insfigsvg{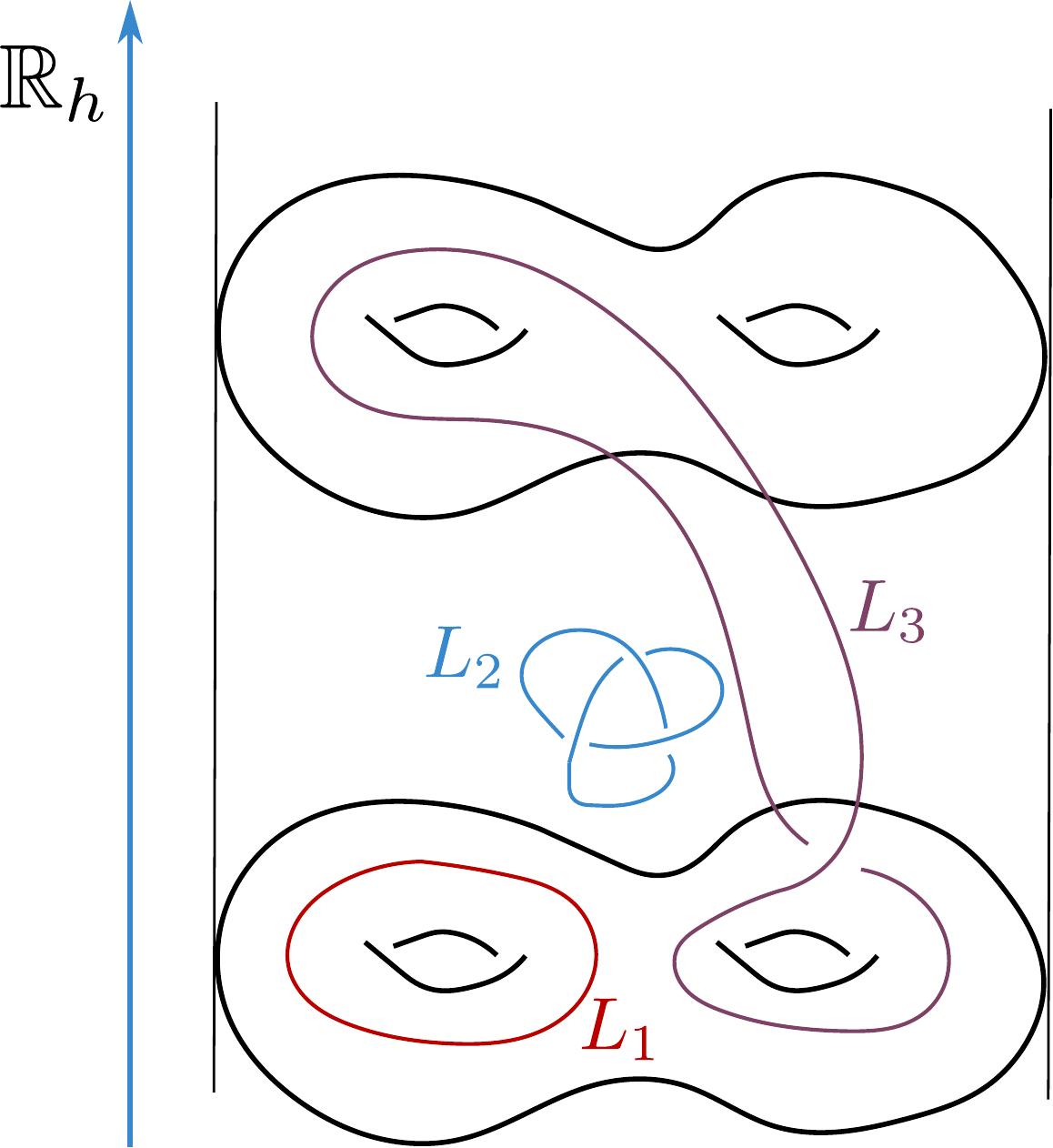}{0.37}{Illustration of the three-manifold $M=C\times\R_h$. A generic link $L\subset M$, such as the knot $L_3$ in the figure, has a finite extent along the $\R_h$ direction that can't be isotoped away. For special links that can be isotoped to simple closed curves on $C$, such as the knot $L_1$ in the figure, $F(L)$ computes the generating function of protected spin characters for the corresponding $1/2$-BPS line defects. There are also special links which are contained in a 3-ball, such as the trefoil knot $L_2$ in the figure. In this case $F(L)$ is expected to compute the 1-parameter limit of HOMFLY polynomial (\ref{eq:homfly-intro}).}

The skein algebra $\Sk(M,\fgl(N))$ is generated by links $L\subset M=C\times \R$. In the context of the 6d $(2,0)$ theory of type $\fgl(N)$, we consider two-dimensional surface defects wrapping $L$ while extending along the time direction, which descend to defects in class $S$ theories after the compactification on $C$. Similarly $\Sk(M,\fgl(1))$ is generated by links $\tL\subset \tM=\tC\times\R$, corresponding to surface defects in the abelian 6d $(2,0)$ $\fgl(1)$ theory. In this context the quantum UV-IR map can be thought of as a map sending a surface defect in the 6d $\fgl(N)$ theory into a combination of surface defects in the 6d $\fgl(1)$ theory. We refer the readers to \cite{Neitzke:2020jik} for a more detailed description of this 6d picture.

If the link can be isotoped to a simple closed curve on $C$ at a fixed point along the $\R_h$-axis, such as $L_1$ in \autoref{fig:qUVIR}, then the corresponding surface defect in 6d descends to a $1/2$-BPS line defect $\mathbb{L}_1$ in the 4d class $S$ theory. In this case $F(L_1)$ computes the generating function of protected spin characters for $\mathbb{L}_1$. In general, though, the link $L$ will have a finite extent along the $\R_h$-direction that can not be isotoped away, as for $L_3$ shown in \autoref{fig:qUVIR}. In this case, as we will describe in \autoref{sec:skeinalgebras}, the corresponding defect $\mathbb{L}_3$ in class $S$ theory is effectively a $1/4$-BPS line defect, which breaks the spatial rotation to $U(1)_P$ while preserving a $U(1)_R\subset SU(2)_R$ R-symmetry. For such $1/4$-BPS line defects, we can still define the protected spin characters similar to (\ref{eqn:PSC}), whose generating function is once again computed by the map $F$.

Before we continue we would like to comment on our notations. Due to the geometric construction of line defects in class $S$ theories, the quantum UV-IR map $f$ as introduced in \autoref{sec:PSC} is equivalent to the homomorphism $F$ in (\ref{eqn:Fhom-intro}). In the following we often also denote the quantum UV-IR map by $F$.

\subsection{Application to HOMFLY polynomials}\label{sec:introHOMFLY}

As we discussed in \cite{Neitzke:2020jik},
the quantum UV-IR map we consider has broader applicability than 
computing the protected spin characters for line defects in 4d $\cN=2$ theories of class $S$. 
In particular, we can take the case of $M = \R^3$ and $\tM$ a disjoint union of $N$ copies of $\R^3$. In that case the IR line defects
are all trivial as $\tM$ only has trivial homology class, and the quantum UV-IR map just assigns to a link $L \subset \R^3$
a polynomial $F(L) \in \Z[q,q^{-1}]$.
From the skein relations obeyed by the 6d surface defects, it follows that
$F(L)$ computes the following 1-parameter specialization of the HOMFLY polynomial:\footnote{More precisely $L$ is a framed oriented link, and (\ref{eq:homfly-intro}) holds if $L$ carries the blackboard framing. We refer the readers to \cite{Neitzke:2020jik} for a more detailed description regarding framing.}
\begin{equation} \label{eq:homfly-intro}
	F(L) = q^{N wr(L)} P_\HOMFLY(L, a = q^N, z = q - q^{-1})
\end{equation}
where $wr(L)$ denotes the writhe of $L$.

More generally, we expect \eqref{eq:homfly-intro} to hold also for a special class of links in generic $M = C \times \R_h$, namely for links contained inside a 3-ball in $M$, such as the trefoil knot $L_2$ shown in \autoref{fig:qUVIR}. We give a simple example of this kind for $N=3$ in \autoref{sec:unknotbp}, and more examples for the $N=2$ case can be found in \cite{Neitzke:2020jik}.

\subsection{Our main results and outlook} \label{sec:results-outlook}

As we have described, the quantum UV-IR map 
unifies the computation of protected spin characters for line defects
in class $S$ theories and the 1-parameter limit (\ref{eq:homfly-intro}) of the HOMFLY polynomials for links in $\R^3$.
Thus it is desirable to understand this map better, and in particular 
to have a concrete way of computing it.

In \cite{Neitzke:2020jik} we set out a general scheme for computing the
quantum UV-IR map. This scheme is motivated by the physics of 5d $\cN=2$ $U(N)$ super
Yang-Mills theory; concretely it is a geometric recipe mapping a link $L \subset M$ to a linear combination of links $\tL \subset \tM$.
The links $\tL$ consist partly of
lifts of segments of $L$ and partly of extra contributions from {\it BPS webs}
attached to $L$. We also described in \cite{Neitzke:2020jik} the detailed
implementation of this scheme in the case of $N=2$. In particular, we determined
the precise $q$-dependent prefactors which weigh the contributions from each $\tL$
to $F(L)$, and gave sketch proofs that our scheme indeed produces a map $F$ 
with all the expected properties.

In this paper we continue the construction of the quantum UV-IR map 
to the case $N=3$. This case is considerably
more complicated than $N=2$, because the BPS webs which occur can have more complicated topologies. For $N=2$, each BPS web consists only of a single segment;
for $N=3$ they can be general trivalent graphs, with or without loops.
Part of the problem of describing the UV-IR map is to 
determine the $q$-dependent factors 
accompanying BPS webs with all of these different topologies.
In \autoref{sec:qab-general} we give detailed formulas for the
factors associated to webs without loops, as well as for webs containing
a single loop, and all of the other $q$-dependent factors entering
the map.

These formulas are already enough to compute $F(L)$ in various examples,
and we obtain results which pass all tests we have available; some of
these are described in \autoref{sec:examples}.
In particular, the examples in \autoref{sec:link-1}-\autoref{sec:link-3}
illustrate how the quantum UV-IR map leads to a new recipe for computing 
the specialized HOMFLY polynomial \eqref{eq:homfly-intro}, and
in \autoref{sec:once-punctured-torus}-\autoref{sec:four-punctured-sphere}
we give new results for protected spin characters in a few class $S$
theories of type $\fgl(3)$.

Most of our computations were done with the aid of computer programs,
described in \autoref{app:algorithms}.
With the arXiv version of this preprint 
we include Mathematica and Python code which implements
our computations in detail, and which can be used to do other similar
computations.

We indicate here a few open questions:
\begin{itemize}
	
	\item For the case of $M=C \times \R_h$, the quantum UV-IR map generically depends on the Coulomb branch data of the corresponding class $S$ theory. As we move around on the Coulomb branch, the map is locally constant; however, it is expected to jump 
	across certain codimension-one walls, 
	according to the {\it framed wall-crossing formula} \cite{Gaiotto:2010be,MR2567745,Kontsevich:2008fj,Dimofte:2009tm}.
	By examining the quantum UV-IR map before and after the jump, one should be
	able to extract the BPS spectrum with spin in the bulk class $S$ theory; similar computations were done in \cite{Gaiotto:2010be,Galakhov:2014xba}, using interfaces between surface defects instead of line defects. One can also ask whether there is a way to construct an invariant of $L \subset C \times \R_h$ which does \ti{not} suffer wall-crossing. An interesting possibility is to compose the quantum UV-IR map\footnote{Here we mainly mean the quantum UV-IR map for the effective $1/4$-BPS line defects.} with the IR formula for computing the line defect Schur index developed in \cite{Cordova:2016uwk}.
	
	\item The quantum UV-IR map gives an IR effective description for supersymmetric line defects in class $S$ theory at a point on its Coulomb branch. In other words, it tells the behavior of line defects when the bulk theory is deformed onto its Coulomb branch.\footnote{This scenario is different from the defect RG flow, where the defect undergo nontrivial RG flows while the bulk is little affected far away from the defect, as explored e.g. in \cite{PhysRevLett.67.161,Friedan:2003yc,Casini:2016fgb,Cuomo:2021rkm,Gaiotto:2014gha,Jensen:2015swa,Wang:2020xkc,Wang:2021mdq}.} It would be very interesting to understand this map more directly from a 4-dimensional field theory point of view, especially at generic points of  
	the Coulomb branch (the UV-IR map near infinity has been computed by semiclassical means
	in \cite{Moore:2015szp}).

	\item It should be possible to complete our construction by giving explicit 
	formulas for the weights associated to 
	webs with arbitrary numbers of loops. However, this may require different
	techniques than the brute-force isotopy-invariance methods we have used so far.
	Ideally one should be able to determine the weights by a direct computation in 5d $\cN=2$ super Yang-Mills theory. We would like to pursue this in the future.

	\item As we have remarked, when applied to a 
	link $L \subset \R^3$, our construction of $F(L)$ 
	computes the specialized HOMFLY polynomial \eqref{eq:homfly-intro}.
	In a special ``almost-degenerate'' limit our computation of $F(L)$ 
	reduces to a conventional sort of vertex
	model; see \autoref{sec:exchange-clusters} for this.
	Away from this limit, it seems to give a
	new way of computing \eqref{eq:homfly-intro} by summing over
	webs attached to the link $L$, and it would
	be interesting to know whether this new method can be used to expose
	any new structures in this link invariant.

	\item Our construction of $F(L)$ for class $S$ theories 
	is applicable at any
	point of the Coulomb branch, and thus it determines the protected spin characters
	at any point (as long as multi-loop webs do not appear).
	So far, we have done explicit computations at special ``almost-degenerate''
	regions of the Coulomb branch,
	where the enumeration of the BPS webs 
	simplifies considerably (see \autoref{sec:fg-foliations}).
	It would be very interesting to use our method to compute 
	$F(L)$ at more general loci of the Coulomb branch.

	\item Although we have made various checks of our description of $F$,
	we do not have a general mathematical proof that it has the 
	expected isotopy invariance properties or that it respects the skein relations. 
	(In the $N=2$ case we were able to give a sketch proof by considering an appropriate list of ``Reidemeister-type''
	moves \cite{Neitzke:2020jik}; for $N=3$ this kind of approach does not seem feasible at 
	present.) It would be very
	interesting to find such a proof. One intriguing possibility would be to
		somehow realize $F$ in Floer thory, 
		which has recently been related to skein relations \cite{Ekholm2019}, and then
		appeal to general properties of that theory.

	\item Up to minor differences in conventions, we expect that the quantum 
	UV-IR map for $M = C \times \R$ in almost-degenerate regions of the Coulomb branch 
	should be
	equal to the ``quantum trace''
	introduced in \cite{Bonahon2010} for $N=2$ and \cite{douglas2021quantum} for $N=3$,
	as well as with the computation of protected spin characters given in \cite{Gabella:2016zxu}.
	For $N=2$, it was shown in \cite{kim2020rm} that the approach of  \cite{Gabella:2016zxu} indeed reproduces the quantum trace of
	\cite{Bonahon2010}, up to
	some adjustments (see also \cite{Korinman-Quesney} for more on the relation between 
	$q$-nonabelianization and the quantum trace).
	For $N=3$, we have checked in some examples, with the help of Daniel Douglas, that our formulas agree with those of
	\cite{douglas2021quantum} --- see \autoref{sec:once-punctured-torus} and \autoref{sec:four-punctured-sphere}.
	It would be desirable to establish this equality directly along the lines of \cite{kim2020rm}.

\end{itemize}

\subsection*{Acknowledgements}
We thank Dylan Allegretti, J\o rgen Andersen, Ibrahima Bah, Sungbong Chun, Clay C\'ordova, Anindya Dey, Daniel Douglas, Sergei Dubovsky, Thomas Dumitrescu, Davide Gaiotto, Sergei Gukov, Jonathan Heckman, Po-Shen Hsin, Max H\"{u}bner, Saebyeok Jeong, David Jordan, Ahsan Khan, Zohar Komargodski, Pietro Longhi, Rafe Mazzeo, Gregory Moore, Hirosi Ooguri, Du Pei, Pavel Putrov, Fabian R\"{u}hle, Shu-Heng Shao, Yifan Wang, Itamar Yaakov, and Masahito Yamazaki for interesting discussions. The work of AN on this project was supported 
in part by NSF grants DMS-1711692 and DMS-2005312.
FY is supported by DOE grant DE-SC0010008.

\section{The quantum UV-IR map}

As described in \autoref{sec:introduction}, the computation of framed BPS spectrum for line defects in class $S$ theories and the computation of HOMFLY polynomials are unified through the quantum UV-IR map between the following two skein algebras:
\begin{equation}\label{eqn:Fhom}
F: \Sk(M,\fgl(N)) \to \Sk(\tM,\fgl(1)).
\end{equation}
We recall the definitions of these two skein algebras in \autoref{sec:skeinalgebras}, motivated from considerations of line defects OPE. The map $F$ depends on certain foliation data on the surface $C$; this is described in \autoref{sec:wkb-foliations}.

Our approach to construct the map $F$ is a combination of physical considerations and bootstrap-like methods. We outline the strategy in \autoref{sec:strategyF}, and then give a concrete description of $F$ in \autoref{sec:qab-general}.

\subsection{Line defect OPE and skein algebras} \label{sec:skeinalgebras}

The set of line defects in a 4d $\cN=2$ theory is equipped with a natural algebra structure given by the line defect OPE, defined by considering the theory with the insertion of two line defects. Concretely we consider two $1/2$-BPS line defects $\mathbb{L}_1$ and $\mathbb{L}_2$, extending along the time direction while inserted at two points $x_1$ and $x_2$ in the spatial $\R^3$, preserving a common half of the 
supercharges in the bulk 4d $\cN=2$ theory. The dependence of the correlation function on $x_1-x_2$ is $Q$-exact where $Q$ is one of the common preserved supercharges. Up to $Q$-exact terms the limit $x_1\to x_2$ is nonsingular, and by locality this limit should be equivalent to another line defect \cite{Gaiotto:2010be}, which we denote as $\mathbb{L}_1 \cdot \mathbb{L}_2$. In this way the line defect OPE defines a ring multiplication on the space of line defects.\\

\insfigsvg{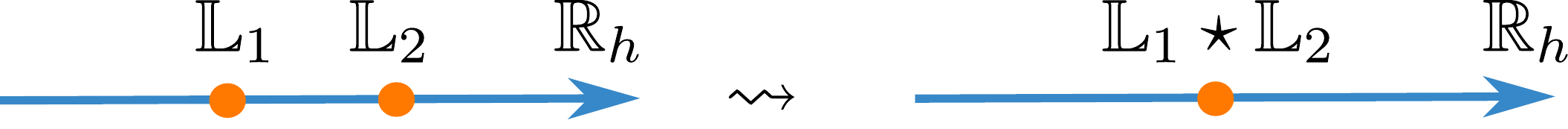}{0.4}{Turning on the Nekrasov-Shatashvili limit of the Omega background restricts the insertion points of line defects to a real line $\R_h$ inside the spatial $\R^3$. The resulting line defect OPE $\star$ is noncommutative.}

The line defect OPE admits a distinguished noncommutative deformation, or 
quantization \cite{Gaiotto:2010be,Cordova:2013bza}. Physically this quantization can be understood as arising from turning on the Nekrasov-Shatashvili limit \cite{Nekrasov:2009rc} of Omega background along a spatial $\R^2$-plane \cite{Ito:2011ea,Yagi:2014toa,Brennan:2018yuj}.\footnote{More precisely, one considers spacetime as the product of a real line $\R_h$ and a twisted $\R^2$-bundle over $S^1$ where the line defect wraps $S^1$.} After turning on such a background supersymmetric line defects are restricted to sit at the origin of the $\R^2$-plane, or equivalently, the insertion points are restricted to a real line $\R_h$ in spatial $\R^3$, perpendicular to the $\R^2$-plane. Due to this restriction of the insertion points, the OPE of line defects $\mathbb{L}_1$ and $\mathbb{L}_2$ becomes noncommutative; we denote it as $\mathbb{L}_1\star \mathbb{L}_2$. This is illustrated in \autoref{fig:LDOPE}.

In a generic 4d $\cN=2$ theory, the noncommutative $\star$ product could be complicated. However, the situation is better in abelian gauge theories, where $1/2$-BPS line defects could be explicitly written down as in \eqref{eqn:abelianline}. Denoting a line defect with electromagnetic charge $\gamma$ as $\mathbb{L}_\gamma$, the noncommutative OPE is concretely given by the following {\it quantum torus} relation:
\begin{equation}\label{eqn:quantumtorus}
\mathbb{L}_{\gamma_1}\star \mathbb{L}_{\gamma_2}=(-q)^{\langle \gamma_1,\gamma_2 \rangle} \mathbb{L}_{\gamma_1+\gamma_2},
\end{equation}
where $\langle \cdot,\cdot\rangle$ denotes the Dirac-Schwinger-Zwanziger pairing on the electromagnetic charge lattice. This quantum torus relation also has a nice physical interpretation. The line defects $\mathbb{L}_\gamma$ could be thought of as worldlines of heavy BPS dyons with charge $\gamma$. Bringing together two dyons with charges $\gamma_{1,2}$ produces a dyon with charge $\gamma_1+\gamma_2$, but one needs to take into account the angular momentum stored in the electromagnetic field sourced by the two dyons, which gives rise to the extra prefactor in \eqref{eqn:quantumtorus}. 

In this paper we focus on the case of class $S$ theories, where line defects and their OPEs have an extra geometric meaning. For a sampling of references see \cite{Drukker:2009tz,Alday:2009fs,Drukker:2009id,Fock-Goncharov,Gaiotto:2010be,Xie:2012dw,Xie:2012jd,Xie:2013lca,Saulina:2014dia,Tachikawa:2015iba,Coman:2015lna,Gabella:2016zxu}. To produce a line defect in a class $S$ theory, one starts with a two-dimensional surface defect in the 6d $(2,0)$ theory of type $\mathfrak{g}$ and wrap it along $\ell \times \R_t$, where $\ell$ is a certain topologically nontrivial one-dimensional trajectory on the Riemann surface $C$ and $\R_t$ is the time direction in 4d spacetime. The surface defect in 6d carries a representation of $\mathfrak{g}$; throughout our discussion in this paper we take $\mathfrak{g}=\fgl(3)$ and consider surface defects in the fundamental representation. Moreover we restrict ourselves to the case where $\ell\subset C$ is a simple closed curve, as indicated in \autoref{fig:classSLD}. The cases where $\ell$ is a {\it lamination} or contains junctions are deferred to future work.

\insfigsvg{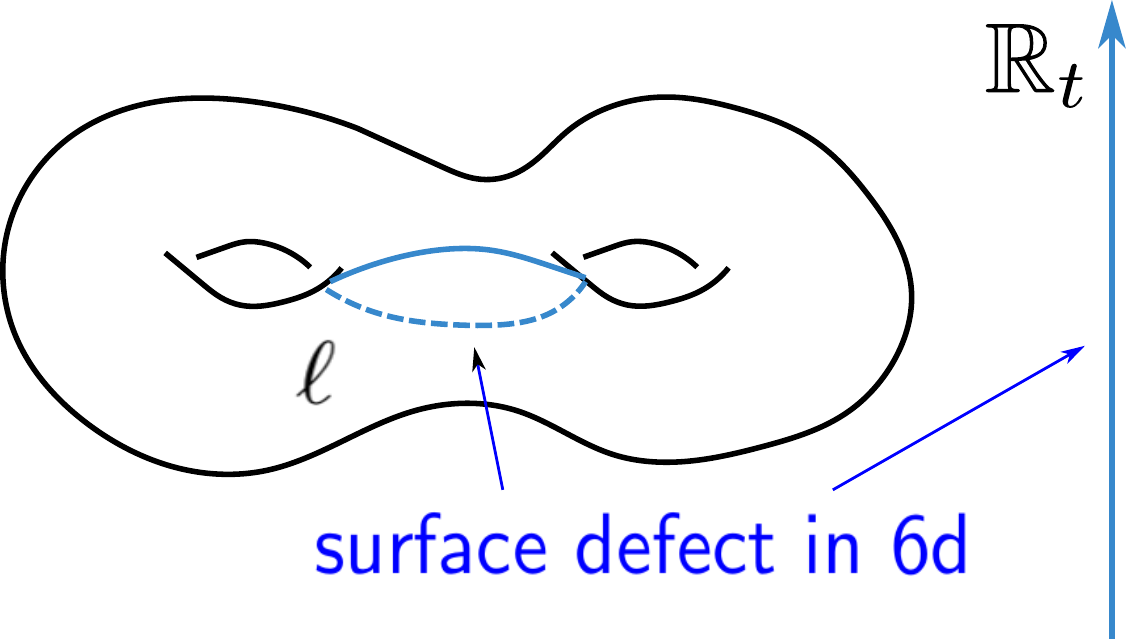}{0.56}{We consider a surface defect in the 6d $(2,0)$ $\fgl(3)$ theory, wrapping a simple closed curve $\ell$ on the Riemann surface $C$ and extending along the $\R_t$ direction. After the compactification on $C$, this produces a line defect extending along the $\R_t$ direction in the 4d class $S$ theory of type $\fgl(3)$.}

At a point on the Coulomb branch of a class $S$ theory, the low energy effective theory is an abelian gauge theory with couplings described by the Seiberg-Witten curve $\tC\subset T^*C$, where $\tC$ is generically a branched covering of the Riemann surface $C$ \cite{Seiberg:19941,Seiberg:19942,Gaiotto:2009we,Gaiotto:2009hg}. The 4d IR effective theory can be viewed as the compactification of 6d $(2,0)$ $\fgl(1)$ theory on $\tC$. In this context, IR line defects can be obtained from wrapping surface defects in the 6d abelian theory along loops $\tilde{\ell}\subset\tC$. From this point of view, the UV-IR map for line defects corresponds to a procedure which takes a loop $\ell$ on the Riemann surface $C$ and lifts it to a combination of loops $\tilde{\ell}_i$ on the branched covering Seiberg-Witten curve $\tC$, where generically $i$ runs over a finite set. This is illustrated in \autoref{fig:UVIRclassical}. This geometric picture suffices for obtaining the integer coefficients / framed BPS indices in the UV-IR decomposition for line defects; indeed this was the strategy used in \cite{Gaiotto:2012rg}. However to obtain the $\Z[q,q^{-1}]$-valued protected spin characters, or equivalently to construct the quantum UV-IR map (\ref{eqn:qUVIR}), this picture is not enough. It turns out instead of just thinking about loops $\ell\subset C$ and $\tilde{\ell}\subset\tC$, we need to consider links $L\subset C\times\R$ and $\tL\subset \tC\times\R$, as we will describe next.

\insfigsvg{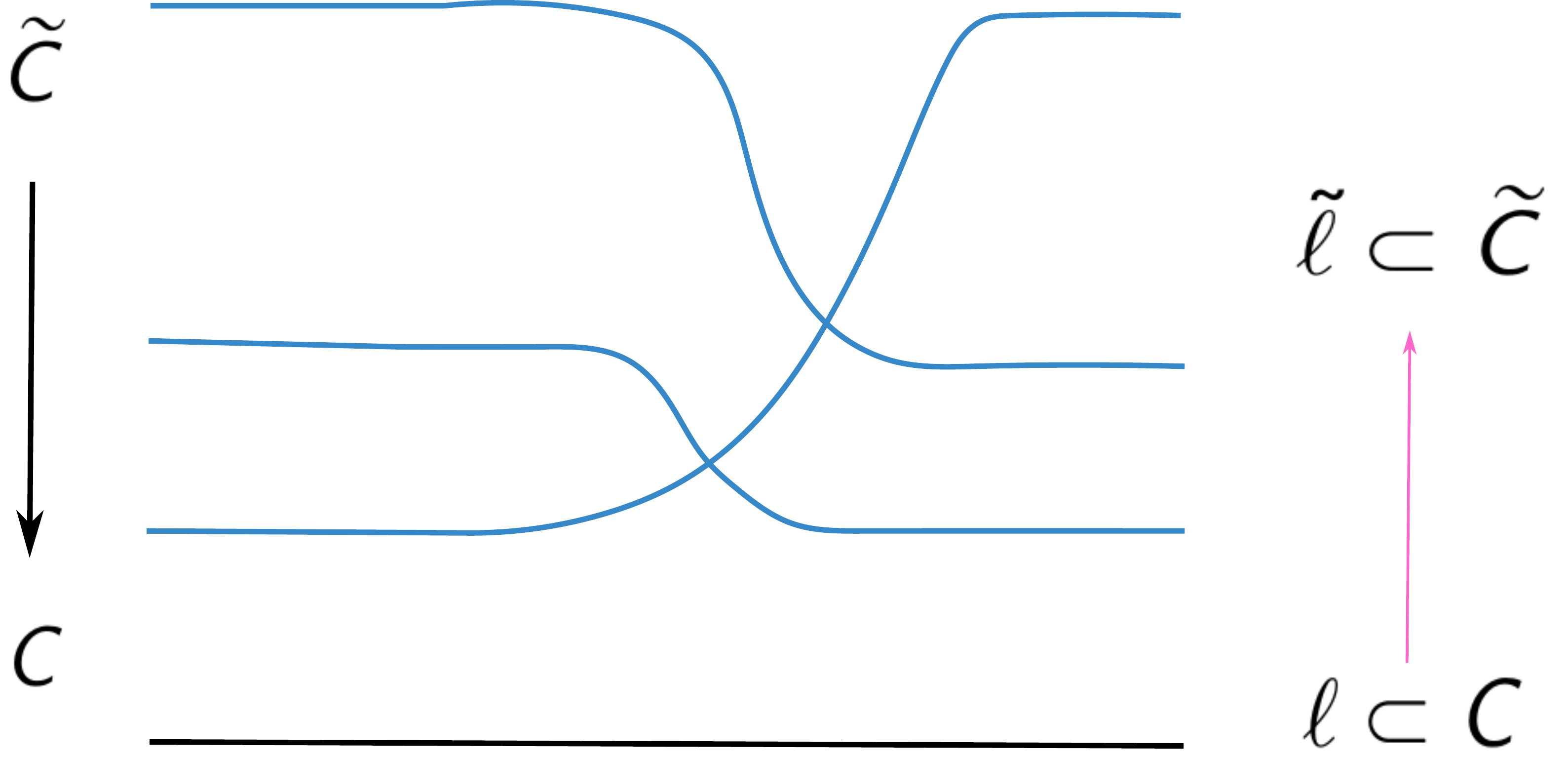}{0.25}{Here we show the Seiberg-Witten curve $\tC$ as a branched covering over the Riemann surface $C$ associated with a 4d class $S$ theory. $1/2$-BPS UV line defects correspond to topologically nontrivial loops $\ell\subset C$ while $1/2$-BPS IR line defects are represented by loops $\tilde{\ell}\subset \tC$. The classical limit ($q=-1$) of the quantum UV-IR map is a procedure which takes $\ell\subset C$ and uplifts it to combinations of $\tilde{\ell}\subset \tC$.}

A crucial ingredient in formulating the quantum UV-IR map is the compatibility with line defects OPE. Geometrically the commutative line defects OPE correspond to the concatenation of loops $\ell\subset C$ in the UV and the concatenation of loops $\tilde{\ell}\subset\tC$ in the IR. On the other hand, the noncommutative line defects OPE are described by {\it skein algebras}. Intuitively speaking, to incorporate the non-commutativity we need to introduce an ordering of loops $\ell \subset C$ and $\tilde{\ell}\subset \tC$ respectively. Instead of simply considering loops on $C$ and $\tC$, we are led to think about links in 3-manifolds with a product structure, namely $M=C\times \R_h$ for the UV lines and $\tM=\tC\times \R_h$ for the IR lines respectively. The non-commutative algebra structure is then given by stacking the links along the $\R_h$ direction. The quantum UV-IR map should send a link $L\subset M$ to combinations of links $\tL\subset \tM$, compatible with the algebra structures on both sides. In other words, the quantum UV-IR map should correspond to a homomorphism of skein algebras.

In the following we recall the definition of skein algebras and skein modules; for a more detailed description we refer to \cite{Neitzke:2020jik}. Fixing an oriented 3-manifold $M$, the $\fgl(N)$ HOMFLY skein module $\Sk(M,\fgl(N))$ is defined as the free $\Z[q^{\pm1}]$-module generated by ambient isotopy classes of framed oriented links in $M$, modulo the submodule generated by the skein relations shown in \autoref{fig:gln-skein-relations}. On the IR side, we consider another oriented 3-manifold $\tM$, where $\tM$ is an $N$-fold covering of $M$ branched along a codimension-2 locus $\cF$. The $\fgl(1)$ skein module with branch locus, 
$\Sk(\tM,\fgl(1))$, is the free $\mathbb{Z}[q^{\pm 1}]$-module generated by ambient isotopy classes of framed oriented links in $\tM \setminus \cF$, modulo the submodule generated by skein relations shown in \autoref{fig:gl1-skein-relations}.

\insfigsvg{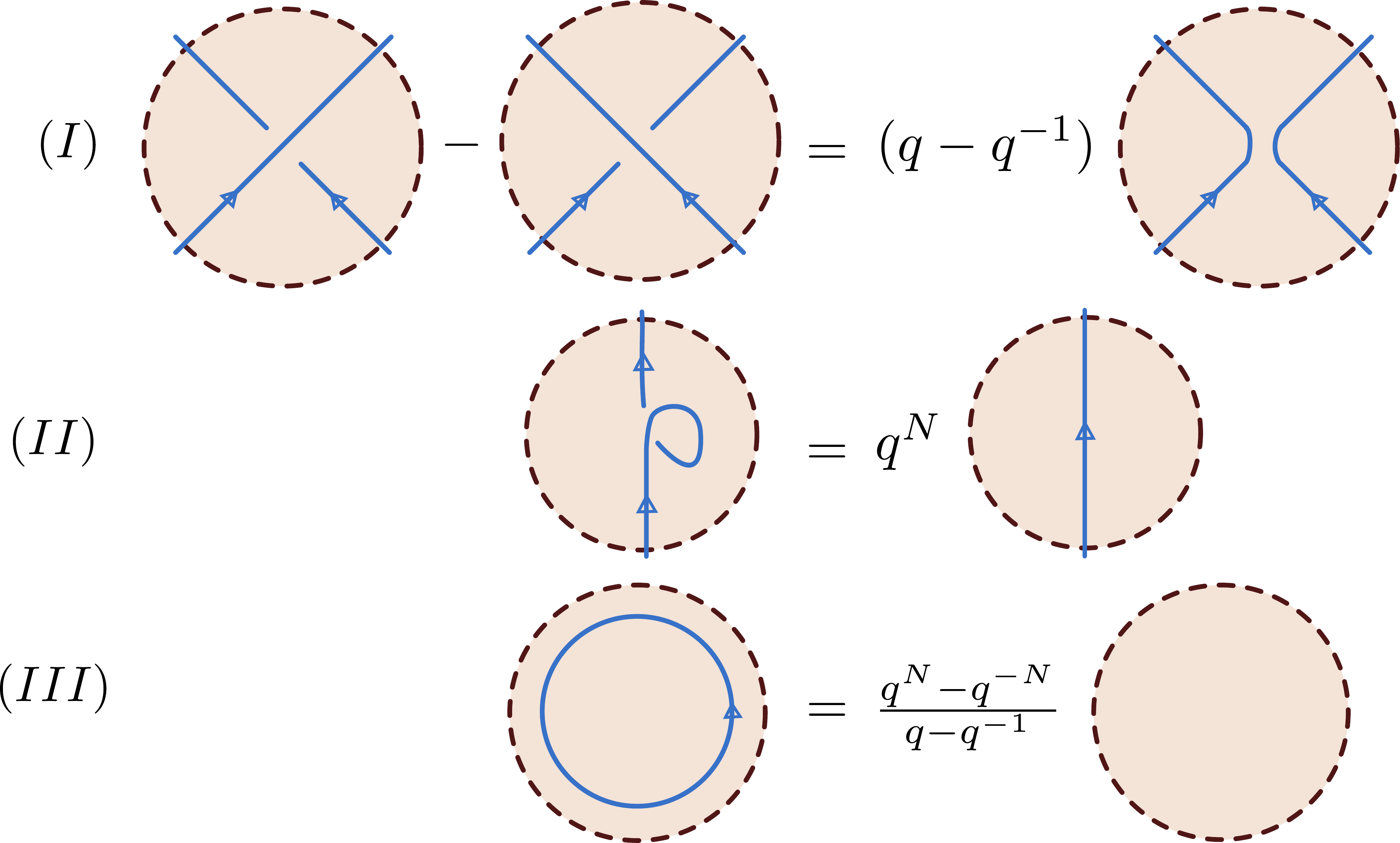}{0.23}{The skein relations defining $\Sk(M,\fgl(N))$ for links in blackboard framing.}

\insfigsvg{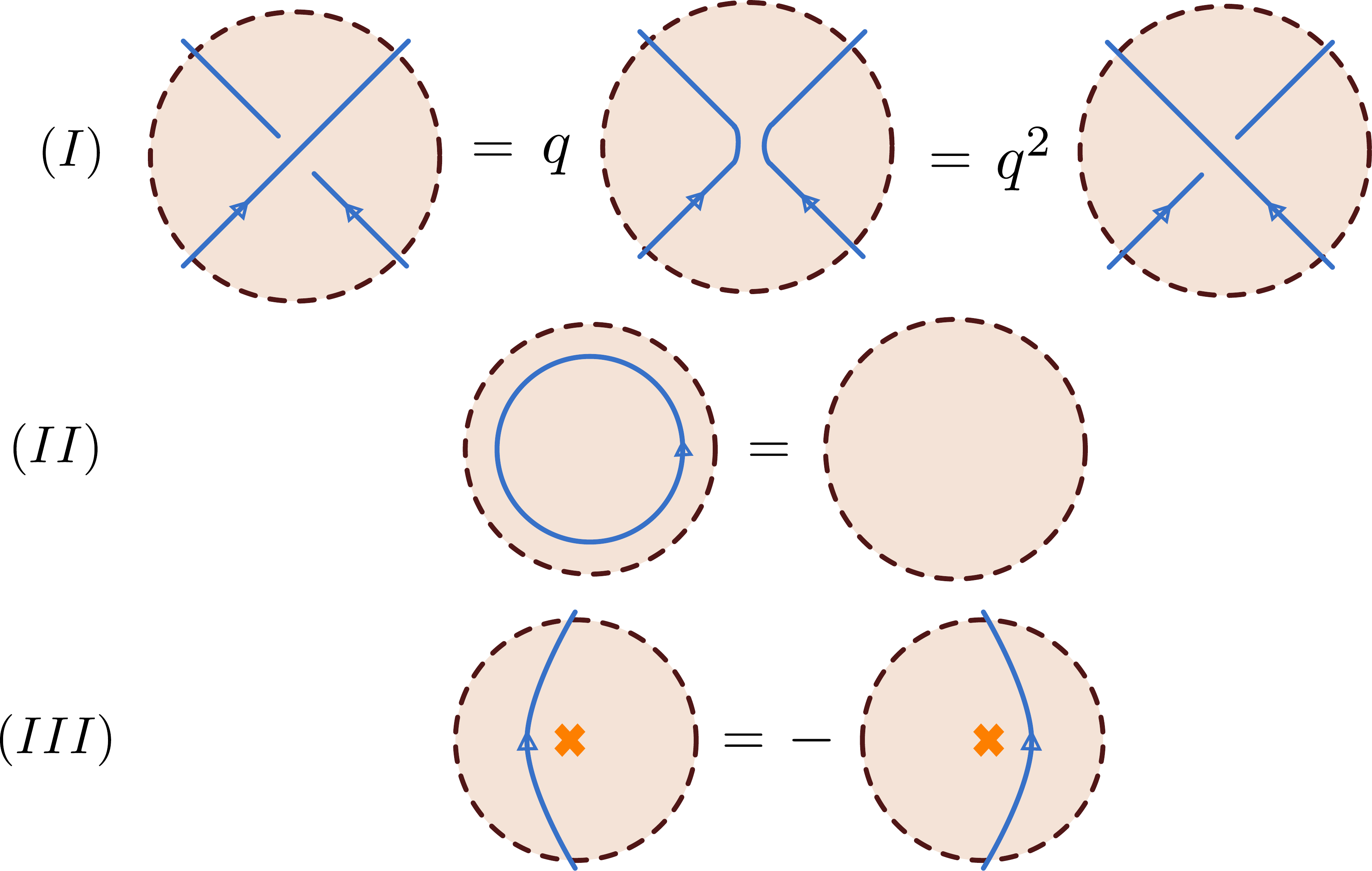}{0.23}{The skein relations defining $\Sk(\tM,\fgl(1))$ for links in blackboard framing. The orange cross represents the codimension-2 branch locus $\mathcal{F}$.}

As described above, in this paper we take $M=C\times\R_h$ and $\tM=\tC\times\R_h$, where $C$ is the Riemann surface associated with a class $S$ theory and $\tC$ corresponds to its Seiberg-Witten curve at a point in the Coulomb branch. Both $C$ and $\tC$ are oriented surfaces; we take the orientation of $M$ (resp. $\tM$) to be the one induced from the orientation of $C$ (resp. $\tC$) and the standard orientation of $\R_h$. Then the skein modules $\Sk(M,\fgl(N))$ and $\Sk(\tM,\fgl(1))$ defined above are actually algebras over $\Z[q^{\pm1}]$, where the multiplication is given by stacking links along the $\R_h$ direction, with a sign twist as explained in \cite{Neitzke:2020jik}. 

The IR skein algebra $\Sk(\tM,\fgl(1))$ is isomorphic to a quantum torus $Q_\Gamma$, where $\Gamma=H_1(\tC,\Z)$ is the IR electromagnetic charge lattice, and the skew bilinear intersection pairing $\IP{\cdot,\cdot}$ on $\Gamma$ 
is identified with the Dirac-Schwinger-Zwanziger pairing.
The quantum torus $Q_\Gamma$ is a $\Z[q^{\pm1}]$-algebra with a basis $\{X_\gamma\}_{\gamma\in\Gamma}$ obeying the product law
$$ X_\gamma X_{\gamma'} = (-q)^{\IP{\gamma,\gamma'}} X_{\gamma + \gamma'}. $$
Each $X_\gamma$ is represented by a certain loop on $\tM$ in homology class $\gamma$;
we refer the readers to Section 3 of \cite{Neitzke:2020jik} for details. This isomorphism is consistent with the physical picture we described above: the noncommutative line defect OPE is described by the skein algebra; 
identifying $X_\gamma$ with the IR line defect $\mathbb{L}_\gamma$, this product law exactly matches the noncommutative IR dyonic line defect OPE (\ref{eqn:quantumtorus}).

We started this section describing $1/2$-BPS line defects in class $S$ theories corresponding to loops $\ell$ on the Riemann surface $C$; considerations regarding their noncommutative OPE led us to skein algebras generated by links $L$ in the 3-manifold $M=C\times\R_h$. If $M$ were a compact hyperbolic 3-manifold, this would be the setup to study line defects in the 3d $N=2$ theory $T[M]$ obtained by the twisted compactification of 6d $(2,0)$ theory on $M$, where the line defects descend from two-dimensional surface defects in the 6d theory wrapping links $L \subset M$ \cite{Ooguri:1999bv,Dimofte:2011ju,Gadde:2013wq,Dimofte:2013lba,Chun:2015gda,Gang:2015bwa,Gukov:2015gmm,Gukov:2017kmk}. 

However, here we are taking the 3-manifold $M$ to be $C\times\R_h$ and performing the twisted compactification of the 6d $(2,0)$ theory on $C$ to obtain 4d $\cN=2$ class $S$ theories. A reasonable question to ask is: what are the defect objects in class $S$ theories that geometrically correspond to links $L$ in $M=C\times \R_h$? Strictly speaking such defects are strips with a finite width, as the links $L$ have a finite extent along the $\R_h$-direction, which is one of the spatial directions in 4d. However if we go to a length scale which is much larger than the finite extent of the strips, we effectively obtain line defects $\mathbb{L}$ in 4d class $S$ theories.\footnote{It would be great to have a gauge-theoretic description of these objects in class $S$ theories that admit a Lagrangian description. In this paper we do not pursue this direction. We thank Yifan Wang for helpful discussions regarding this point.} Such line defects explicitly break the $SU(2)_P$ spatial rotation symmetry to $U(1)_P$ and only preserve $U(1)_R\subset SU(2)_R$ R-symmetry. Moreover, these effective line defects $\mathbb{L}$ preserve 2 supercharges of the bulk theory, i.e. they are $1/4$-BPS. For these $1/4$-BPS effective line defects, we can still define the protected spin character $\fro(\mathbb{L},u,\gamma,q)$ as in (\ref{eqn:PSC}), counting  ground states with spin of the bulk-defect system. Moreover, the generating function of the protected spin characters is again computed by the quantum UV-IR map.

\subsection{Strategy to construct the quantum UV-IR map}\label{sec:strategyF}

Thanks to the geometric construction of supersymmetric line defects in class $S$ theories, the quantum UV-IR map also has a physical meaning in 6d, as a UV-IR map for supersymmetric surface defects in two different $(2,0)$ theories. In the UV, we have 6d $(2,0)$ superconformal field theory of type $\fgl(N)$ on $M\times \R^{2,1}$, with insertion of a supersymmetric surface defect (in the fundamental $\fgl(N)$-representation) on $\R^{0,1} \times L$, where $L$ is a link in $M=C\times\R$. Deforming to the tensor branch of the $\fgl(N)$ theory, in the IR we then have 6d $(2,0)$ theory of type $\fgl(1)$ on $\tM\times \R^{2,1}$, with insertion of supersymmetric surface defects on $\R^{0,1}\times \tL$, where $\tL$ are links in $\tM=\tC\times\R$ and $\tC$ is the Seiberg-Witten curve. From this point of view, we expect the quantum UV-IR map to be a map sending links $L\subset M$ to combinations of links $\tL\subset \tM$.

It is illuminating to think about the 5d picture. We Euclideanize and compactify the time direction to $S^1$ with an extra insertion of $(-q)^{2J_3}q^{2I_3}$, and we reduce on the $S^1$. In the UV we then have a $\Omega$-deformed\footnote{The relation between $q$ and the $\Omega$-deformation parameter $\epsilon$ is $q=\text{e}^{R\epsilon}$ where $R$ is the radius of $S^1$.} 5d $\cN=2$ $U(N)$ super Yang-Mills (SYM) on $(C\times\R) \times \R^2_\epsilon$ with the class $S$ twist on $C$ (as introduced in \cite{Gaiotto:2009hg}), with the insertion of a fundamental Wilson line along $L$. Going to a point in the Coulomb branch corresponds to turning on VEV for a complex adjoint scalar $\Phi:=\Phi_1+\text{i}\Phi_2$, where $\Phi_{1,2}$ are two real adjoint scalars in the 5d $\cN=2$ $U(N)$ SYM charged under the $so(2)_R\subset so(5)_R$ R-symmetry that gets identified with the $so(2)_C$ holonomy algebra in the class $S$ twist. In particular, eigenvalues of $\Phi$ correspond to the $N$ sheets of the Seiberg-Witten curve $\tC$; generically $\Phi$ breaks the 5d gauge group $U(N)$ to $U(1)^N$.

In this context, the quantum UV-IR map $F(L)$ in principle could be understood through computing the partition function of the $\Omega$-deformed twisted 5d $N=2$ $U(N)$ SYM on $(C\times\R)\times\R^2_\epsilon$, in presence of a fundamental Wilson line along the link $L$, and in the symmetry-breaking background determined by the Seiberg-Witten curve $\tC$. It would be great to derive the UV-IR map using localization methods in this setup. We defer this to future work.

Our current strategy is to combine the above physical picture with the fact that $F$ is a homomorphism between two skein algebras as written in (\ref{eqn:Fhom}). Our construction can be understood as a two-step process:
\begin{itemize}
	\item Step 1: Given a link $L\subset M$, we enumerate all possible $\tL\subset\tM$ that contribute in the UV-IR map. Physically speaking, $\tL$ correspond to Wilson lines in the IR 5d theory. Constructing all possible $\tL$ requires semi-classical analysis plus consideration of instanton-like massive $W$-boson corrections.\footnote{From the 6d (2,0) theory point of view, these $W$-bosons correspond to dynamical BPS strings on the tensor branch of the (2,0) theory.}
	
	\item Step 2: To each $\tL$, we assign a weight factor $\alpha(\tL) \in \Z[q^{\pm 1}]$. These weight factors are worked out by requiring that the quantum UV-IR map $F$ preserves the skein relations and should be isotopy-invariant.
\end{itemize}

We report the details of our construction in \autoref{sec:qab-general}, with a brief description in \autoref{sec:bootstrapweightfactor} of the bootstrap-like methods to determine weight factors $\alpha(\tL)$. The construction of quantum UV-IR map requires an important ingredient. In the 5d $\cN=2$ theory context, the massive $W$-bosons have to be mutually BPS with the fundamental Wilson line inserted along $L$. As a result, the $W$-bosons have to travel along particular trajectories called the {\it WKB leaves}. We will describe the WKB leaves and the associated WKB foliation structure on $C$ in \autoref{sec:wkb-foliations}.

\subsection{WKB foliations} \label{sec:wkb-foliations}

The quantum UV-IR map depends on some extra structure on $C$, which we call WKB foliation data. We fix a complex structure on $C$ and a N-fold covering $\tC \to C$ given in the form
\begin{equation} \label{eq:spectral-curve}
\tC = \left\{\lambda:\lambda^N+\sum\limits_{k=1}^N\phi_k\lambda^{N-k}=0 \right\} \subset T^*C,
\end{equation}
where each $\phi_k$ is a meromorphic $k$-differential on $C$. Moreover we choose generic $\phi_k$ such that all branch points are simple.

Locally on $C$ we then have $N$ $1$-forms $\lambda_i$, given by the sheets
of $\tC$, i.e. the branches of solutions to \eqref{eq:spectral-curve}. 
We define $\binom{N}{2}$ foliations locally on $C$
using pairs of these one-forms: for example $ij$-leaves are the paths on $C$ along which $\lambda_i-\lambda_j$ is real.\footnote{In general one can specify a phase $\vartheta$, and the condition changes to requiring $\text{e}^{-\text{i}\vartheta}(\lambda_i-\lambda_j)$ is real. We stick to $\vartheta = 0$ in this paper.} In the context of the 5d $\cN=2$ theory described previously in \autoref{sec:strategyF}, these $ij$-leaves are potential trajectories of the $ij$-type $W$-bosons, where the above conditions is equivalent to the requirement that such $W$-bosons preserve the supercharges preserved by the fundamental Wilson line inserted along $L$.\footnote{In this paper, without loss of generality we have fixed a choice of the supercharges preserved by the Wilson line along $L$.} Although these leaves are not naturally oriented, choosing a sheet
induces an orientation as follows: for an $ij$-leaf suppose we choose sheet $i$, then the positive direction is the direction for 
which the pairing between $\lambda_i-\lambda_j$ and the tangent vector is negative. Thus the lift of an $ij$-leaf to either sheet $i$ or sheet $j$ is naturally oriented.
\insfigsvg{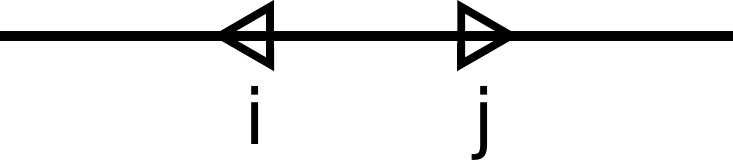}{0.42}{An $ij$-leaf with its two orientations.}

Recall that for $N=2$, locally at a branch point there is a three-pronged singularity as shown in \autoref{fig:foliation-zero}:\footnote{To understand this three-pronged structure
note that around a branch point at $z=0$ we have $\lambda^{(i)} - \lambda^{(j)} \sim c z^{\frac12} \, \de z$, so $w^{(ij)} \sim c z^{\frac32}$.}
\insfigsvg{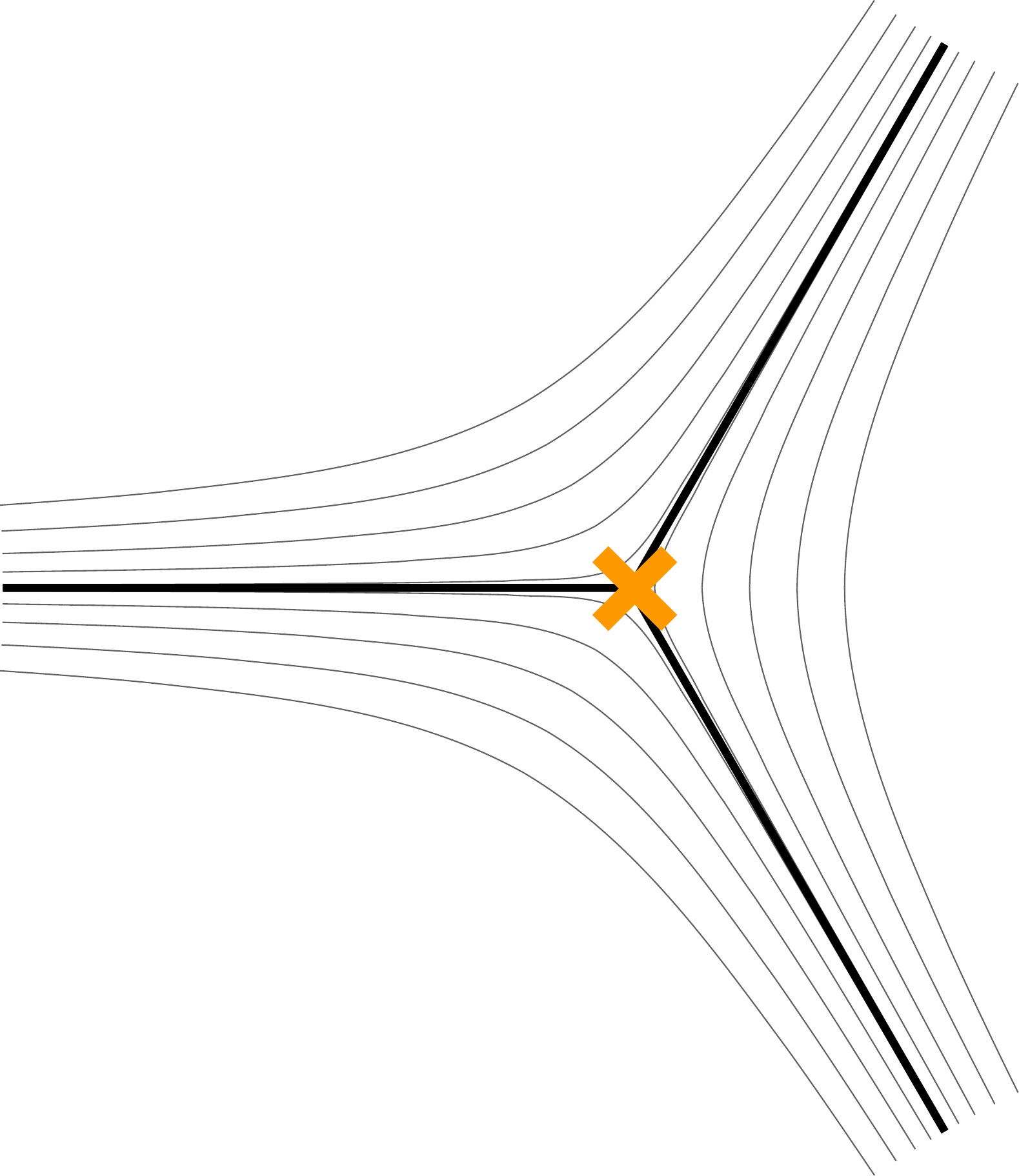}{0.25}{The local structure of the WKB foliation for N=2 around
a branch point of $\tC \to C$. The branch point is represented by an orange cross. 
The dark lines represent critical leaves, while the lighter lines are generic leaves.}

The WKB foliations around a simple branch point for $N>2$ 
look considerably more complicated than for $N=2$.
A typical picture in the neighborhood
of a simple branch point for the $N=3 $ case is shown in \autoref{fig:branch-foliations-N3} below.
\insfigsvg{branch-foliations-N3}{0.05}{Left: The WKB foliations near a simple branch point when $N=3$. The branch point is the orange cross in the center of the figure. The three black curves emanating from the branch point are the critical leaves. The red line marks the ``caustic'' where the three local foliations become tangent. The blue and yellow leaves were generated by fixing a 7 by 8 rectangular grid of points, drawing the three leaves through each of these 56 points, then coloring them blue or yellow according to their $ij$-types. Right: A zoomed-out view showing the larger-scale behavior of these leaves.}

Suppose locally we label the sheets $i = 1,2,3$ so that the branch point is of type $(12)$.
Then the $12$-leaves, colored blue in \autoref{fig:branch-foliations-N3}, give a single well defined foliation around the branch point, looking just like the foliation around a branch point in the $N=2$ case. The $13$-leaves and $23$-leaves, on the other hand, mix with one another under
monodromy around the branch point. There is also another important
phenomenon: there is a curve emanating from the branch point, called a ``caustic'',
along which the $\lambda_i - \lambda_j$ all have the same phase.
Therefore along the caustic all three $ij$-foliations are all tangent to 
one another (though not tangent to the caustic itself!)

A good local model for this behavior is
\begin{equation}
\lambda_1 = a + \sqrt{z}, \quad \lambda_2 = a - \sqrt{z}, \quad \lambda_3 = b
\end{equation}
for constants $a, b \in \C$; then
\begin{equation}
\lambda_1 - \lambda_2 = 2 \sqrt{z}, \quad \lambda_1 - \lambda_3 = (a-b) + \sqrt{z}, \quad \lambda_2 - \lambda_3 = (a-b) - \sqrt{z}.
\end{equation}
In particular, to first approximation near the branch point
the $13$-leaves and $23$-leaves are parallel, and point in a generic 
direction in the plane. In a sense, this local model is similar to the $N=2$ case. The difference is that we now have two more types of leaves.\\

For $N>2$ the global topology of WKB foliations is more complicated than for $N=2$, and we will not try to 
discuss it in full detail. Instead we just point out some general features:
\begin{itemize}
	\item In a neighborhood of a generic point of $C$, there are $\binom{N}{2}$ distinct
	foliations. Globally, these
	foliations mix with one another, since the monodromy around branch points permutes the
	sheets. Around a branch point where sheets $i$ and $j$ collide, the
	local structure of the $ij$-foliation is three-pronged as in \autoref{fig:foliation-zero}.

	
	\item For any $i$, $j$, $k$ the local structure around a generic point of $C$ looks 
	topologically (but not necessarily conformally) like
	\autoref{fig:ijk-foliations}.
	\insfigsvg{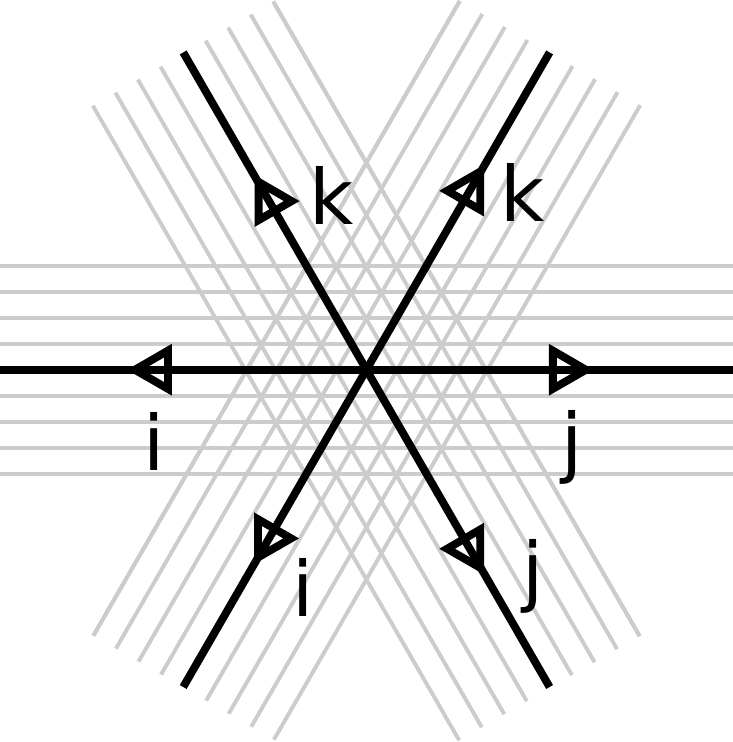}{0.5}{The local structure of the WKB 
	$ij$-, $jk$- and $ki$-foliations around a generic point of $C$.}

	\item 
	Locally on $C$,
	dividing by the equivalence relation that identifies points lying on the same $ij$-leaf,
	one obtains the \ti{leaf space} $C_{ij}$. Thus locally we have $\binom{N}{2}$ 
	projection maps
	$p_{ij}: C \to C_{ij}$, where each $C_{ij}$ is a 1-manifold.
	Each $C_{ij}$ also comes with a natural Euclidean structure induced by the 1-form 
	$\abs{\mathrm{Im}(\e^{-\I \vartheta} \dot{w}^{(ij)})}$.
	Globally, the $C_{ij}$ mix with one another. 
\end{itemize}

The WKB foliations of $C$ induce foliations of $M = C \times \R$, again with 1-dimensional leaves: the leaves on $M$ have the form $\ell\times \{x^3=c\}$ where $c$ is any constant and $\ell\subset C$ is a leaf.
As before, the leaf spaces 
$M_{ij} = C_{ij} \times \R$ inherit natural Euclidean structures, so locally
they look like patches of $\R^2$. Locally choosing a sheet $i$ induces an orientation on $M_{ij}$. The induced orientation is determined by the orientation of $ij$-leaves on sheet $i$, together with the ambient orientation of $M$. We take it to be opposite to the quotient orientation. 

\subsection{The quantum UV-IR map for \texorpdfstring{$N =3$}{N =3}} \label{sec:qab-general}

Following the strategy described in \autoref{sec:strategyF}, we have constructed the quantum UV-IR map in the case of $N=2$ in \cite{Neitzke:2020jik}. Here we describe the quantum 
UV-IR map in the case of $N=3$.

For any framed oriented link $L$ in $M=C\times\R_h$ with standard framing\footnote{Throughout this paper we adopt the standard framing for links where the framing vector points along the $\R_h$ direction.}, $F(L)$ is given by a sum of the form
\begin{equation}\label{q-ab-eqn}
F(L)=\sum\limits_{\tL} \alpha(\tL) \tL.
\end{equation}
In this sum, 
$\tL$ runs over all framed oriented links in $\tM=\tC\times\R_h$ built out of the following types of local pieces:
\begin{itemize}
	\item {\it Direct lifts} of a strand of $L$ to a single sheet of $\tM$, equipped with the standard framing. Each sufficiently short strand admits $N$ direct lifts.
	\insfigsvg{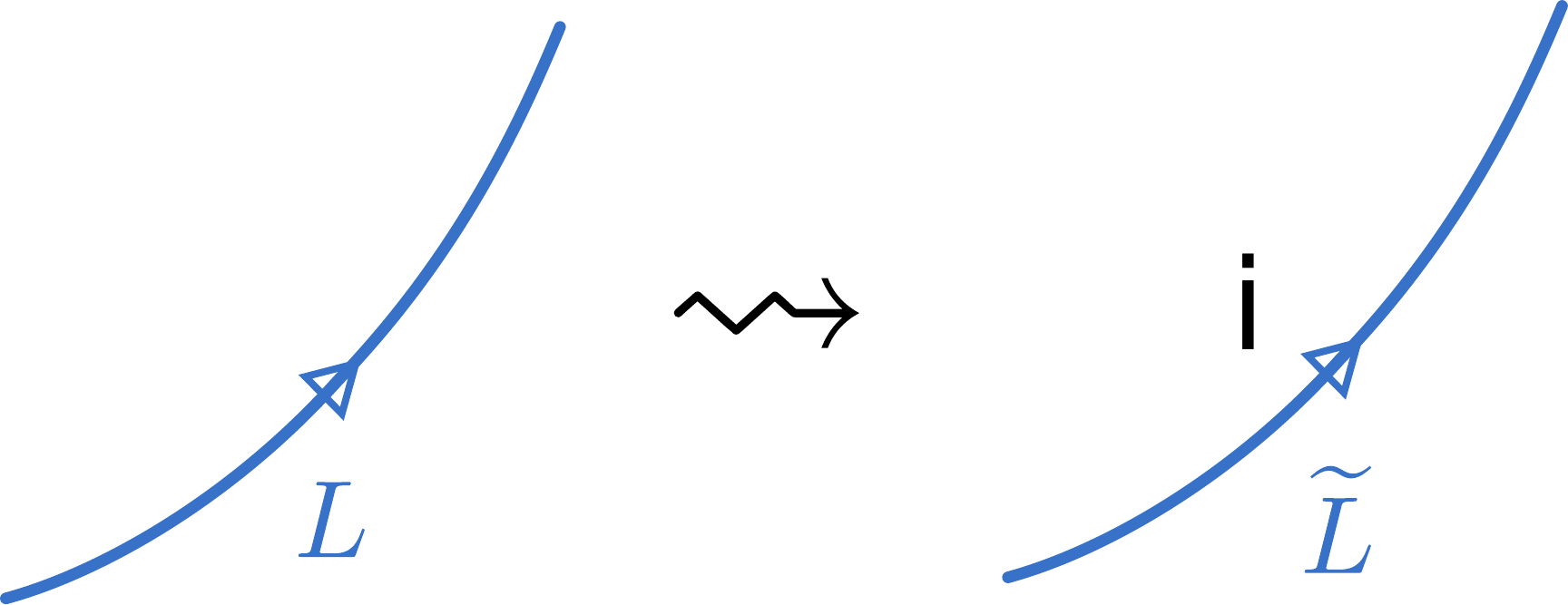}{0.27}{The direct lift of a segment of $L$ to sheet $i$ of the covering $\tM$.} 
In the 5d $N=2$ SYM setup, these lifts can be understood as follows. Locally away from the branch locus of $\tC\to C$, the gauge group $U(N)$ is broken to $U(1)^N$. Correspondingly the fundamental representation of $U(N)$ is split into its $N$ weight spaces. Therefore locally a fundamental Wilson line in the $U(N)$ SYM is expected to decompose into $N$ Wilson lines in the $U(1)^N$ theory.

\item \ti{Lifted webs} defined as follows. Let a \ti{web} be a collection of 
	$ij$-leaf segments on $M$, where each $ij$-segment ends either at an $ij$-branch point, 
	at a point of the link $L$, or at a trivalent junction, as shown in \autoref{fig:segment-ends}.

\insfigsvg{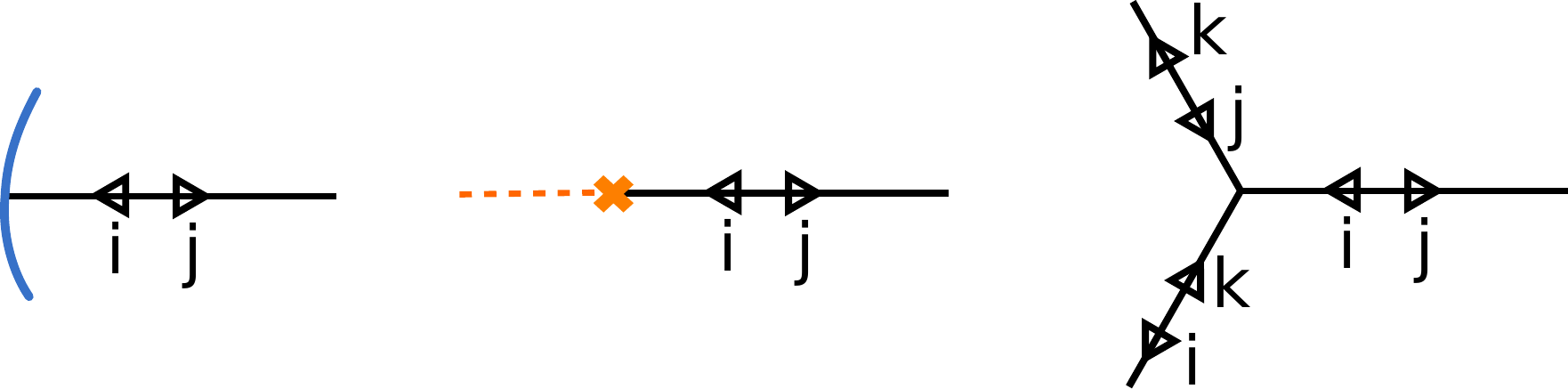}{0.5}{Three ways in which an $ij$-segment of a web can end:
on a link segment, a branch point, or a junction.}

A lifted web is a $1$-chain on $\tM$, 
obtained by lifting each $ij$-segment in the web to sheets $i$ and $j$ of $\tM$, equipping each lift with its canonical orientation and standard framing. 

In the 5d $\cN=2$ SYM setup, lifted webs correspond to trajectories of massive $W$-bosons, which had been integrated out in reducing to the abelian description. To be mutually BPS with the Wilson lines, such $W$-bosons must travel along $ij$-leaves as described in \autoref{sec:wkb-foliations}. Lifted webs represent corrections from massive $W$-bosons to the semiclassical picture involving only direct lifts.

Simple examples of lifted webs are the detours and exchanges which already appeared in the $N=2$ case in \cite{Neitzke:2020jik}:
\begin{itemize}
\item If the web consists of a single $ij$-segment connecting an $ij$-branch point to the link $L$,
then we get a detour as shown in \autoref{fig:detour-web}.
	\insfigsvg{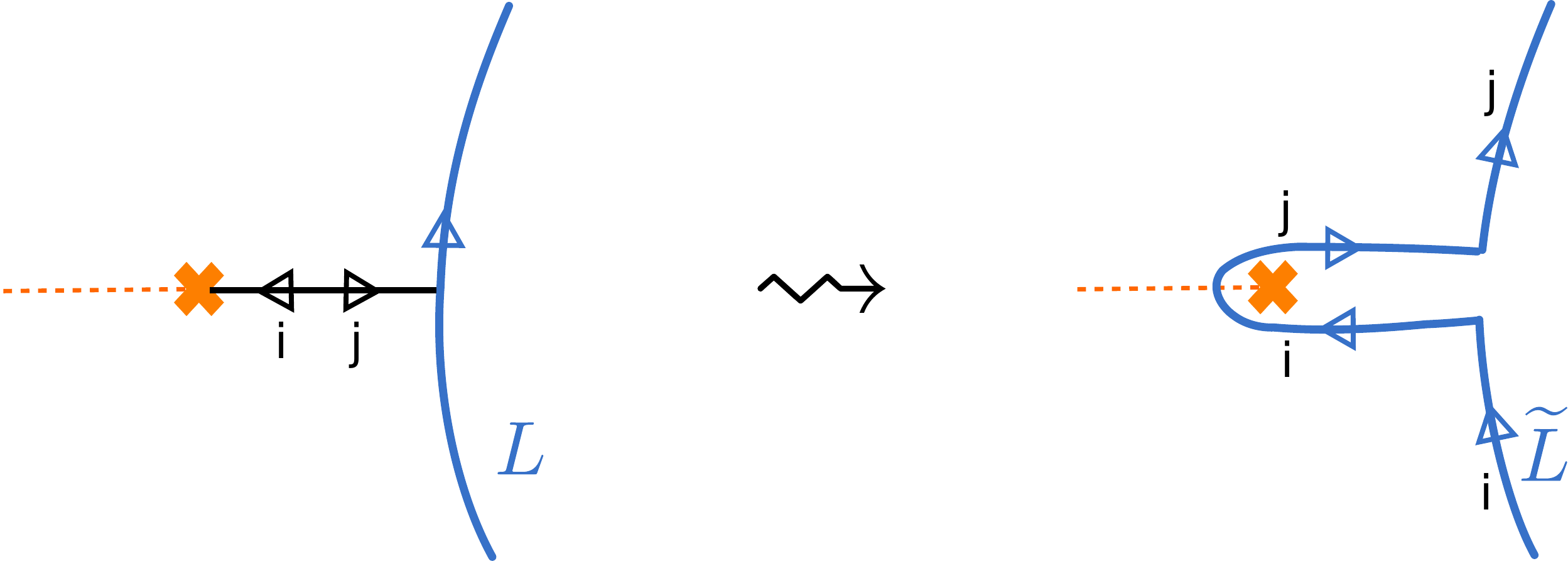}{0.3}{Left: A web (shown in black) consisting of a single $ij$-segment with one end
	on a branch point and one end on the link $L$. Right: A lift $\tL$ containing
	the lift of the web.}
	
\item If the web consists of a single $ij$-segment connecting two different points of the link $L$,
then we get an exchange as shown in \autoref{fig:exchange-web}.
(The term ``exchange'' comes from the fact that the two strands of $\tL$ exchange
their labels $i \leftrightarrow j$ as they go across the web.)
	\insfigsvg{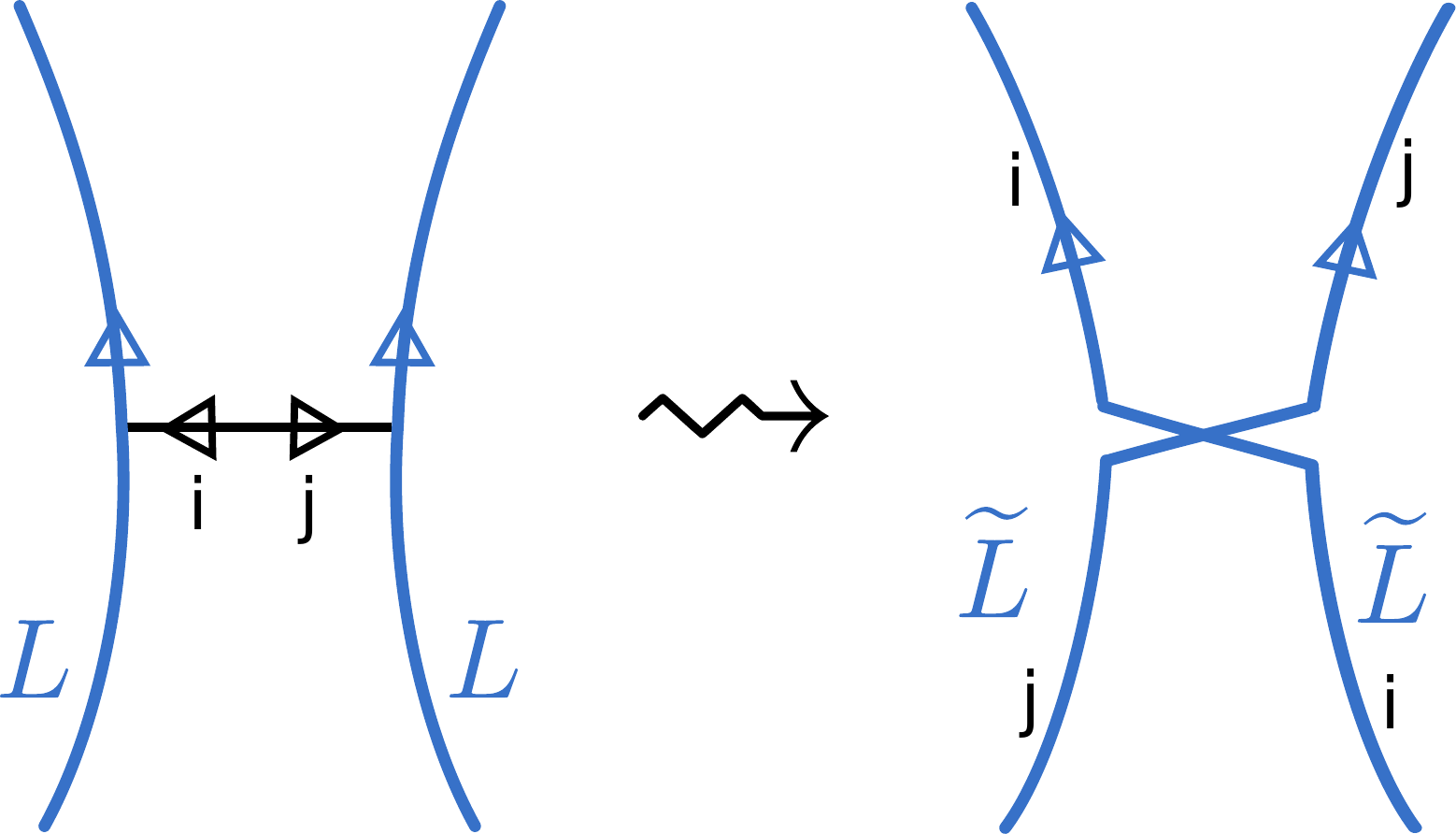}{0.3}{Left: A web (shown in black) consisting of a single $ij$-segment with both 
	ends on the link $L$. Right: A lift $\tL$ containing
	the lift of the web.}

\end{itemize}
An example of a more generic web is illustrated in \autoref{fig:web-example}.

\insfigsvg{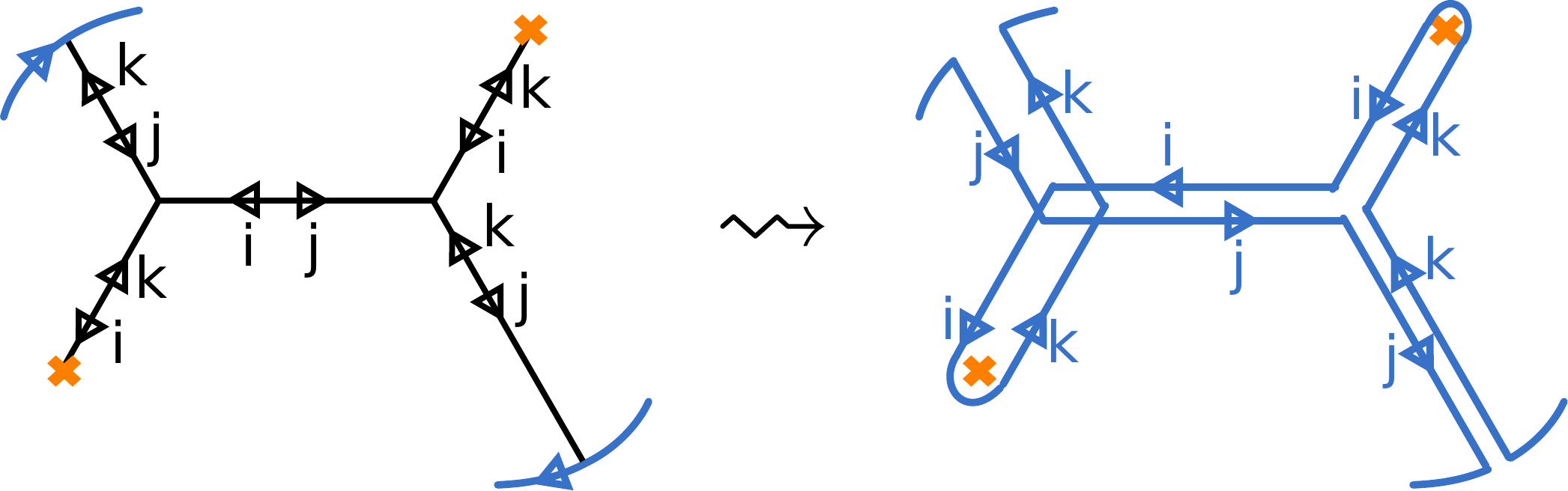}{0.45}{Left: A more elaborate web, with two endpoints on branch points and two on the link $L$. Right: A portion of a lift $\tL$ of $L$ containing the lifted web.}
\end{itemize}

After we enumerate all possible $\tL$ by assembling direct lifts and lifted webs, we move on to the second and more complicated step of our construction: to each $\tL$, we associate a weight factor $\alpha(\tL)\in\Z[q^{\pm1}]$.
The weight $\alpha(\tL)$ is 
built as a product of local factors of various kinds, which we describe
in \autoref{sec:tangency-factors}-\autoref{sec:loop-factors} below.

In \cite{Neitzke:2020jik} we have also formulated the quantum UV-IR map for $N=2$ in a covariant way, which would make sense on a general 
3-manifold $M$, without the product structure $M=C\times\R_h$. It would be desirable to formulate the quantum 
UV-IR map for $N=3$ in a similarly covariant 
way, but we do not do that in this paper. In the following we let $x_3$ denote the coordinate along $\R_h$.

\subsubsection{Tangency factors} \label{sec:tangency-factors}

At every place where the projection of $\tL$ onto $C$ becomes tangent to an $ij$-leaf, if $\tL$ is on sheet $i$ or $j$, we get a contribution $q^{\pm\frac{1}{2}}$ to $\alpha(\tL)$, with the sign determined by  \autoref{fig:framing-factor-gln}. 
An exception arises for segments of $\tL$ consisting of lifted $ij$-leaves;
such a segment is everywhere tangent to the leaf, and we do not include any
tangency factor associated to such a segment. 

Unlike the $N=2$ case considered in \cite{Neitzke:2020jik}, these tangency factors really depend on the particular lift $\tL$ we consider; for example, for a strand of $\tL$ on sheet $1$, we only consider contributions from the tangencies to the $12$- and $13$-leaves, not the
	$23$-leaves.
	\insfigsvg{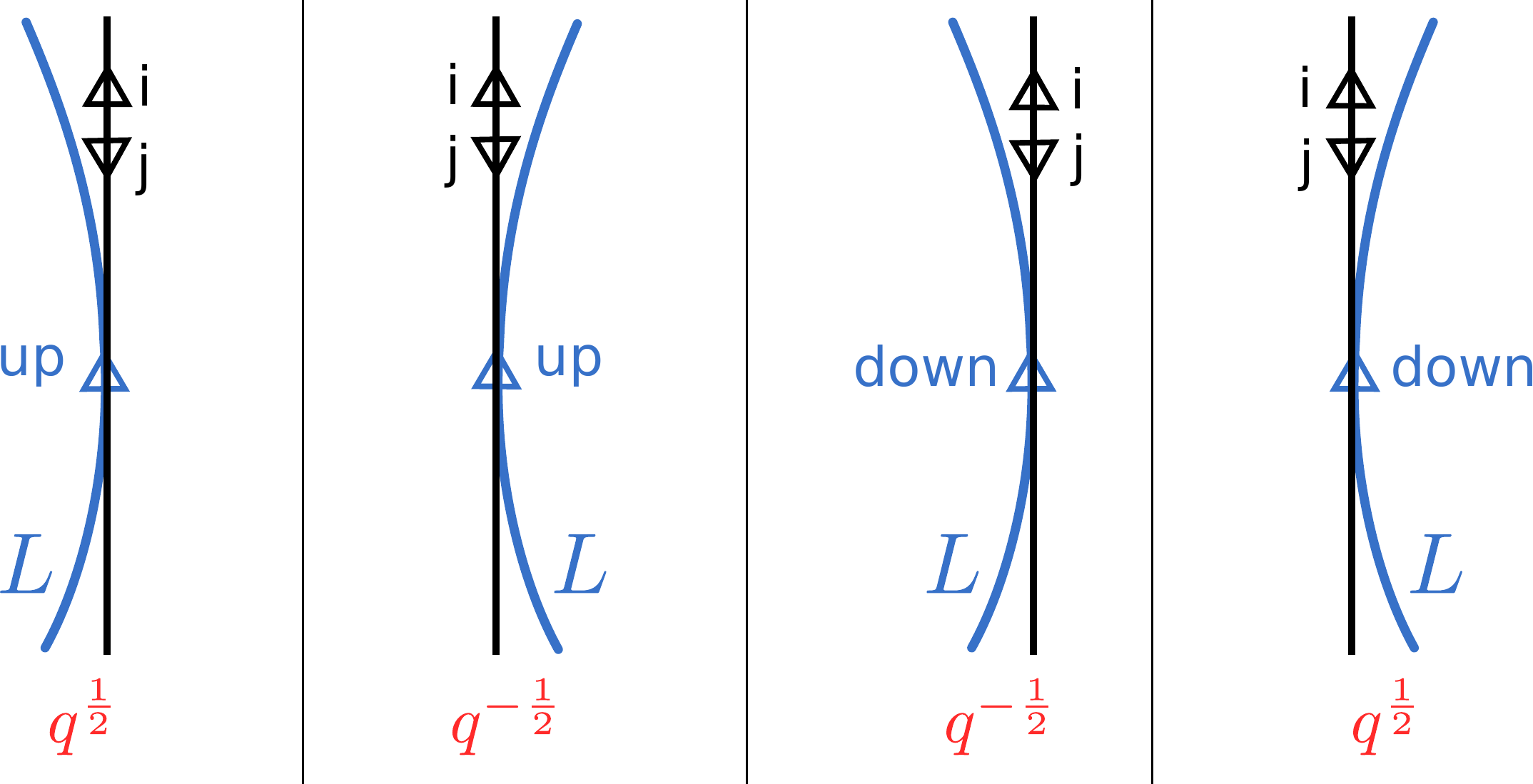}{0.3}{Tangency factors contributing to the overall weight $\alpha(\tL)$. The black line denotes a WKB leaf of type $ij$. Here the word ``up'' or ``down'' next to a segment of $L$ indicates the behavior
	in the $\R_h$-direction, which is perpendicular to the paper.}

\subsubsection{Winding factors} \label{sec:winding-factors}

There is a contribution $q^{w(\tL)}$ to $\alpha(\tL)$, where $w(\tL)$ denotes
	the total leaf space winding of $\tL$, defined as follows. We divide $\tL$
	into small arcs; let $a$ denote such an arc. Suppose arc $a$ is on sheet $i$. The arc $a$ projects to an arc in each of the $2$ leaf spaces $M_{ij}$ 
	with $i \neq j$. In a neighborhood of the image of $a$,
	each $M_{ij}$ is equipped with an orientation and 
	flat metric, as we have discussed in \autoref{sec:wkb-foliations} above. 
	Thus we can define the winding $w_{ij}$ of $a$ in each of the leaf spaces $M_{ij}$. We define
	the total winding of $a$ to be $\sum_{i \neq j} w_{ij}$. Finally, summing up the total winding over all the arcs $a$ we get the total winding $w(\tL)$ of $\tL$.

\subsubsection{Weight factors for exchanges}\label{sec:exchange-weights}

If $\tL$ includes the lift of an exchange which does not cross a caustic, 
then there is an additional
contribution to $\alpha(\tL)$, of the form $\pm q^m( q - q^{-1})$ with $m\in\Z$.
The precise factor is the product of two pieces, as follows.

	\begin{itemize}
		\item
	The first factor depends on two pieces of data: 
	\begin{itemize}
		\item whether the two
	legs of $L$ cross the exchange in the same direction or in opposite
	directions when viewed in the standard projection $C \times \R \to C$,
	\item 
	whether the crossing in the $ij$-leaf space projection of $L$ is an overcrossing
	or an undercrossing.
	\end{itemize} 
	Sample configurations of the four possible 
	types, and the corresponding factors, are listed in \autoref{fig:exchange-factor}. 

	This piece of the exchange factor is very similar to the exchange factor for $N=2$
	described in \cite{Neitzke:2020jik}; the only difference is that the prefactors $q^{\pm 1}$ which appeared there are replaced here by $q^{\pm 2}$. 

	To avoid confusion, we remark that in \autoref{fig:exchange-factor} we have only shown the $ij$-leaf space projection up to certain rotations. In particular one can adjust the height tendencies (indicated as the $e_3$-direction in \autoref{fig:exchange-factor}) of the two strands, as long as the handedness of the crossing doesn't change; this kind of adjustment does not change the assigned factor. Another remark is that \autoref{fig:exchange-factor} only shows the projections for link strands in $C\times \R$; after applying our path-lifting rules there are link strands in $\tC\times \R$ going along lifted $ij$-leaves as indicated in  \autoref{fig:exchange-web}. We also refer to Figure 19 of \cite{Neitzke:2020jik} for projections in $\tC\times\R$ of the configurations shown here in \autoref{fig:exchange-factor}.
	\insfigsvg{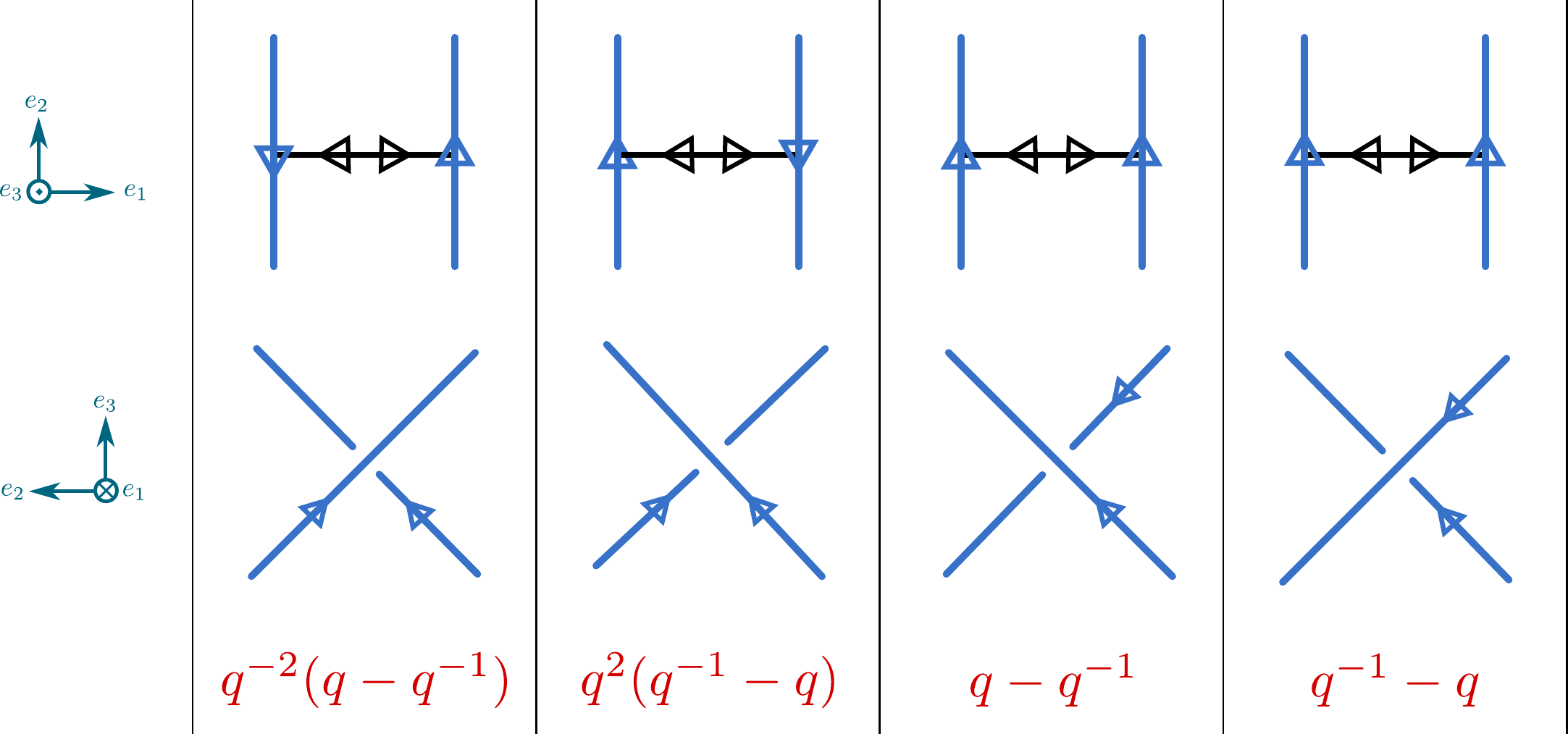}{0.45}{The first part of exchange factors contributing to the weight $\alpha(\tL)$. We show the standard projection on top and the $ij$-leaf space projection below.}
	
	\item
	The second factor is an extra correction $q^\delta$, 
	where $\delta \in \{0,2,-2\}$, determined as follows. (This factor has no direct
	analog in the $N=2$ case.)

	For each of the points $a$ where the link attaches to the exchange,
	let $v_a$ denote the oriented tangent vector to the link at $a$.
	We have $\delta = 0$ unless the two $v_a$ have opposite components
	in the height direction (one up, one down). So from now on assume they are opposite. In this case the exchange has a natural
	 orientation, pointing toward the leg which is going up.
	Let $v_{exch}$ be a tangent vector to 
	the exchange, compatible with this orientation. Let 
	$k \notin \{i,j\}$ be the third sheet label, and let $v_{k}$ be a 
	vector tangent to either the $ik$ or $jk$ foliation,
	with the orientation induced by choosing sheet $k$.
	Then we consider the $x^3$-component of $w = v_{exch} \times v_k$,
	evaluated at some point along the exchange. If $w$ points up, and the leaf space crossing is an overcrossing,
	then $\delta = 2$. If $w$ points down, and the leaf space crossing is an undercrossing, then $\delta = -2$. In all other
	cases $\delta = 0$.

The above formula for $\delta$ also has another interpretation in an 
important special case. Consider the case where the $ij$-, $ik$- and $jk$-leaves are almost degenerate. Then $\delta\neq 0$ exactly when the following happens: in addition to lifts on sheet $i$ and sheet $j$ that contain the lifted $ij$-exchange, there is also a small loop on sheet $k$ containing lifted $ik$- and $jk$-exchanges. This is illustrated in \autoref{fig:exchange-factor-extra}. (Note that there is another possible configuration of foliations given by swapping the $kj$- and $ik$-foliations; the analysis works similarly in that case.) 

\insfigsvg{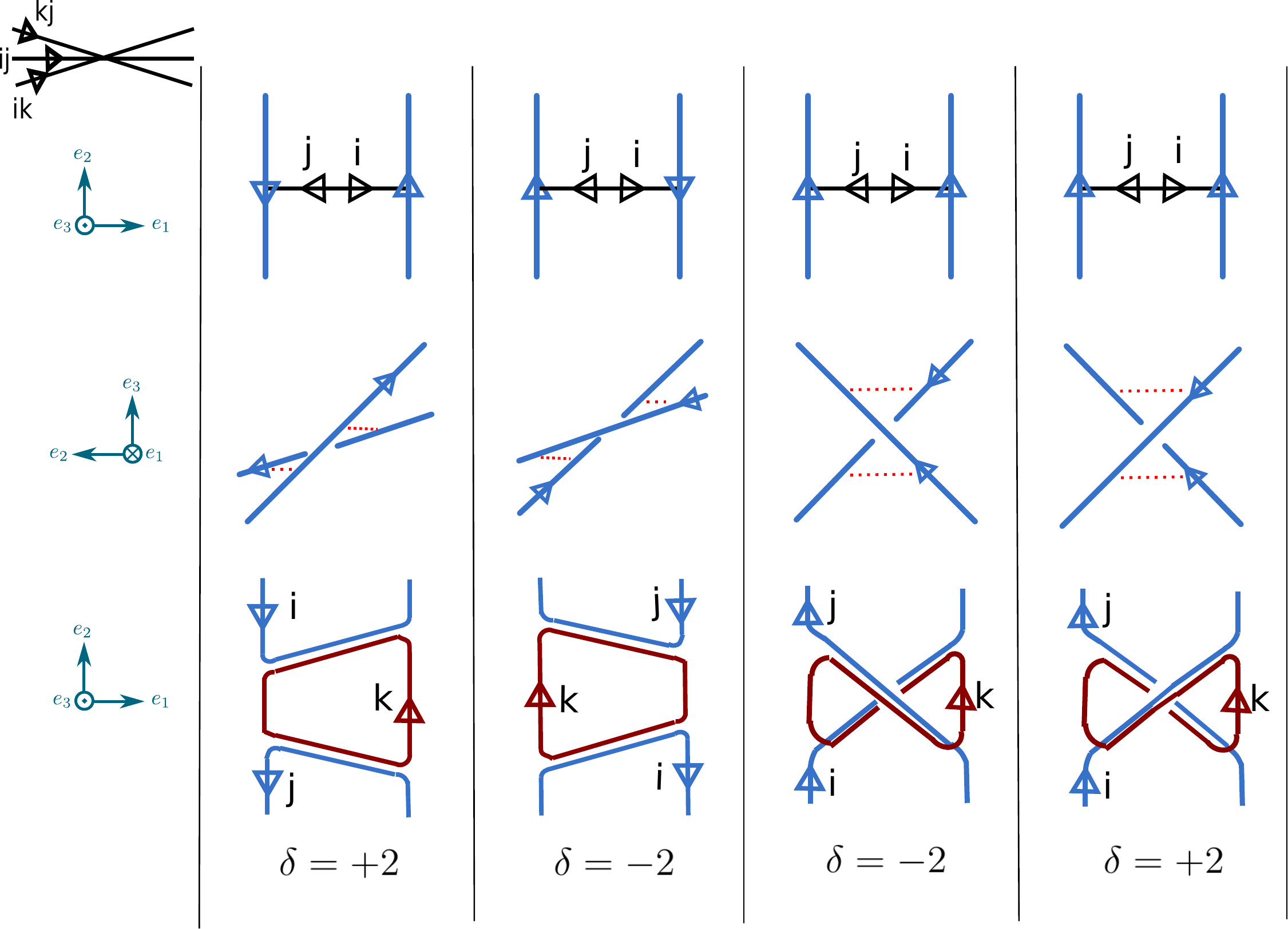}{0.5}{The extra correction factors $q^\delta$ contributing to the exchange weights. The format here is similar to \autoref{fig:exchange-factor}. In the first row we show the standard projection onto $C$. The $ij$-leaf space projections are shown in the second row, where we use dotted red lines to indicate locations of the $ik$- and $kj$-exchanges near the $ij$-exchange. In the third row we show the standard projection of the extra lift containing a small loop (in red) on sheet $k$. }

We remark that $\delta$ depends both on the link configuration and the almost-degenerate foliation. For example, suppose we change the foliation in \autoref{fig:exchange-factor-extra} by swapping the $kj$- and $ik$-foliations. Then the link configuration shown in the figure
would have $\delta=0$; indeed it would be impossible to have a small loop on sheet $k$ compatible with the new foliation directions. 

\end{itemize}


Above we have restricted to the case of exchanges which 
do not cross caustics. We leave the case of an exchange which does cross a caustic
to future work. (One can see that some modification of the rules above will be 
required: our description of the factor $q^\delta$ relies on the
property that $v_{exch} \times v_k$ is has the same height tendency at every point along the
exchange, which would be violated if the exchange crosses a caustic.
Another way to see that this case should be more subtle is to note
that when there is an exchange crossing a caustic there is also
a continuous 1-parameter family of webs.)

\subsubsection{Weight factors for detours} \label{sec:detour-weights}

The weight factor $q^{\pm \frac12}$
associated with a detour is shown in \autoref{fig:detour-factor}.
This is the same as the detour factor in the $N=2$ case, given in \cite{Neitzke:2020jik}.

\insfigsvg{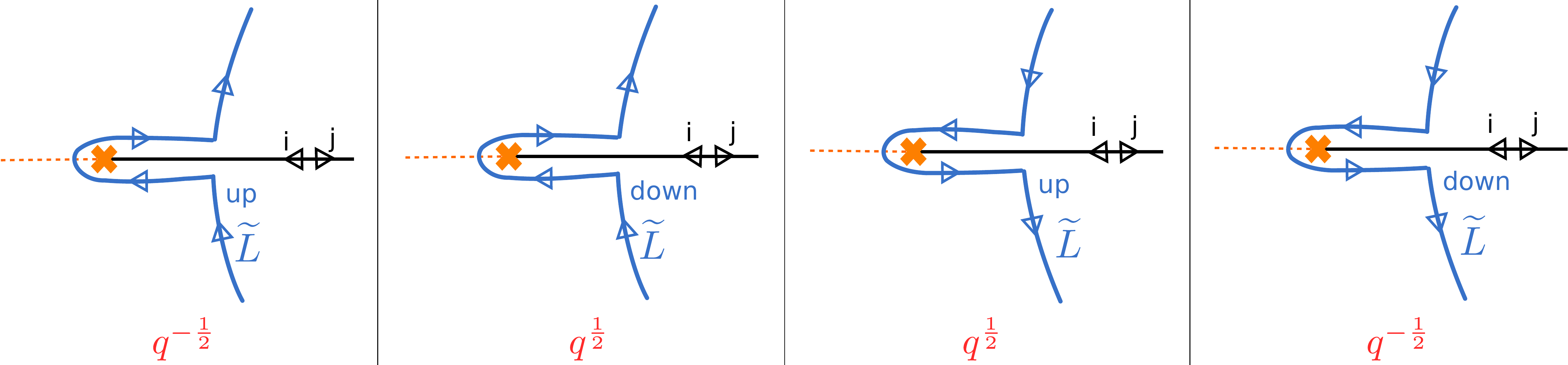}{0.35}{Detour factors contributing to the overall weight factor $\alpha(\tL)$. The notation is the same as in \autoref{fig:framing-factor-gln}.}

We remark that this factor can also be interpreted in another way:
its effect is to cancel a factor $q^{\pm \frac12}$ coming from the
winding of the detour in the $ij$-leaf space. Thus, alternatively,
one could adopt the rule that we do not count the winding of the detour
in the $ij$-leaf space (while still counting the winding in other leaf spaces),
and then omit the detour factor.

\subsubsection{Weight factors for trees with all ends on the link} \label{sec:tree-weights}

\insfigsvg{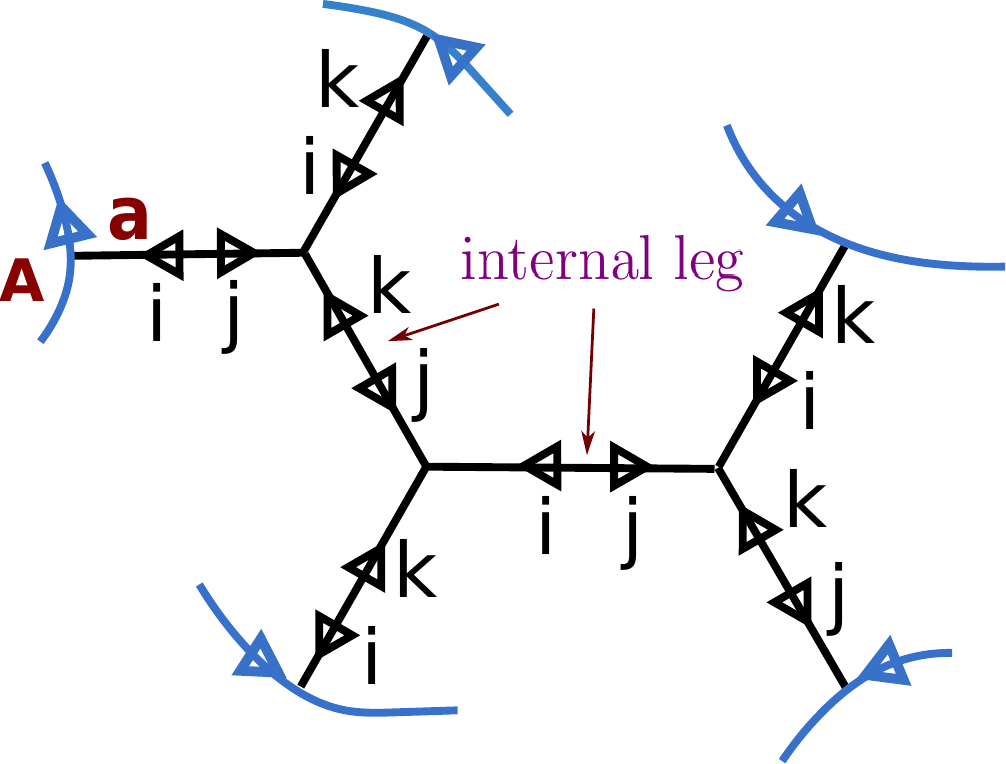}{0.6}{An example of a trivalent tree (shown in black) with five legs attached to the link $L$ (shown in blue). This trivalent tree has in total seven legs, two of which are internal in the sense that they are not attached to the link $L$.}

Now let us consider a web with the topology of a trivalent tree, with all external legs (i.e. legs which have at least one end point not being an internal vertex) 
attached to the link $L$. Let $K$ denote the number of external legs, and assume $K > 2$
(the case $K=2$ corresponds to an exchange, which we have already considered above.). In \autoref{fig:trivalent-tree} we show an example of such a trivalent tree. The web factor associated with such a tree is of the form
\begin{equation}\label{eqn:treefactor} 
 \left( \prod_{\ell=1}^{2K-3} \sigma_\ell \right) q^{-\sum_{a=1}^K m_a} (q - q^{-1})^{K-1}.
\end{equation}
Here the index $a$ runs over
the $K$ external legs of the tree that are attached to the link $L$, $\ell$ runs over the $2K-3$ total legs (including the internal legs),
and the quantities $m_a$, $\sigma_\ell$ are defined below.

\begin{itemize}

\item The quantity $m_a \in \Z$ for an external leg $a$ is defined as follows. Let
$t$ be a positively oriented tangent vector to the link, and let $l$ be a tangent vector to the external leg $a$,
oriented into the tree. Then $m_a = +1$ if $t$ and $t \times l$ have the same sign for their
$x^3$ components (both pointing out of or into the paper), and $m_a = -1$ otherwise. As an example consider the external leg $a$ in \autoref{fig:trivalent-tree} that is attached to the link strand labeled as A. If the link strand A points out of (into) the paper, we then have $m_a=-1$ ($m_a=-1$).

\item The sign $\sigma_\ell \in \{\pm 1\}$ for a leg $\ell$ is defined as follows.
We first realize $\ell$ as a transverse 
intersection of two surfaces (which we call {\it thickenings} in the following) in $M$, associated to the two ends of $\ell$. 
Each thickening is a union of $ij$-half-leaves. 
For an end $p$ of $\ell$ which lies on the link $L$, the corresponding thickening is just the union of
all $ij$-half-leaves beginning at points $p'$ near $p$ on $L$. For an end $p$ 
which lies on an internal vertex $v$, the thickening is defined by induction, as follows. 

We consider 
the other two legs $\ell'$, $\ell''$ incident on $v$, and we denote their common endpoint on $v$ as $q$. Let the ends of $\ell'$ be $q$ and $q'$, and
likewise let the ends of $\ell''$ be $q$ and $q''$.
We assume we already know the thickening of $\ell'$ associated to the end $q'$,
and likewise for $\ell''$.
The intersection of these two thickenings is a curve passing through $q$; we think of 
this curve as a kind of virtual link segment on which $q$ lies. This procedure is illustrated in \autoref{fig:tree-sign}, where we show a virtual link segment in red, obtained as the intersection of two thickenings associated with $q'$ and $q''$.
We can then proceed as if $q$ were lying on the actual link $L$: 
we obtain a thickening of $\ell$
by taking the union of all $ij$-half-leaves beginning at points near 
$q$ on the virtual link segment. In this way we inductively obtain thickenings 
of all legs, and a virtual link segment passing through each end of each leg.

Each virtual link segment has a canonical orientation: let $v_{in}$ 
denote a vector pointing into $q$ along a leg of type $ij$ with its
$i$-orientation, and let $v_{out}$ denote a vector pointing out of $q$
along a leg of type $ki$ with its $i$-orientation; then
$v_{in} \times v_{out} \cdot t > 0$, 
where $t$ is a positively oriented tangent vector
to the virtual link.

Finally, once we have defined the (actual or virtual) 
oriented link segments passing through the two ends of $\ell$, we are
ready to define the sign $\sigma_\ell$.
Namely, projecting these two link segments to the $ij$
leaf space we get a crossing; the sign $\sigma_\ell = +1$ if this crossing is an overcrossing,
$\sigma_\ell = -1$ if it is an undercrossing. 

\end{itemize}

Note that making the transformation $x^3 \to -x^3$ sends $m_a \to -m_a$,
and also sends $\sigma_\ell \to -\sigma_\ell$ when $\ell$ is one of the $K-3$ legs connecting two internal vertices; thus it changes 
$\prod_\ell \sigma_\ell$ by a factor $(-1)^{K-3}$.
These changes in \eqref{eqn:treefactor} are equivalent to taking $q \to q^{-1}$,
as expected since the map $x^3 \to -x^3$ reverses the orientation of $M$.

\insfigsvg{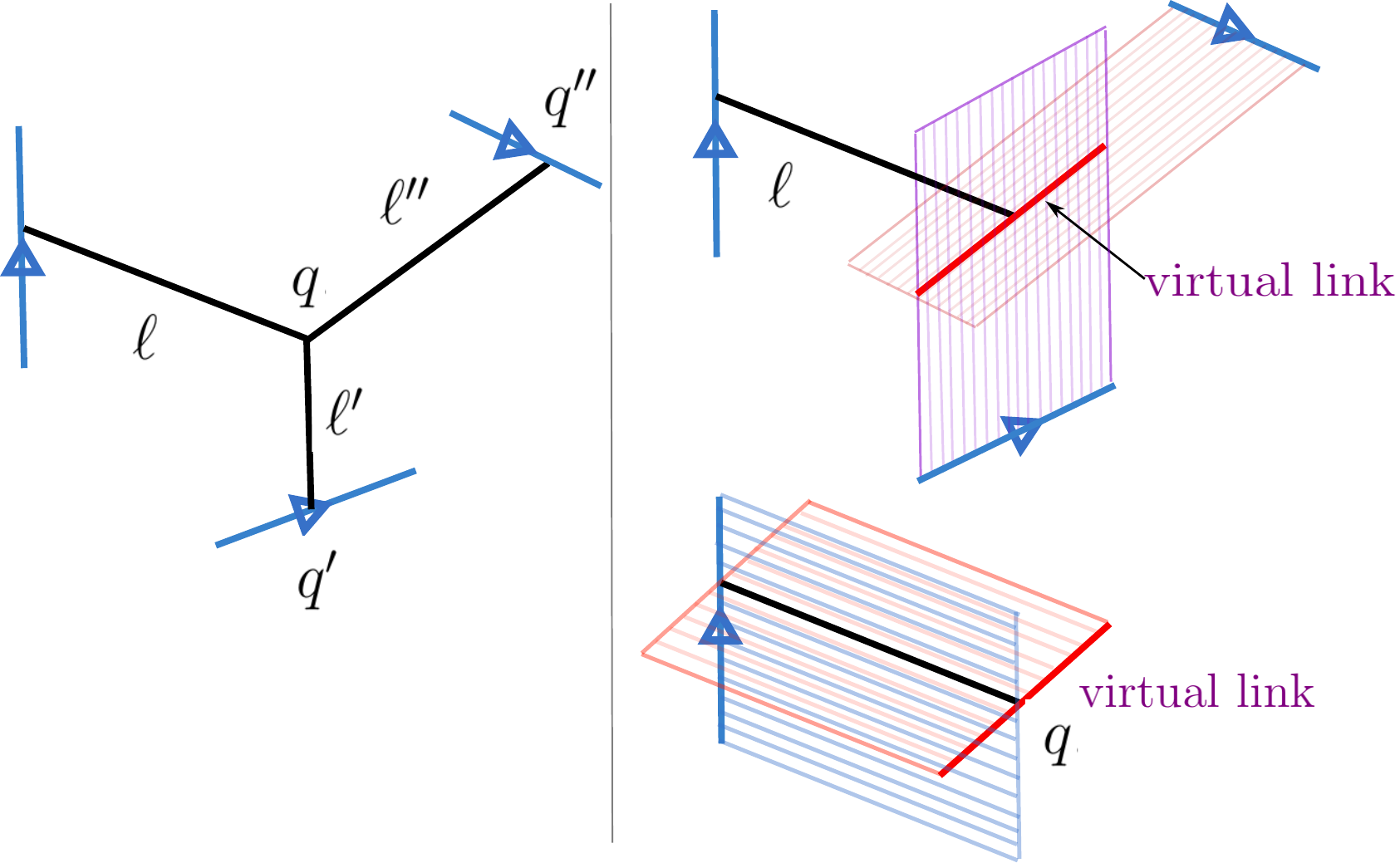}{0.6}{Upper right: the construction of a virtual link (red) passing through the internal vertex $q$ of a trivalent web. The virtual link is obtained as the intersection of two surfaces (called {\it thickenings}) emanating from the link ends of the two legs $\ell'$ and $\ell''$. Lower right: the segment $\ell$ in the web is the 
intersection of two thickenings, one emanating from the link end, the other from
the virtual link.}

\subsubsection{Weight factors for more general trees}

We can also consider webs with the topology of a tree 
which has some ends on branch points and some 
ends on the link $L$, like the one depicted in \autoref{fig:web-example}. 
In this case we obtain a weight factor which is a
hybrid of the detour and tree factors above. Namely, we take the factor
\eqref{eqn:treefactor} as before, where the index $a$ still runs only over
those ends which are attached to the link; for the legs attached
to branch points we adopt the rule mentioned at the end of
\autoref{sec:detour-weights} (namely, 
don't include any explicit extra factor and don't count the $ij$ leaf space winding
along these legs).

In the examples in \autoref{sec:examples} we will not meet this kind of web, 
but we expect that such webs should appear 
when we explore further away from the almost-degenerate
locus of the Coulomb branch.

\subsubsection{Weight factors for webs containing one loop} \label{sec:loop-factors}

\insfigsvg{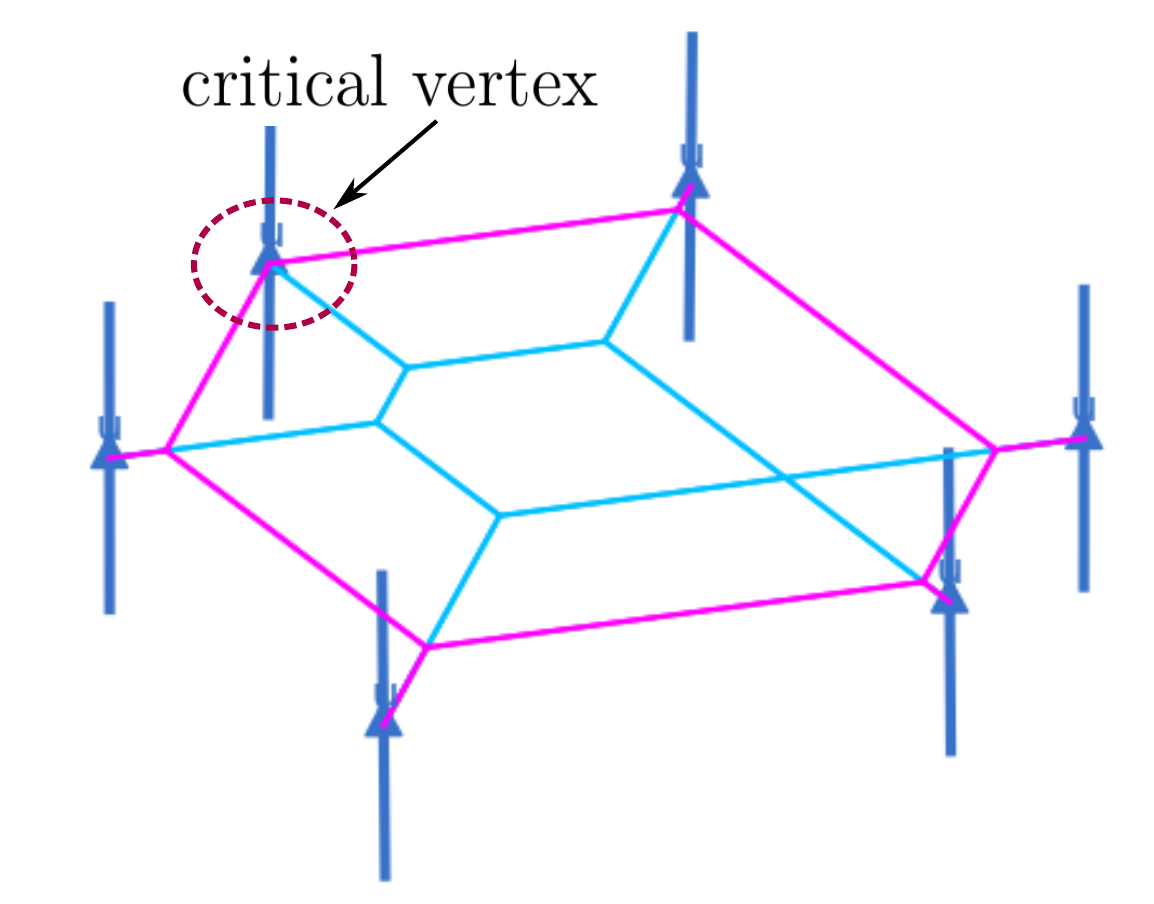}{0.5}{Here we illustrate examples of webs containing a loop with six vertices. Such webs come in a continuous family bounded by two limiting configurations. One limit is a web containing a loop with six vertices where one vertex is directly attached to a link strand; this limit is shown in pink. The other limit is a tree with a self-intersection; 
this limit is shown in light blue. The vertex of the limiting web that is directly attached to a link strand is called the {\it critical vertex} below.}

In simple examples (including all the examples
we will discuss in \autoref{sec:examples} below),
all of the webs that appear are trees. In more
complicated cases, however, there can be webs containing loops.
We have not completely determined the weight factors associated with webs containing loops; we leave that to future work.
However, we did explore the case
of webs which contain exactly one loop. 

Such webs come in continuous families, bounded by two limiting configurations. One limiting configuration is a web containing a loop which is directly attached to one of the link strands. The other limiting configuration is a trivalent tree in which two of the
legs cross one another. An example illustrating both limiting configurations bounding a family of webs with one loop is shown in \autoref{fig:webloop}. 

Since webs with one loop come in a continuous family, 
our usual prescription for the quantum UV-IR map lead to an infinite sum coming from all members in this continuously family. A crucial fact is that all the webs in the interior of such a continuous family are all isotopic, therefore their contributions to the $F(L)$ could be summed together and treated as a single contribution. Concretely we can choose any single representative web from the interior of this continuous family and only use this representative to build the lifted links $\tL$, where the total contribution from the continuous family is absorbed into the weight factor associated with this single representative.

As usual, we need to include a weight factor associated with this web.
By performing an isotopy which kills the continuous family (dragging the link strand where the critical vertex is attached towards the interior), we have determined the necessary weight factor:
\begin{equation}
\alpha_{\text{1-loop}} = q^\epsilon \alpha_{\text{tree}},
\end{equation}
where $\alpha_{\text{tree}}$ is the weight factor associated with the limiting web with a tree topology (e.g. the light blue web illustrated in \autoref{fig:webloop}) using the rules described in \autoref{sec:tree-weights}.\footnote{The limiting web with a tree topology has a self-intersection, unlike the tree webs described in \autoref{sec:tree-weights}. However, we can still apply the rules described there to fix the normalization factor $\alpha_{\text{tree}}$.} The extra factor $q^\epsilon := q^{\epsilon_1\epsilon_2}$ is determined by two pieces of data associated with the critical vertex:
\begin{itemize}
\item $\epsilon_1$ is determined by the link strand attached to the critical vertex: if the tangent vector to the link has a positive (negative) $x^3$-component, 
then $\epsilon_1$ is $+1$ ($-1$). 
\item $\epsilon_2$ is determined by the {\it type} of the critical vertex, as follows. There are two types of trivalent vertices, as shown in \autoref{fig:vertexparity}; they differ in the orientation of leaves corresponding to the three legs. We assign $\epsilon_2$ to be $\pm 1$ according to the type, as indicated in \autoref{fig:vertexparity}. 
\end{itemize}

\insfigsvg{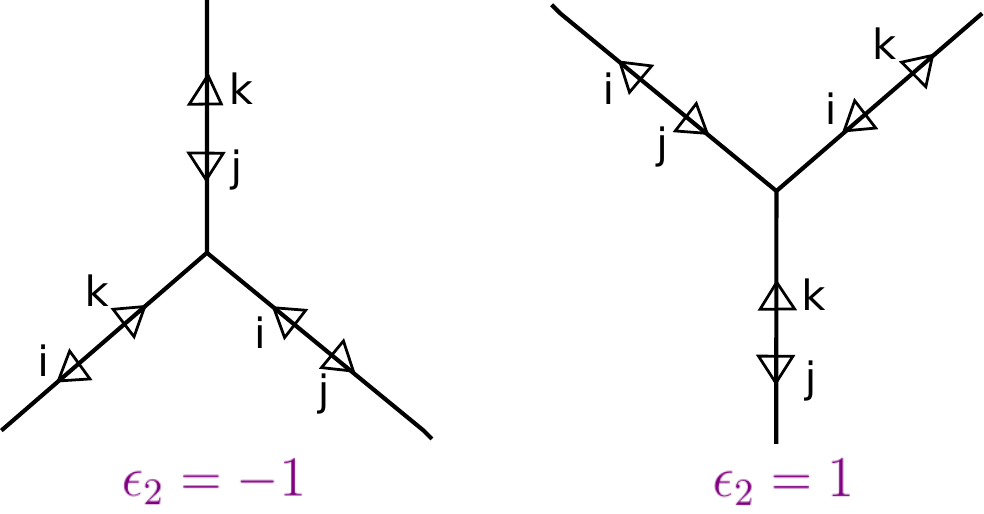}{0.65}{The two types of trivalent vertices and their corresponding $\epsilon_2$. }

\subsection{Bootstrapping the weight factors \texorpdfstring{$\alpha(\tL)$}{alpha(tL)}} \label{sec:bootstrapweightfactor}

In this section we briefly describe the bootstrap-like method which we used to determine the weight factors $\alpha(\tL)$, which are important ingredients in the construction of quantum UV-IR map described in \autoref{sec:qab-general}. We used two properties of the quantum UV-IR map $F$:
\begin{itemize}
	\item Given a framed oriented link $L$ in $M$, $F(L)$ must only depend on the ambient isotopy class of $L$. This puts very strong constraints on $F$, as there are a large number of ambient isotopies which need to be respected by the map $F$.
	
	\item $F$ must be a homomorphism from the UV skein algebra $\Sk(M,\fgl(N))$ to the IR skein algebra $\Sk(\tM,\fgl(1))$: said otherwise, it must respect the UV skein relations,
	up to the IR skein relations.
\end{itemize}

Our general strategy is to cook up isotopies which relates webs with $K$ legs to webs with less than $K$ legs; an example of such isotopies is illustrated in \autoref{fig:isotopy-example}. This kind of isotopies allow us to solve for the weight factors in a recursive way. 

\insfigsvg{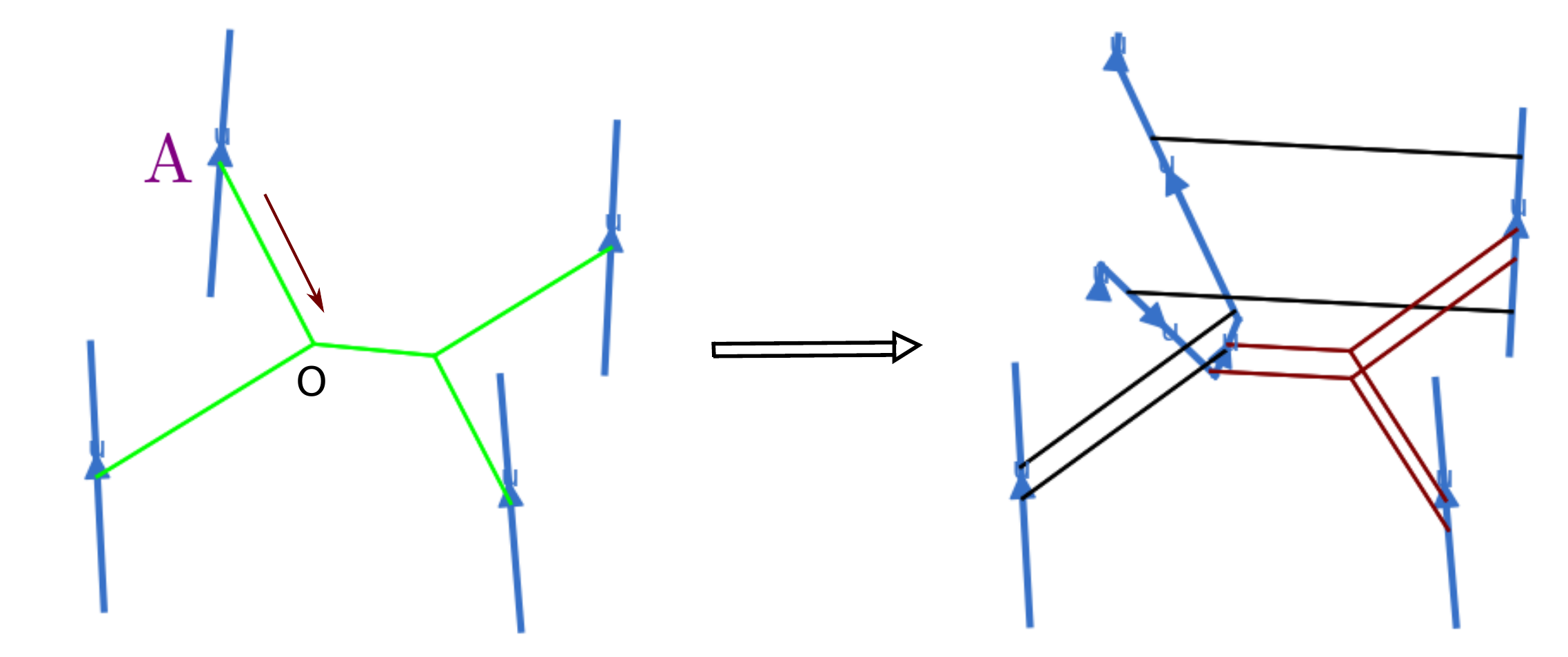}{0.48}{Here we demonstrate an example of an isotopy used to obtain the weight factors associated with webs. We show link segments (in blue) contained in a small contractible region inside $C \times [0,1]$. The relevant isotopy move here is dragging part of a link segment (denoted as {\color{violet} A} in the figure) past an internal vertex (denoted as $O$) along the arrow direction. The configuration before this isotopy is shown on the left, where a single 5-legged web (green) is stretched between the link segments. After the isotopy, the
	5-legged web disappears; instead we have two 3-legged webs (red) and 4 exchanges (black).} 

Concretely, we proceed as follows. We tune the link configuration to single out a $K$-legged web lying in a small contractible region. For example, in \autoref{fig:isotopy-example} we show on the left a 5-legged web attached to four link segments. Then we perform an isotopy which deforms part of a link segment near the point where it is attached to the web, stretching it past an internal vertex of the web. After making such an isotopy, the original $K$-legged web disappears, and we end up with (multiple) webs with fewer legs. This procedure is illustrated in \autoref{fig:isotopy-example}. As $F(L)$ should be the same before and after the isotopy, we obtain an equality relating the local factor for a $K$-legged web to those for webs with less than $K$ legs. This enables us to determine the local web factors in an inductive way.

\subsection{Comments on possible derivations of the weight factors}

As described earlier, we obtained various local contributions to the weight factors $\alpha(\tL)$ using bootstrap-like methods, by exploiting the homomorphism property of $F$ and isotopy invariance. It would be very desirable to give a physical derivation for these factors, perhaps in the context of twisted $\Omega$-deformed 5d $\cN=2$ super Yang-Mills theory on $M\times \R^2_\epsilon$ with insertion of a fundamental Wilson line along $L\subset M$, as described in \autoref{sec:strategyF}. We leave this to future work.

We can also uplift the setup back to 6d, where the quantum UV-IR map is interpreted as the correspondence mapping a surface defect in the strongly-interacting 6d $(2,0)$ $\fgl(N)$ theory to combinations of surface defects in the abelian $(2,0)$ theory at a point of the tensor branch. In this context, the WKB webs stretched between link segments correspond to dynamical BPS strings in the 6d theory. It would be very interesting to derive the weight factors $\alpha(\tL)$ from the 6d point of view, taking into account contributions from the 6d abelian theory as well as contributions from the worldvolume theory of the BPS strings. We leave this derivation to future work.

As one comment, we notice that the weight factor (\ref{eqn:treefactor}) associated with a tree web is proportional to $(q-q^{-1})^{P-Q}$, where $P$ is the total number of legs in the tree web (including the internal legs) and $Q$ is the total number of trivalent vertices. The simplest example of a tree web, namely an exchange, has $P=1, Q=0$ and contributes a weight factor proportional to $(q-q^{-1})$. As already pointed out in \cite{Gaiotto:2011nm} (see also \cite{Galakhov:2016cji}), this factor can be interpreted as counting the ground states on the BPS string stretched between two strands of the link. Concretely this string has two states which differ by 1 in fermion number and angular momentum.\footnote{Recall that our definition (\ref{eqn:PSC}) for protected spin character contains $q^{2J}$ instead of $q^J$} Similarly we can also hope to interpret $(q-q^{-1})^{P-Q}$ as counting the ground states for the BPS string web stretched between link strands. For example, it might be possible to consider an effective quantum mechanical model, which assigns two Majorana fermions to each leg of the string web. Each internal vertex represents a certain interaction lifting off some of the degenerate states. It would be interesting to pursue this direction.

\subsection{Comments on isotopy invariance}

In the case of $N=2$, we gave a sketch proof in \cite{Neitzke:2020jik} that the $F$ we constructed indeed has the expected isotopy invariance
and homomorphism properties, by 
checking that it behaves correctly under a set of Reidemeister-type moves. In so doing we were aided by the fact that we only had to consider
one foliation, and everything was (in an appropriate sense) local in the two-dimensional leaf space of that foliation. 
For $N \ge 3$ we do not have
this kind of locality available in general, and so it seems more difficult to find an appropriate finite set of Reidemeister-type moves. Thus we have not been able to 
prove the full isotopy invariance or the homomorphism property for the $F$
we construct; however, they must be true if our physical picture is 
correct. It would be very desirable to give a proof.

Lacking a full proof, we have conducted various experimental checks, mostly with computer assistance (see \autoref{app:algorithms} for some
discussion of the algorithms). 
For instance, we generated random polygonal knots with 
$n \le 8$ vertices
in $\R^3$ and verified that in all cases 
our recipe produces the expected polynomial invariants.
We also studied various examples of knots on surfaces, again obtaining the expected
results in all cases.
A few of these checks are described in more detail
in \autoref{sec:examples} below.

\section{The limit of almost-degenerate foliations} \label{sec:almost-degenerate-limit}

In this section we consider an important simplifying limit, where
the quantum UV-IR map becomes much easier to compute.
This limit was previously considered in \cite{Gaiotto:2012db}.

Recall that the 1-forms $\lambda_i$ which determine the WKB foliations
are determined by an equation of the form
\begin{equation}
	\lambda^3 + \phi_1 \lambda^2 + \phi_2 \lambda + \phi_3 = 0,
\end{equation}
where $\phi_r$ is a meromorphic $r$-differential on $C$.
Now suppose we take $\phi_1 = 0$ and take $\phi_3$ to be much smaller in norm than $\phi_2$.
Then the $\lambda_i$ can be written locally, to first order in $\phi_3$, as
\begin{equation}
	\lambda_1 \approx \sqrt{\phi_2} + \frac{\phi_3}{2 \phi_2}, \qquad \lambda_2 \approx - \frac{\phi_3}{\phi_2}, \qquad \lambda_3 \approx -\sqrt{\phi_2} + \frac{\phi_3}{2 \phi_2}. 
\end{equation}
Thus we have (writing $\lambda_{ij} = \lambda_i - \lambda_j$)
\begin{equation} \label{eq:approx-lambda}
	\lambda_{12} \approx \sqrt{\phi_2} + \frac{3 \phi_3}{2 \phi_2}, \quad \lambda_{23} \approx \sqrt{\phi_2} - \frac{3 \phi_3}{2 \phi_2}, \quad \lambda_{13} \approx 2 \sqrt{\phi_2}.
\end{equation}
In particular, all three of these have approximately the same phase (that of $\sqrt{\phi_2}$), so that the
three local foliations collapse approximately to one foliation  $\fapprox$.
If we take $\phi_3 = 0$, then they collapse exactly.


\subsection{The case of \texorpdfstring{$\R^3$}{R3}} \label{sec:exchange-clusters}

Let us consider the case where $C = \C$, so $M = \C \times \R \simeq \R^3$.
As we have stated in \eqref{eq:homfly-intro}, in this case we must have
\begin{equation} \label{eq:homfly}
	F(L) = q^{3 wr(L)} P_\HOMFLY(L, a = q^3, z = q - q^{-1}).
\end{equation}
This must hold no matter what we choose for the WKB foliations,
and in particular it must hold if
the WKB foliations are nearly degenerate as just discussed.
In this section we will describe more explicitly how this happens.

To be definite, we take
\begin{equation} \label{eq:differentials-state-sum}
	\phi_1 = 0, \qquad \phi_2 = \de z^2, \qquad \phi_3 = \epsilon \, \de z^3. 
\end{equation}
From \eqref{eq:approx-lambda} we then have 
\begin{equation}
	\lambda_{12} \approx \left(1 + \frac{3}{2}\eps\right) \de z, \quad \lambda_{23} \approx \left(1 - \frac{3}{2}\eps\right) \de z, \quad \lambda_{13} \approx 2 \, \de z.
\end{equation}
Thus the three WKB foliations approximately coincide
with the foliation $\fapprox$ whose leaves point in the $x^1$-direction,
where $z = x^1 + \I x^2$. See \autoref{fig:near-coincident-foliations} for the picture in the
$x^1$-$x^2$ plane.
\insfigsvg{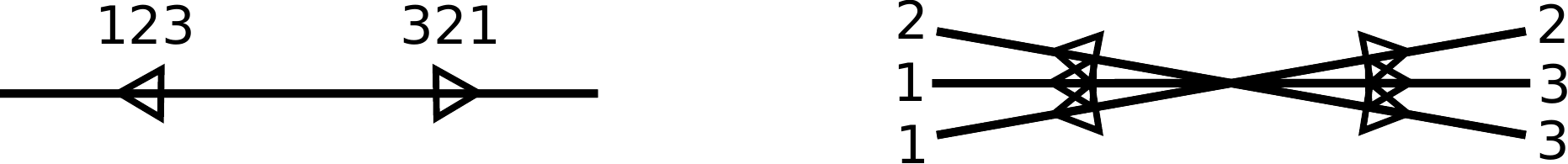}{0.35}{Left: when $\im \epsilon = 0$ the three
WKB foliations collapse to a single foliation $\fapprox$, 
pointed in the $x^1$-direction.
Right: leaves of the three WKB foliations, when $\im \epsilon < 0$.}

This choice leads to a simplification of the UV-IR map, as follows.
For a generic link $L$,
the only type of webs which occur are exchanges.
Moreover, these exchanges occur in clusters:
if $L$ crosses a leaf of $\fapprox$ at two points, there is a
cluster of 3 possible exchanges near the crossing points. See
\autoref{fig:exchange-cluster-simple} for a sample picture.

\insfigsvg{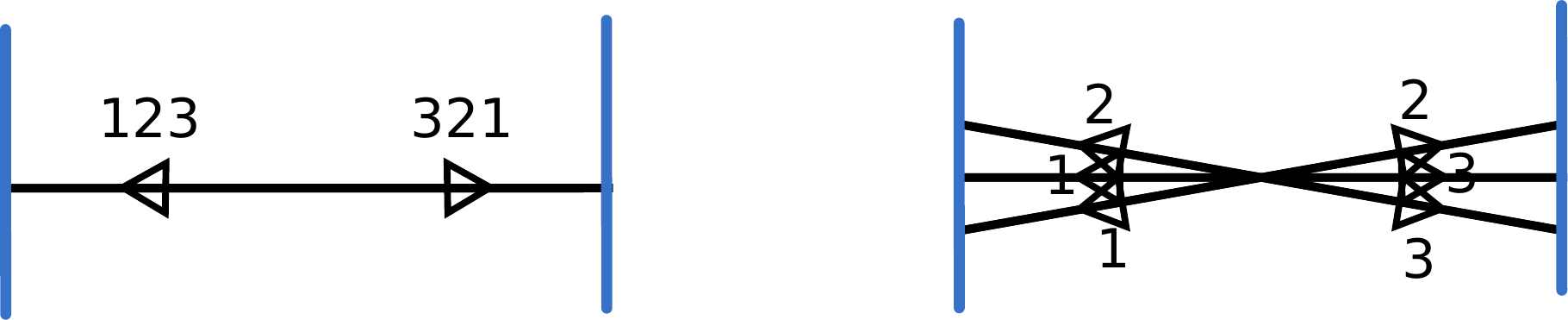}{0.4}{Left: a leaf of 
$\fapprox$ which meets $L$ at two points.
Right: the corresponding cluster of exchanges.}

The possible lifts involving 
such a cluster of exchanges depend on which way the link
segments are oriented and on the sign of $\im \epsilon$.
Fortunately, though, by enumerating all cases, one finds that they
can be captured by a simple
``effective'' rule which is independent of this
sign. Thus in what follows we will not need to worry about
the sign of $\im \epsilon$.\footnote{Concretely, this effective
rule can be described as follows.
We allow lifts including a single detour 
along any one of the $3$ exchanges. However, instead of the $N=3$
exchange weight from \autoref{sec:exchange-weights}, we use the simpler
$N=2$ exchange weight given in \cite{Neitzke:2020jik}. This
means that the prefactors $q^{\pm 2}$ in the first two columns
of \autoref{fig:exchange-factor} are replaced by $q^{\pm 1}$,
and the extra prefactor $q^\delta$ is omitted. Moreover, we do not include
any extra tangency factors associated to the detour; thus in this effective
description the only tangency factors come from places where $L$ itself
is tangent to a WKB foliation. 
We include the winding of the lifts $\tL$ as usual.}

Now suppose given a framed link $L$ in $M = \C \times \R \simeq \R^3$, such 
that the projection of $L$ to the $x^2$-$x^3$ 
plane is a smooth
link diagram, with all crossings transverse.
By a small isotopy we may standardize the restriction of $L$ to a 
domain $\R \times D_c$, where $D_c$ denotes a small disc in the
$x^2$-$x^3$ plane containing the crossing $c$: we require that it enters
and exits pointing in the positive $x^2$-direction, as shown in
\autoref{fig:overcrossing-undercrossing}.
\insfigsvg{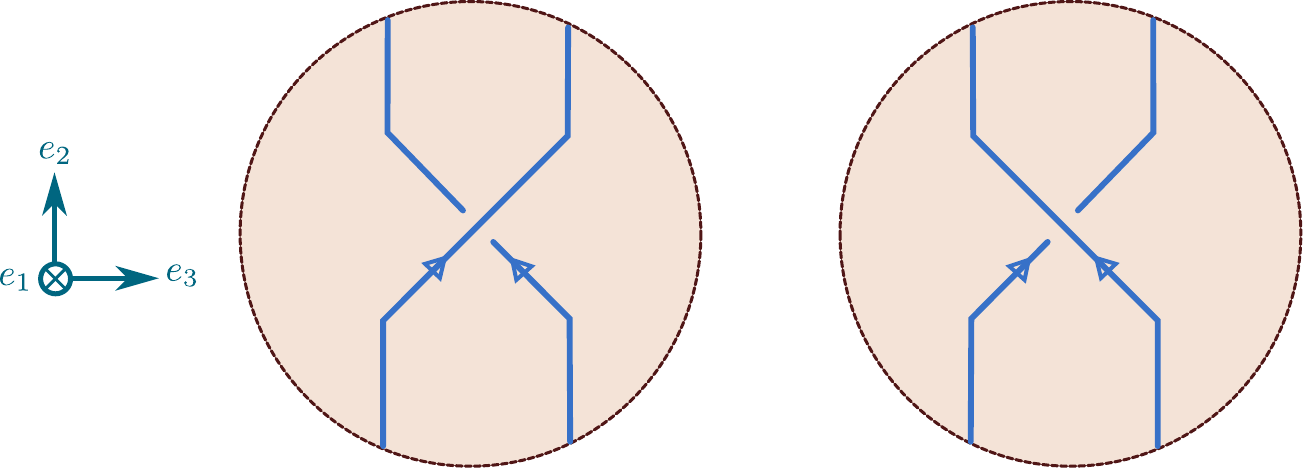}{0.48}{Standard forms for
the projection of the link $L$ to the $x^2$-$x^3$ plane, restricted to 
a domain $\R \times D_c$. Left: overcrossing. Right: undercrossing.}
We decompose $M$ into 
a collection of cylinders $\R \times D_c$ and the complement $\R \times P$.
Importantly, all of the leaves of the WKB foliations are oriented
approximately in the $x^1$ direction, which runs parallel to the domain boundaries,
and thus the computation of $F(L)$ does not involve any 
webs or exchanges which cross from one
domain to another. (There are exchange contributions to $F(L)$, 
but they are localized
around the crossings in the link diagram, so each is contained 
in a single domain.)
This fact means that the desired 
$F(L)$ can be computed by 
first applying the UV-IR map in each domain independently,
then gluing together the resulting lifted paths in all
possible ways, subject only to 
the rule that the sheet labels on the lifts $\tL$ have to match at the 
boundaries.

The results from the individual domains are as follows:
\begin{itemize}
	\item
In each cylinder $D_c \times \R$ the result is given by one of the lines
of \autoref{fig:lifts-one-crossing}.
\insfigsvg{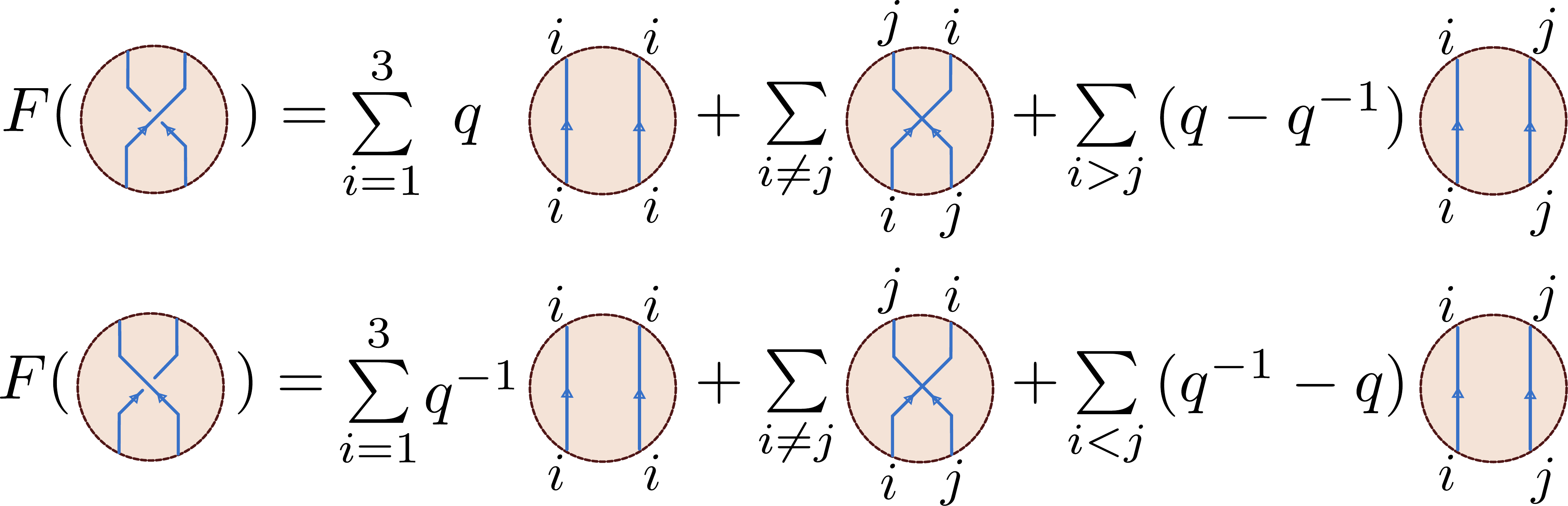}{0.25}{Top: the result
of applying $F$ to the indicated overcrossing. The right side
represents a formal linear combination of open paths on $\widetilde M$.
The first two terms come from the simple lifts of the link; in the
first term we have resolved using the $\fgl(1)$ skein relations.
The last term involves contributions from the cluster of exchanges
localized near the crossing.
Bottom: the result of applying $F$ to the indicated undercrossing.}

\item
For the complement $P \times \R$ the result is given by the sum over direct lifts
$\tL$,
with no contributions from webs or exchanges. The direct lifts correspond to
labelings of the components 
of $L \cap (P \times \R)$ by sheets $i \in \{1,2,3\}$.
Each lift $\tL$ comes with a weight $\alpha(\tL)$, which is a power of $q$,
determined as follows. For each component $s$,
let $w(s)$ be the counterclockwise winding number of $s$
in the $x^2$-$x^3$ plane,\footnote{$w(s)$ is an integer, since the strands are either closed or else
their initial and final tangent vectors are both pointing in the positive $x^2$-direction.} and let $i(s)$ be the label on $s$.
Then $\alpha(\tL) = \prod_s q^{2 (2-i(s)) w(s)}$.
\end{itemize}

We now describe the result of gluing together these local 
contributions directly.
Let a labeling $\ell$ of the knot diagram
be an assignment of a label $i \in \{1, 2, 3\}$ to each arc.
For any labeling $\ell$ and crossing $c$,
let $\beta(\ell,c)$ denote the weight factor
with which the labeled crossing $c$ appears in \autoref{fig:vertex-model},
or $\beta(\ell,c) = 0$ if this crossing does not appear in the figure.
Then define an overall weight for the labeling $\ell$ by
\begin{equation}
	\alpha(\ell) = q^{2(w_1(\ell) - w_3(\ell))} \prod_c \beta(\ell,c) \quad  \in \quad \Z[q,q^{-1}]
\end{equation}
where $w_i(\ell)$ is the sum of the windings of the arcs labeled $i$.
Finally, $F(L)$ is the sum over all labelings,
\begin{equation} \label{eq:sum-labelings}
	F(L) = \sum_{\ell} \alpha(\ell).
\end{equation}
\insfigsvg{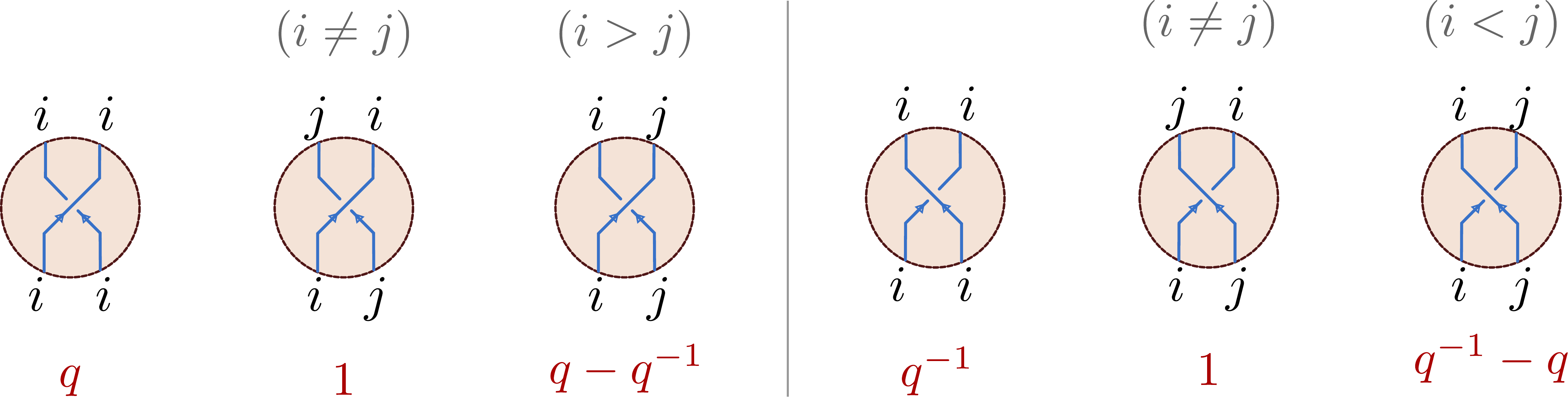}{0.25}{Weights for labeled crossings. Left: overcrossing. Right:
undercrossing. The weight $\beta(\ell,c)$ 
for each labeled crossing is shown in red. For labeled
crossings which are not of any of the types shown, the weight $\beta(\ell,c) = 0$.}

The formula \eqref{eq:sum-labelings} for $F(L)$ is similar to the 
vertex model for the Jones polynomial.
This is not surprising since we already know that $F(L)$ should be computing
a specialization of the HOMFLY polynomial,
\eqref{eq:homfly}.
We have made various experimental checks that \eqref{eq:sum-labelings} 
indeed computes
\eqref{eq:homfly}, some of which we will see in
\autoref{sec:examples} below.
One way to derive this relation would be to connect
\eqref{eq:sum-labelings} to the standard construction 
of knot invariants
from quantum groups \cite{MR1358358}, in the case of $U_q(\fgl_3)$.
In this interpretation, the labels $i$ on the arcs run over a basis 
for the fundamental representation of $U_q(\fgl_3)$,
and the contributions from crossings in \autoref{fig:vertex-model}
come from the $R$-matrix.

\subsection{Fock-Goncharov type foliations} \label{sec:fg-foliations}

Next we consider the case where $C$ is a compact Riemann surface and 
$\phi_2$, $\phi_3$ are meromorphic, again with $\phi_3 \ll \phi_2$ in norm,
and $\phi_1 = 0$.
We assume the setup is generic, in the sense that all zeroes of $\phi_2$ are simple,
no period $\oint \sqrt{\phi_2}$ is real, and $\phi_2$ has at least one pole of order
at least $2$.
In this case $\phi_2$ induces an ideal triangulation of the surface $C$, 
the WKB triangulation described in \cite{Gaiotto:2009hg}; 
the vertices of the triangles are poles of $\phi_2$, and 
each triangle has
exactly one zero of $\phi_2$ in its interior.

The 3-sheeted covering $\Sigma \to C$ 
and the WKB foliations in this case are discussed in \cite{Gaiotto:2012db}. Here we briefly
recall the main points.
When $\phi_3 = 0$ the branch points are the zeroes of $\phi_2$, with order-2
monodromy around each. Turning on
a small $\phi_3$ causes each branch point to split into a group of $3$ nearby branch points
(to see this note that the branch points are zeroes of the discriminant,
$\Delta = 4\phi_2^3 - 27\phi_3^2$.)
The $3$ foliations are essentially independent near the branch points, while 
far from the branch points they are closely aligned, and approximately
make up a single foliation which we call $\fapprox$.

An example of critical leaves before and after such a splitting is 
shown in \autoref{fig:one-triangle}. 
In the following, we will make the assumption that the 
perturbation $\phi_3$ is chosen in such a way
that the picture in each triangle looks as in that figure.
(In each of the explicit examples we consider below, we exhibit an explicit $\phi_3$ 
which does the job.)
\insfigsvg{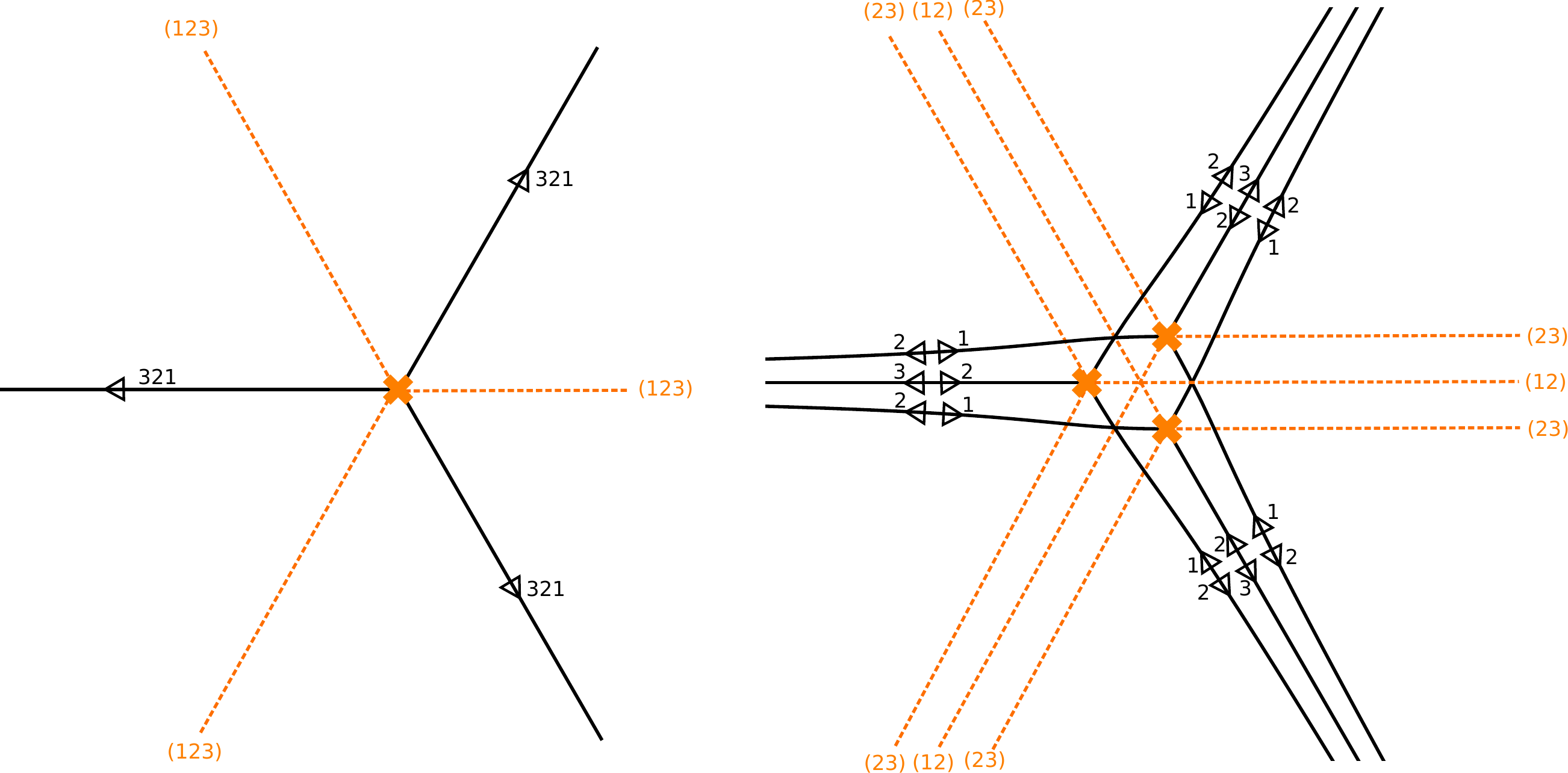}{0.36}{The covering $\Sigma \to C$ and the 
critical leaves around a simple zero of $\phi_2$. Left: the 
picture when $\phi_2 = z \, \de z^2$ and $\phi_3 = 0$. 
Right: the picture with a small nonzero $\phi_3$.}
Some non-critical leaves are shown in \autoref{fig:one-triangle-with-foliation}.
\insfigsvg{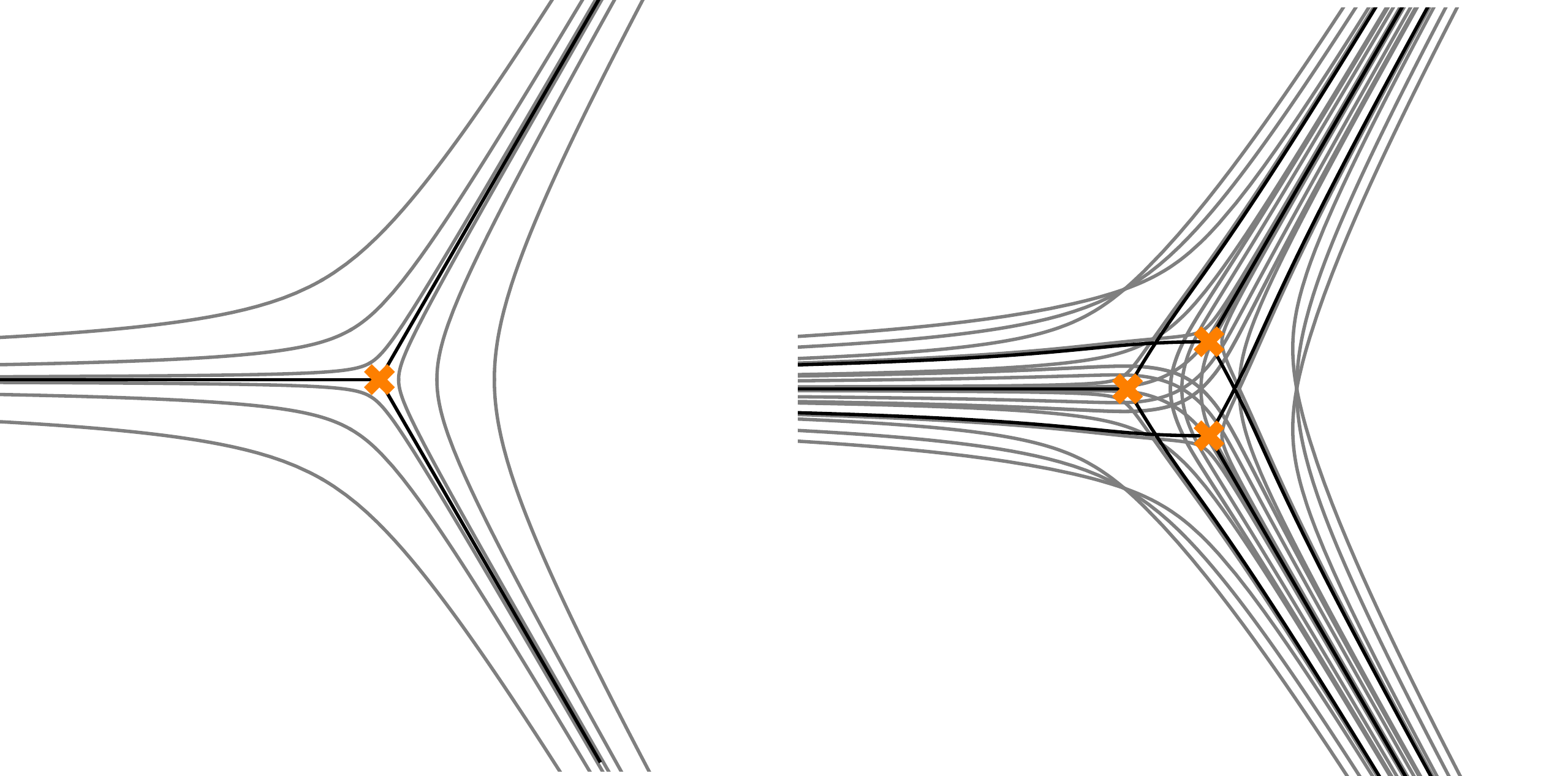}{0.36}{Some examples of non-critical leaves.
The two pictures here correspond to the two pictures in \autoref{fig:one-triangle} above.
In the left picture, we only have one foliation. In the right figure, this one foliation
has been perturbed into three distinct foliations. Far from the branch points, we
have $\phi_3 \ll \phi_2$ and the 
three foliations are close to the single foliation 
$\fapprox$; near the branch points, the three foliations are far from parallel.}

The charge lattice in these examples has a natural basis \cite{Gaiotto:2012db},
which we now recall. 
It has two elements $\gamma_E^1$, $\gamma_E^2$ associated to each edge
$E$ of the triangulation, and one element $\gamma_T$ associated to each triangle $T$. 
To fix the choice of which element to call $\gamma_E^1$ and which to call $\gamma_E^2$, we make
an auxiliary choice of orientation of $E$.
See \autoref{fig:two-triangles}.
\insfigsvg{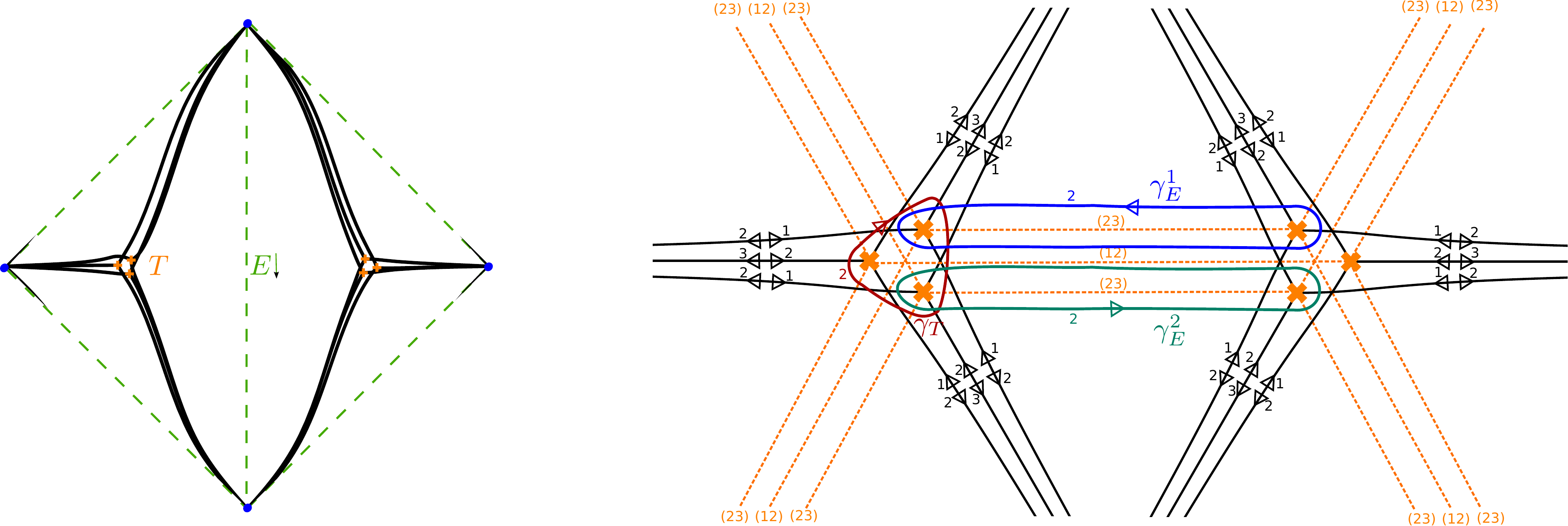}{0.38}{Left: two triangles in the triangulation of $C$.
One triangle $T$ and one edge $E$ are marked. The edge $E$ is decorated by an 
auxiliary choice of
orientation. Right: the corresponding cycles $\gamma_T$,
$\gamma_E^1$, $\gamma_E^2$ on $\tC$.}
This basis is adapted to the cluster structure on moduli spaces of flat
$SL(3,\C)$-connections \cite{Fock-Goncharov}: indeed, as explained 
in \cite{Gaiotto:2012db}, the spectral coordinates $X_\gamma$ associated to the basis 
elements give the cluster coordinates.

Now we consider a link $L \subset C \times \R$. By an isotopy if necessary,
we can arrange that $L$ stays away from the branch points.
In this case, if $L$ is in generic enough position 
there will be no contributions from BPS webs involving trivalent junctions; the only
BPS webs we have to consider are either exchanges or detours.
The structure of the exchanges and detours is determined by how
the link $L$ sits relative to the foliation $\fapprox$, as follows.

First, suppose $L$ meets a non-critical leaf of $\fapprox$
at two distinct points. In this case there is a cluster of $3$ possible exchanges,
as indicated in \autoref{fig:exchange-cluster} below. 
Here we do not show all the lifts explicitly; we just remark that
the effect of this cluster can be efficiently 
captured by the effective rules described in
\autoref{sec:exchange-clusters}, just as for the clusters 
that appear in the case $M = \R^3$.

\insfigsvg{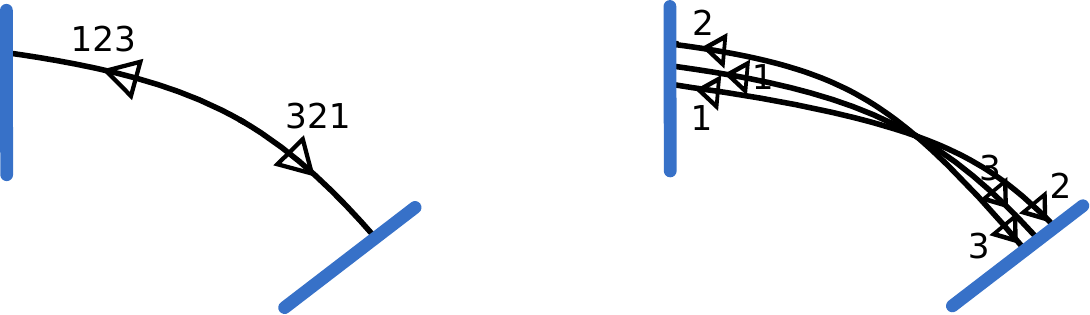}{0.6}{Left: a non-critical leaf of $\fapprox$ which is 
crossed by $L$ at two distinct points. Right: a cluster of $3$ exchanges in a neighborhood
of the leaf.}

Second, suppose $L$ crosses a critical leaf of $\fapprox$.
In this case there will be a cluster of $3$ possible detours
near the crossing, as indicated in \autoref{fig:detour-cluster}.
\insfigsvg{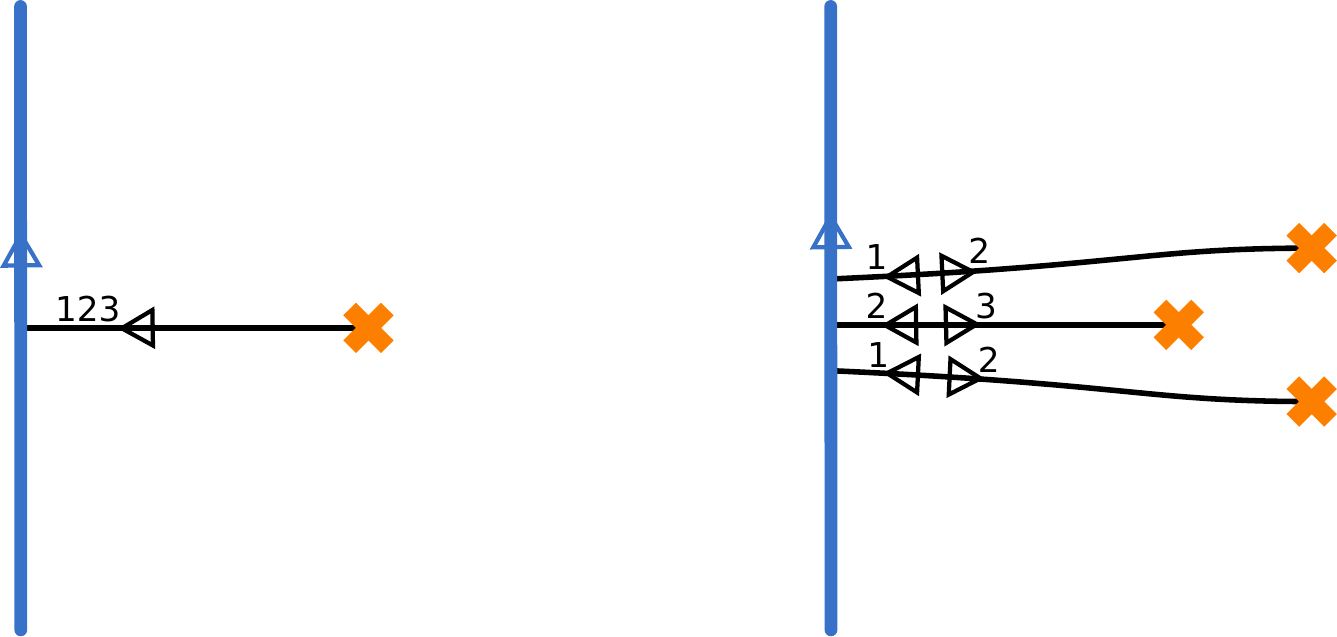}{0.4}{Left: a link segment $L$ crossing a critical leaf
of $\fapprox$. Right: a cluster of $3$ detours in a neighborhood of this crossing.}
This cluster gives rise to various possible lifts, shown in \autoref{fig:cluster-lifts}.
\insfigsvg{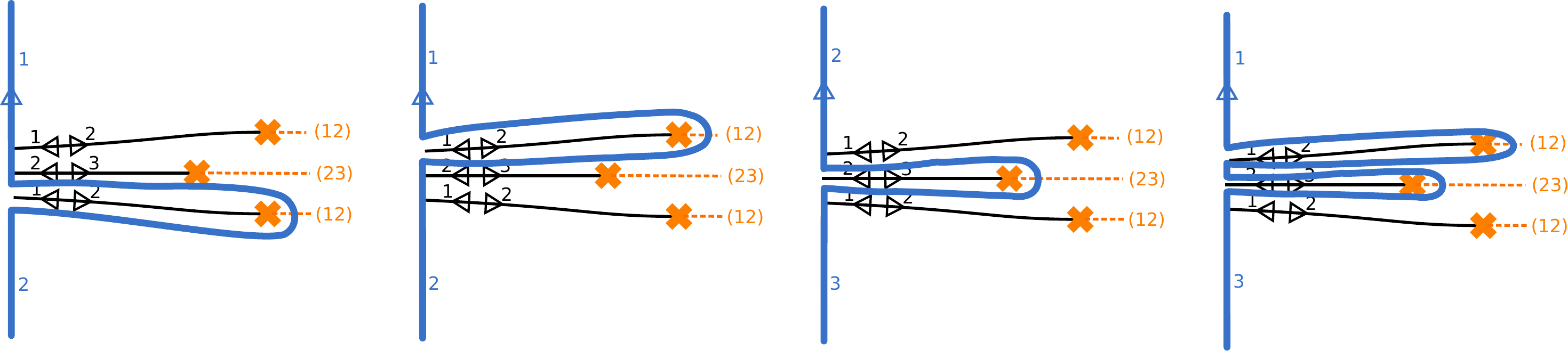}{0.4}{Four lifts of a link segment crossing a critical
leaf of $\fapprox$, using the detours shown in \autoref{fig:detour-cluster}.}

Summing up all the contributions from these exchange and detour clusters, and
including all the factors according
to the general scheme set out in \autoref{sec:qab-general}, one obtains 
$F(L)$. 

We remark that this recipe for computing $F(L)$ closely resembles the rules 
of \cite{Gabella:2016zxu}, and we expect
that (after suitable minor changes reflecting differences in conventions)
the two computations are the same.
In \cite{Gabella:2016zxu} the detour clusters appeared explicitly, just as 
they appear here. The exchange clusters, on the other hand, did not appear
in \cite{Gabella:2016zxu} explicitly; instead there are R-matrix
factors put in ``by hand'' which play the same role.

\section{Examples} \label{sec:examples}

In this section we illustrate our computation of $F(L)$ for various examples of links $L$. In the first few examples we work out everything by hand, to give a
concrete indication of how our scheme works; in later examples
this is infeasible, so we summarize the results of computer computations.
(To reproduce these computations one can use the code supplied
with the arXiv version of this preprint; see \autoref{app:algorithms}.)

\subsection{A simple unknot} \label{eq:N3-simple-unknot}

We consider the case $N=3$ and $C = \C$, with
\begin{equation}
\phi_1 = 0, \qquad \phi_2 = 0, \qquad \phi_3 = \I \, \de z^3.
\end{equation}
Thus the covering $\tC$ is simply given by
\begin{equation}
\lambda^3 + \I \, \de z^3 = 0,
\end{equation}
so the three sheets are
\begin{equation}
\lambda_i = \e^{\pi \I(-1/2 + 2i / 3)} \, \de z,
\end{equation}
and in particular there are no branch points.
The WKB foliations are as shown in \autoref{fig:simple-foliation-N3}.

\insfigsvg{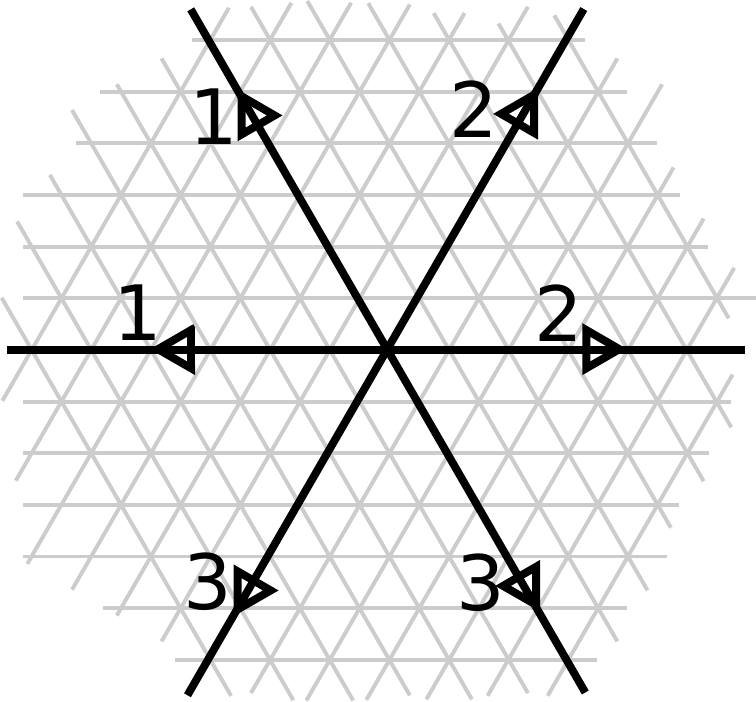}{0.4}{The WKB foliations when $C = \C$ and
	$\phi_1 = 0, \phi_2 = 0, \phi_3 = \I \, \de z^3$. The leaves passing through $z=0$
	are highlighted in black.}

\noindent We consider an unknot $L$ in $M = \R^3$, as shown in \autoref{fig:unknot-3-0}.

\insfigsvg{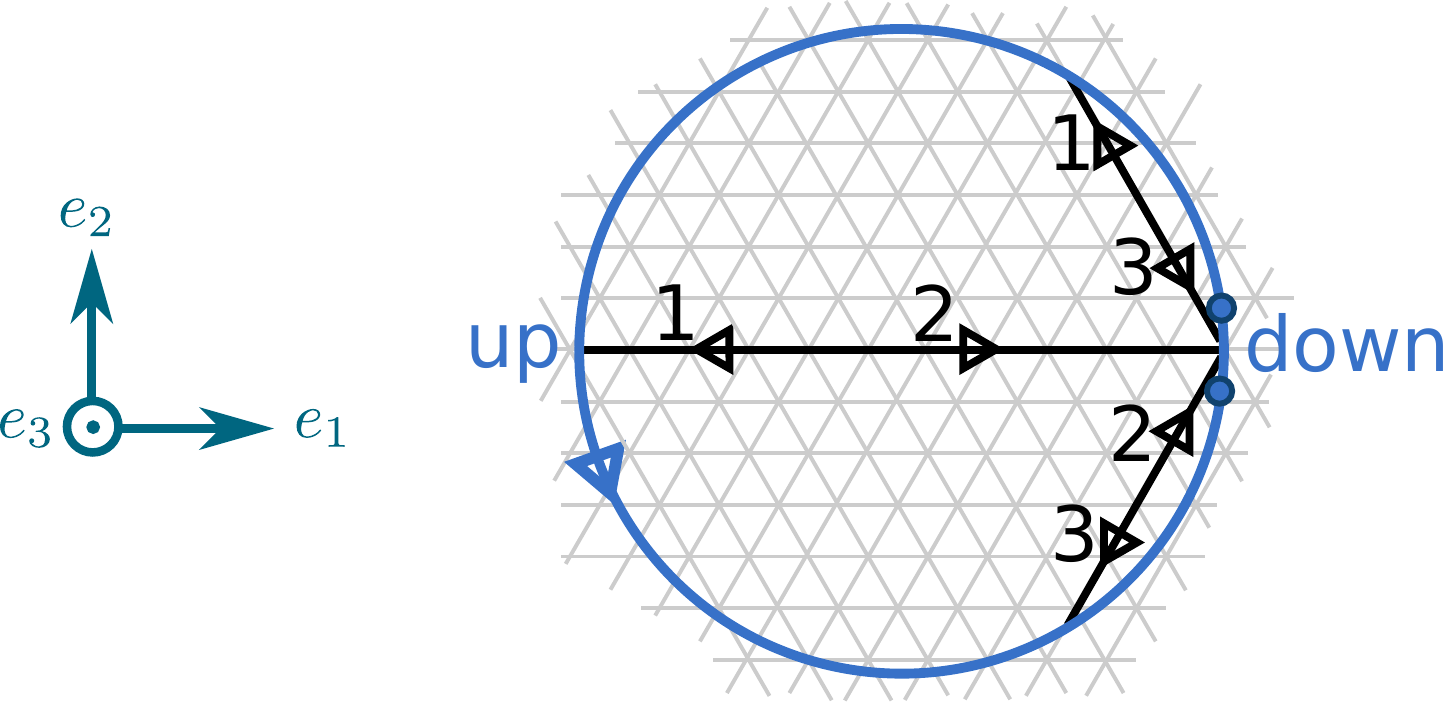}{0.4}{An unknot $L$ in $M = \R^3$. We show the projection to the
$x^1$-$x^2$ plane; the tendency in the $x^3$ direction is indicated by the words
``up'' and ``down.'' $L$ goes up along most of the circle, then goes down steeply
in a small arc; the two marked points are critical points of $x^3$. The
	three possible finite webs are shown in black; they are all exchanges. 
	Generic $ij$-leaves are shown in gray.}

\noindent There are three direct lifts $\tL_1$, $\tL_2$, $\tL_3$, and three lifts involving exchanges, shown in \autoref{fig:unknot-3-1}. 
\insfigsvg{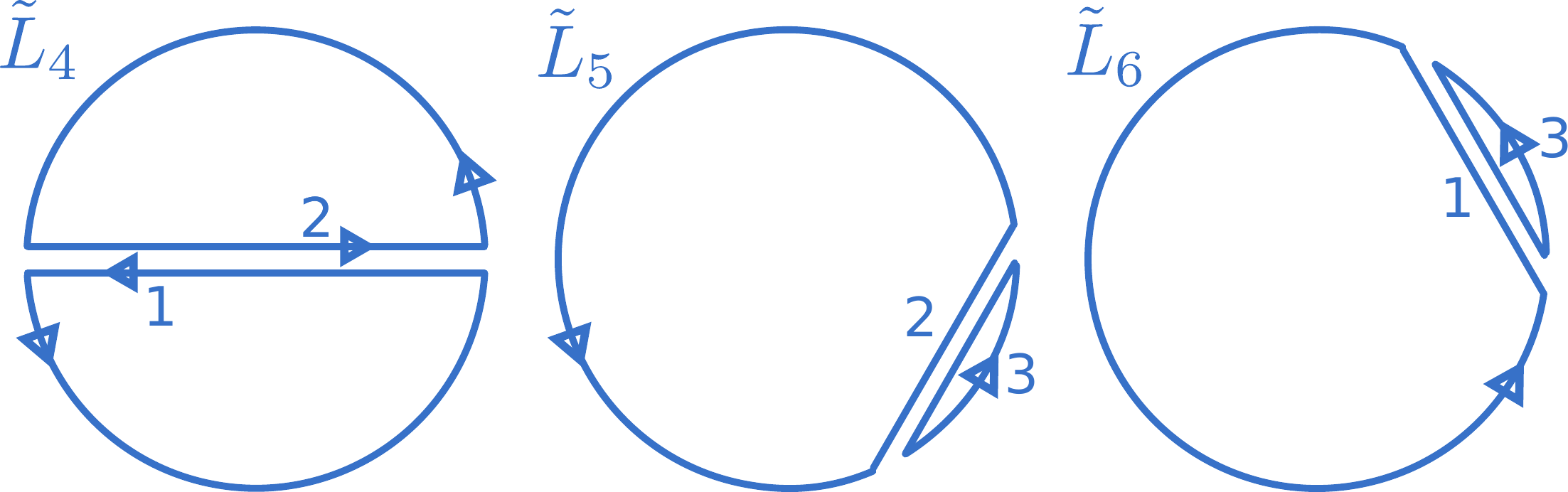}{0.36}{The three lifts of the unknot $L$ involving exchanges.}
The contributions from the lifts are as shown in the table below.
\begin{table}[H]
	\centering
	\begin{tabular}{|c|c|c|c|c|c|c|} \hline
		lift       & tangency    & winding   & exchange           & [lift] & total \\
		\hline \hline
		$\tL_1$    & $q^{2}$     & $1$       & $1$                & $1$    & $q^{2}$ \\
		$\tL_2$    & $q^{2}$     & $1$       & $1$                & $1$    & $q^{2}$ \\
		$\tL_3$    & $q^{2}$     & $1$       & $1$                & $1$    & $q^{2}$ \\
		$\tL_4$    & $q$         & $q^{-4}$  & $q^{2}(q^{-1}-q)$  & $1$    & $q^{-2}-1$ \\
		$\tL_5$    & $q^{2}$     & $q^{-1}$  & $q^{-1}-q$         & $1$    & $1-q^{2}$ \\
		$\tL_6$    & $q^{2}$     & $q^{-1}$  & $q^{-1}-q$         & $1$    & $1-q^{2}$ \\
		\hline
	\end{tabular}
\end{table}
\noindent Summing these contributions gives the expected result,
\begin{equation}
F(L) = q^2 + 1 + q^{-2}.
\end{equation}

\subsection{An unknot around a branch point}\label{sec:unknotbp}

Now we consider a more interesting example,
as follows.
We take $C = \C$ again, but now let
the 1-forms $\lambda_i$ be 
\begin{equation}
	\lambda_1 = -\sqrt{z} \, \de z, \qquad \lambda_2 = \sqrt{z} \, \de z, \qquad \lambda_3 = -\de z. 
\end{equation}
Then we consider an unknot $L \subset M$
whose projection to $C$ is a small loop 
around $z = 0$, with a particular profile
in the $x^3$-direction, as pictured in
\autoref{fig:unknot-around-branch-point}
below.

\insfigsvg{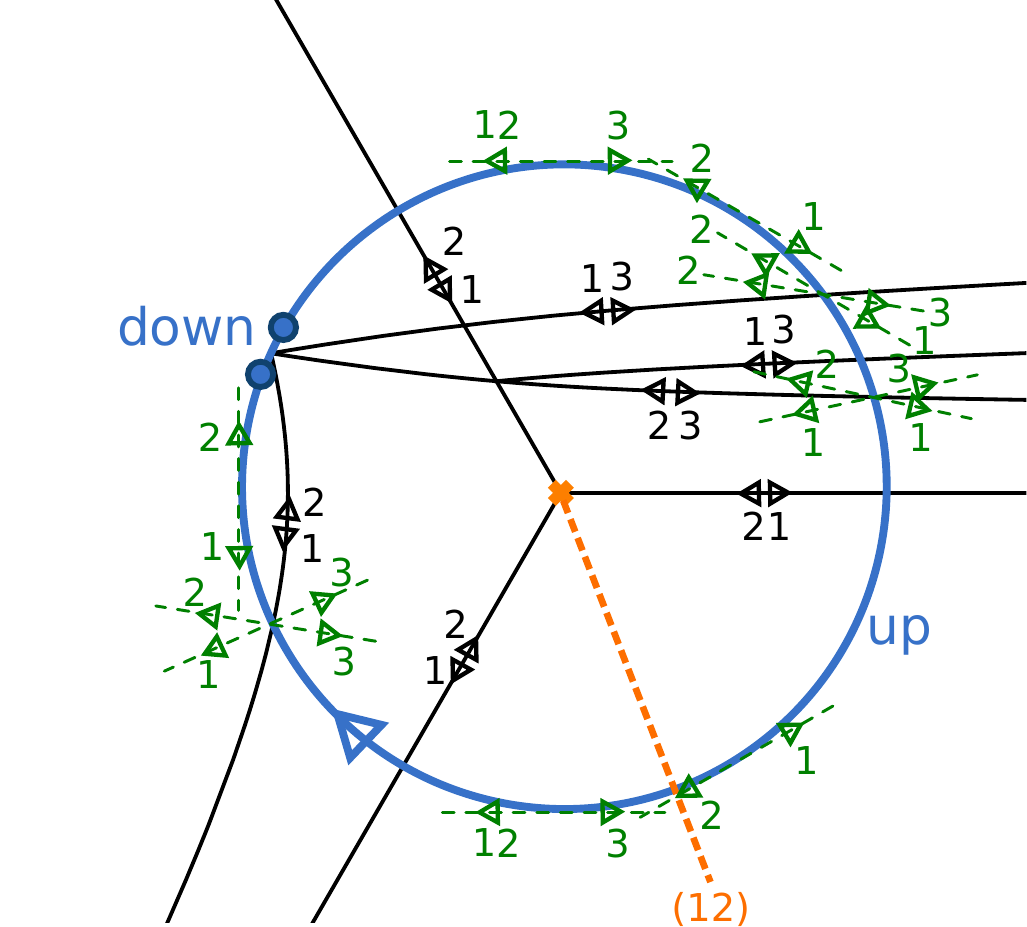}{0.55}{An 
unknot $L \subset M = \R^3$. The $x^3$-coordinate increases as we traverse the loop,
except for a very small region where it decreases. We show, in black, all $ij$-leaves
which participate in BPS webs: this includes in particular 
the three critical leaves emerging
from the branch point at $z = 0$. We also show, in green, the directions of $ij$-leaves
at tangencies to $L$ or its lifts; these are used in the computations of tangency 
and winding factors.}

The link $L$ has $8$ lifts $\tL_1, \dots, \tL_8$.
For each lift, we show in \autoref{fig:unknot-around-branch-point-lifts}
which BPS webs contribute to the lift, and the sheet label for the lift
of each segment of $L$.
We also show the lifted path $\tL_5$ explicitly in 
\autoref{fig:unknot-around-branch-point-one-lift}.

There are five lifts $\tL_1, \dots, \tL_5$ which only involve sheets $1$ and $2$;
these look just like the lifts in the case of a loop around a branch point 
in the $\fgl(2)$ theory, shown in 
Figures 25-27 of \cite{Neitzke:2020jik}.
There are also additional lifts $\tL_6$, $\tL_7$, $\tL_8$ 
in which some segments of $L$ get lifted to sheet $3$.

\insfigsvg{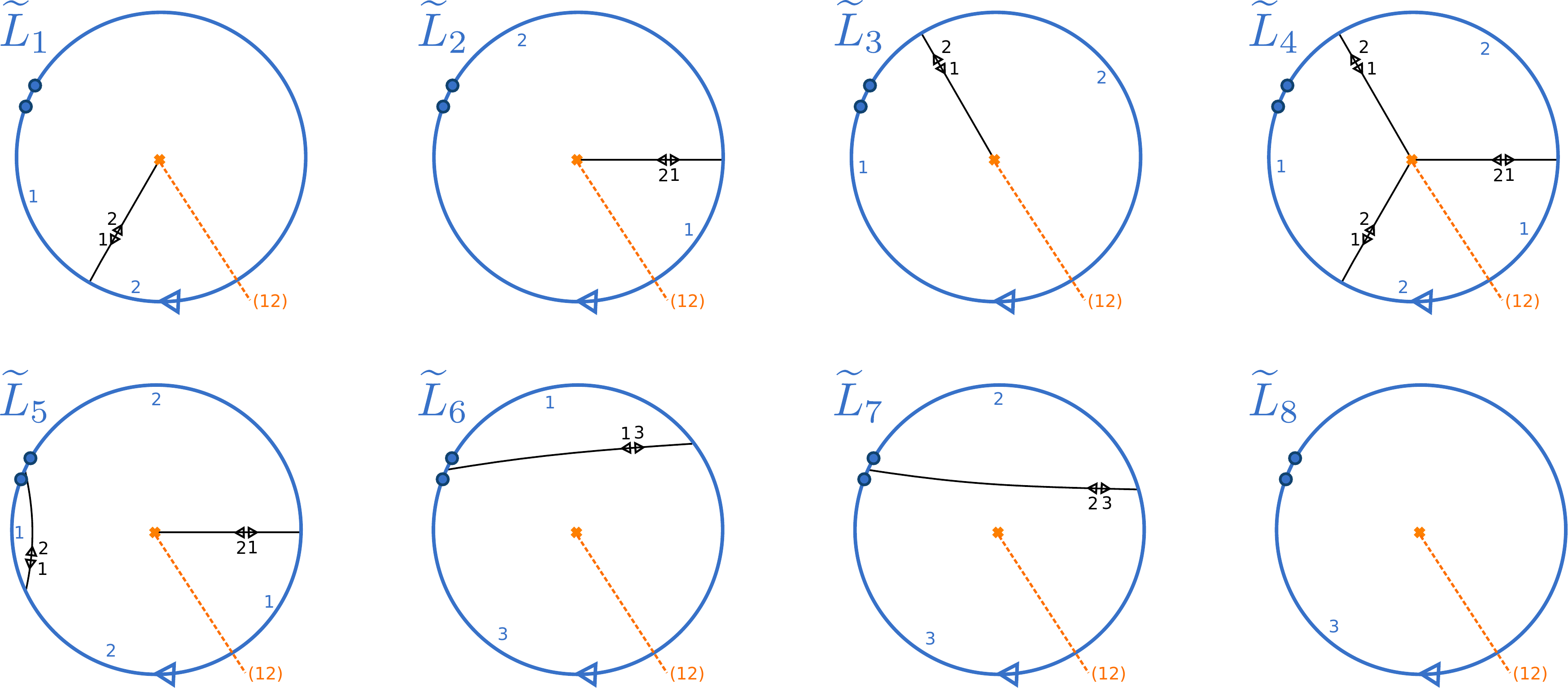}{0.42}{The $8$ lifts
of the unknot $L$. We represent each lift $\tL_k$ 
by specifying the sheet label
for each segment of $L$ and showing the webs which are used. Note that all the webs
which are used are exchanges.} 

\insfigsvg{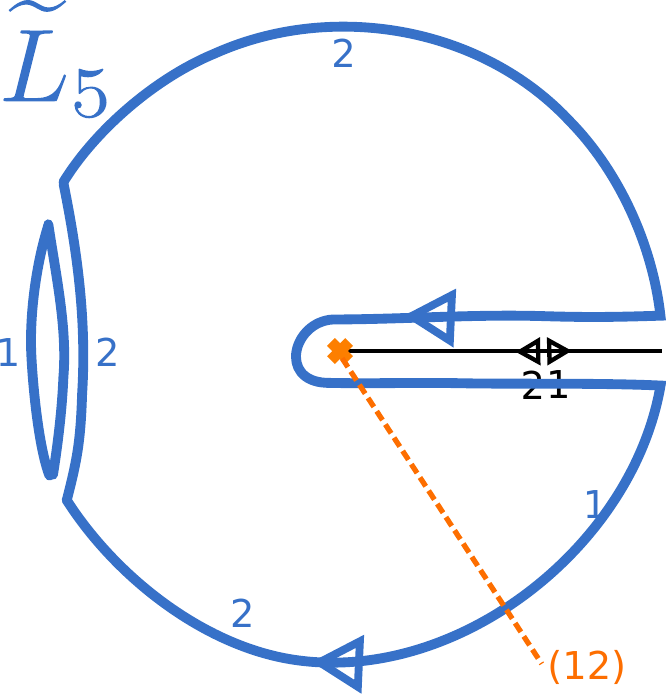}{0.42}{A more explicit picture of the lifted path $\tL_5$.} 

The contributions work out as follows:

\begin{table}[H]
	\centering
	\begin{tabular}{|c|c|c|c|c|c|c|} \hline
		lift       & tangency    &     winding     & exchange           & detour       & [lift]    & total \\
		\hline \hline
		$\tL_1$    & $q^{-5/2}$  &    $1$          & $1$                & $q^{1/2}$   & $1$       & $q^{-2}$ \\
		$\tL_2$    & $q^{-5/2}$  &    $1$          & $1$                & $q^{1/2}$   & $1$       & $q^{-2}$ \\
		$\tL_3$    & $q^{-5/2}$  &    $1$          & $1$                & $q^{1/2}$   & $1$       & $q^{-2}$ \\
		$\tL_4$    & $q^{-5/2}$  &    $1$          & $1$                & $q^{3/2}$   & $-q^{-1}$ & $-q^{-2}$ \\
		$\tL_5$    & $q^{-5/2}$  &    $q$          & $q-q^{-1}$         & $q^{1/2}$   & $1$       & $1 - q^{-2}$ \\
		$\tL_6$    & $q^{-3}$    &    $q^2$        & $q-q^{-1}$         & $1$         & $1$       & $1 - q^{-2}$ \\
		$\tL_7$    & $q^{-1}$    &    $q^4$        & $q^{-2}(q-q^{-1})$ & $1$         & $1$       & $q^2 - 1$ \\
		$\tL_8$    & $q^{-2}$    &    $1$          & $1$                & $1$         & $1$       & $q^{-2}$ \\
		\hline
	\end{tabular}
\end{table}

\noindent Once again, summing these contributions gives
\begin{equation}
F(L) = q^2 + 1 + q^{-2}.
\end{equation}

\subsection{An unknot with a three-leaf web} \label{sec:link-1}

Our examples so far have only involved exchanges and detours, the same basic 
phenomena one meets in the $N=2$ case.
Next we consider a more interesting example,
the unknot shown in \autoref{fig:unknot-with-one-web3}. To clarify the notation, in the knot examples containing webs (both in the paper and in the auxiliary code files), we denote external legs of the webs as {\it leaves}.
\insfigsvg{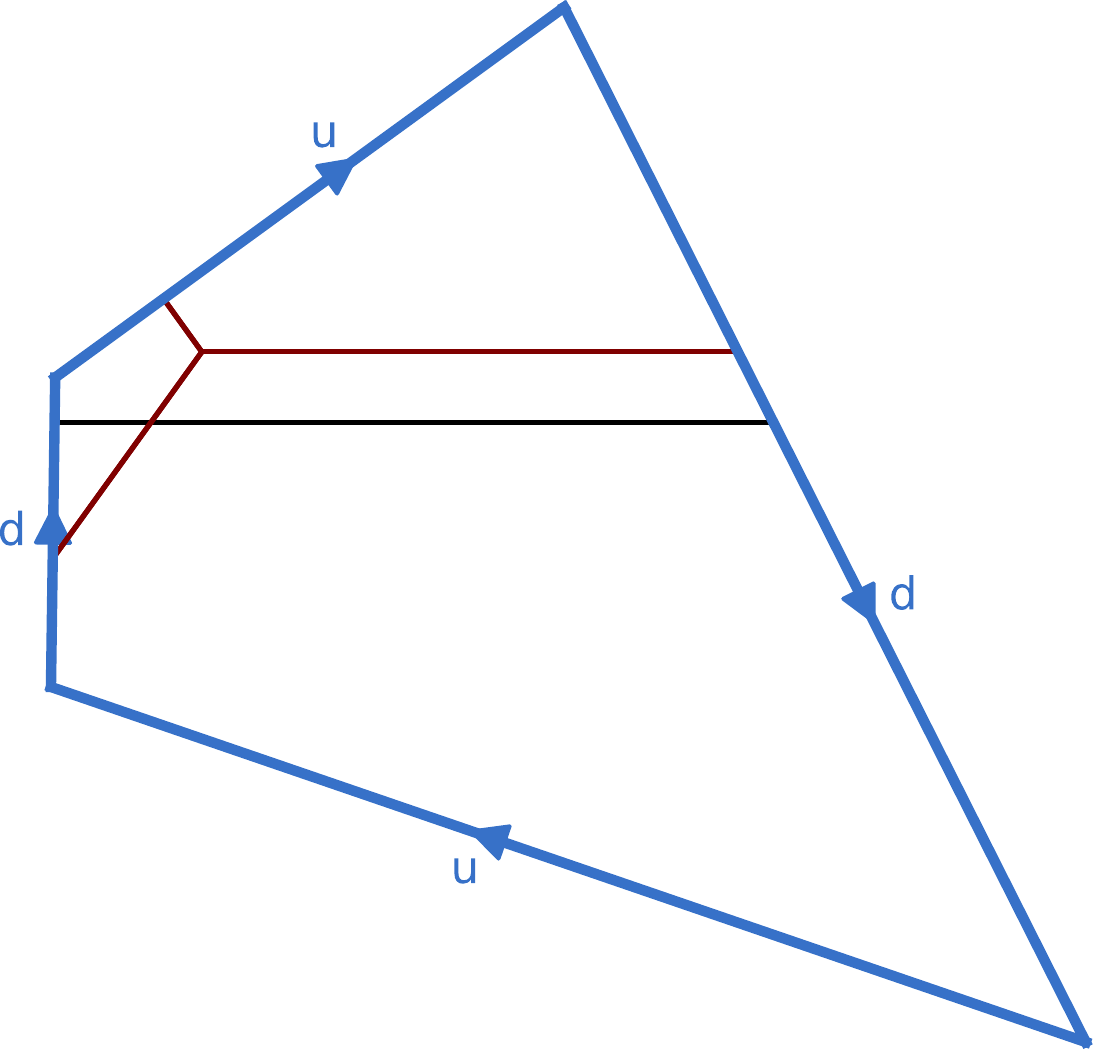}{0.4}{A polygonal unknot $L$ in $M = \R^3$.
We show the projection to the $x^1$-$x^2$ plane;
the labels ``u'' and ``d'' (for ``up'' and ``down'')
next to link segments indicate how
the segment is oriented in the $x^3$-direction.
We also show all possible BPS strings which can attach to $L$: there is one
exchange (black) and one three-leaf web (dark red). The WKB foliations are
as shown in \autoref{fig:simple-foliation-N3}.}
This link has one exchange and one three-leaf web.
There are $5$ lifts: $3$ direct lifts $\tL_1$, $\tL_2$, $\tL_3$
to the $3$ sheets, a lift $\tL_4$ which uses the exchange,
and a lift $\tL_5$ which uses the three-leaf web.
As we have remarked,
the webs in this example were determined using a computer program,
which for computational convenience operates on polygonal links; 
see \autoref{app:algorithms}.
To apply our rules to compute the winding and tangency factors in this situation, 
one should imagine that the corners 
are very slightly rounded off.
Then we obtain the following:

\begin{table}[H]
	\centering
	\begin{tabular}{|c|c|c|c|c|c|c|} \hline
		lift       & tangency   & winding    & web                & [lift]    & total \\
		\hline \hline
		$\tL_1$    & $q$        & $q$        & $1$                & $1$       & $q^2$ \\
		$\tL_2$    & $q$        & $q$        & $1$                & $1$       & $q^2$ \\
		$\tL_3$    & $1$        & $q^{-2}$   & $1$                & $1$       & $q^{-2}$ \\
		$\tL_4$    & $q^{-1}$   & $q^{-2}$   & $-q^2(q-q^{-1})$   & $1$       & $-1+q^{-2}$ \\
		$\tL_5$    & $1$        & $q^{-1}$   & $-q(q-q^{-1})^2$   & $1$       & $-q^2+2-q^{-2}$ \\
		\hline
	\end{tabular}
\end{table}

\noindent Again the sum is
\begin{equation}
F(L) = q^2 + 1 + q^{-2}.
\end{equation}
This example serves as a check on our formula for the contribution from a tree,
in \autoref{sec:tree-weights}.

\subsection{An unknot with a four-leaf web} \label{sec:link-2}

\insfigsvg{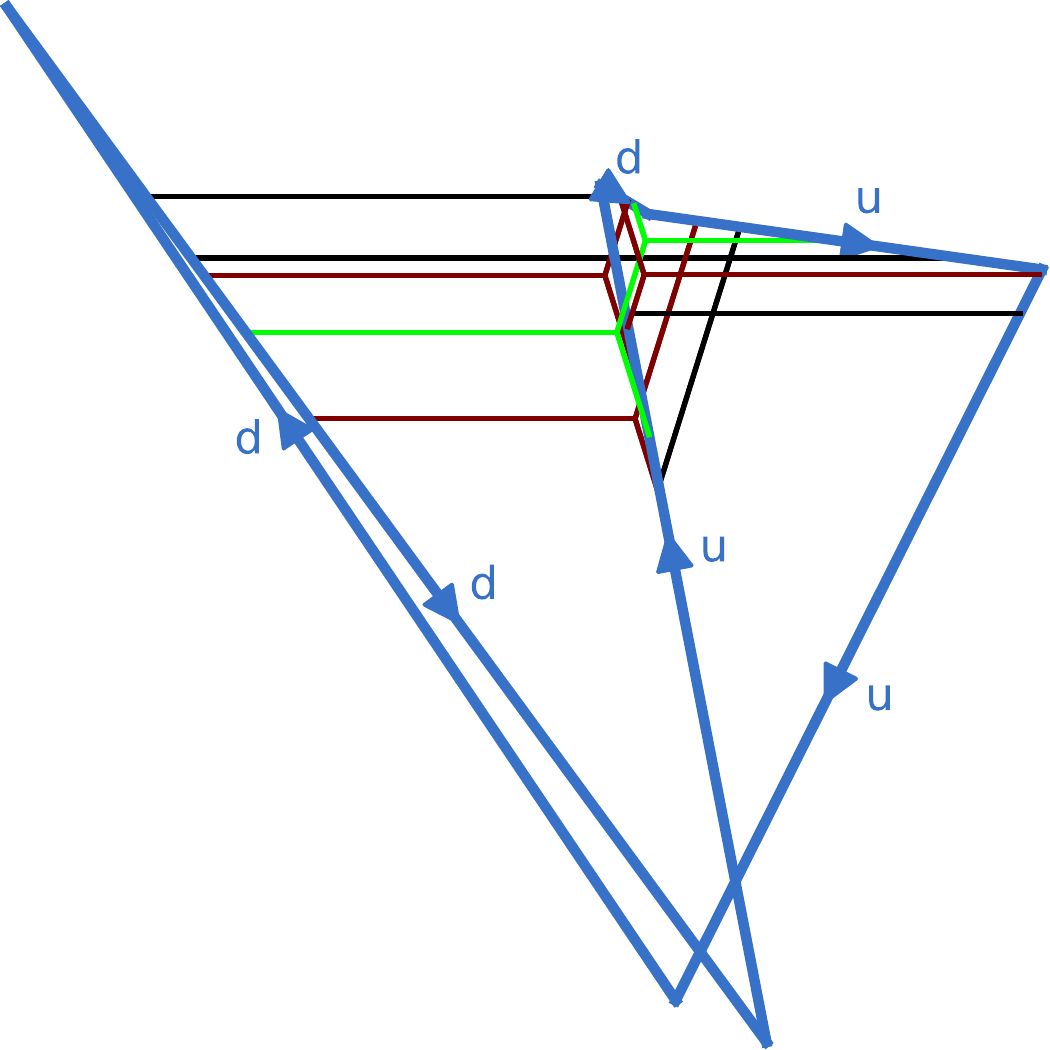}{0.4}{Another polygonal unknot $L$ in
$M = \R^3$. Notation is as in \autoref{fig:unknot-with-one-web3}.
Again we show all BPS strings which can attach to $L$.
In this case there are $4$ exchanges (black), $3$ three-leaf webs (dark red), and $1$
four-leaf web (lime green).
(This example was found by randomly generating polygons in $\R^3$ with $6$ vertices until
we found one which has exactly one four-leaf web and only $11$ lifts.)}

The link $L$ shown in \autoref{fig:unknot-with-one-web4} is an unknot. It has $11$ lifts:
$3$ direct lifts, $4$ lifts which use exchanges, $3$ lifts which use three-leaf
webs, and $1$ lift which uses a four-leaf web.
Applying the rules to evaluate the contribution for each of these
$11$ lifts and summing the results, we obtain the expected $F(L) = q^2 + 1 + q^{-2}$.

\subsection{A trefoil} \label{sec:link-3}

\insfigsvg{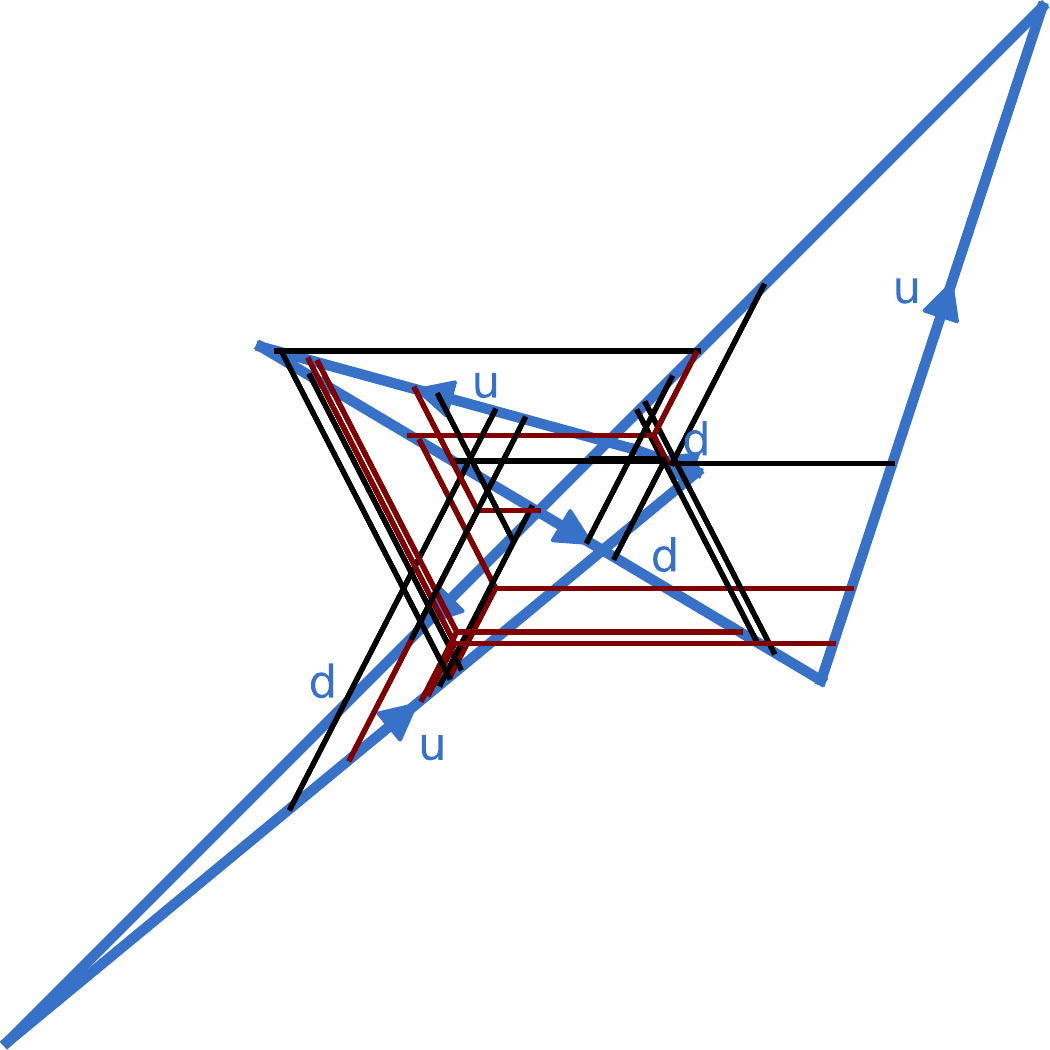}{0.4}{A polygonal trefoil $L$ in $M = \R^3$.
Notation is as in \autoref{fig:unknot-with-one-web3}. As before,
we show all BPS strings which can attach to $L$; in this case there
are only exchanges and three-leaf webs.}

The link $L$ shown in \autoref{fig:trefoil-with-webs} is a trefoil.
It has $23$ webs attached ($16$ exchanges and $7$ three-leaf webs).
Enumerating the possible lifts one finds that there are $43$ in total; 
many of the lifts use multiple webs, and $2$ of the three-leaf webs are not
used in any lift.
Applying the rules and summing the $43$ contributions
we obtain $F(L) = -q^5 - q^3 + q^{-1} + 2 q^{-3} + q^{-5} + q^{-7}$,
matching the expected link polynomial.

\subsection{The once-punctured torus} \label{sec:once-punctured-torus}

Now we let $C$ be a torus with one puncture. We take the puncture to be 
``full'' in the language of \cite{Gaiotto:2009we}, i.e. we consider
$\phi_2$ with a second-order pole and $\phi_3$ with a third-order pole.
We can view $C$ as obtained by gluing in a cylinder to two of the three
punctures on a three-punctured sphere. It follows that
the corresponding field theory of class $S[\fgl_3]$ is (up to the decoupled
$\fgl_1$ part) obtained by gauging a certain $SU(3)$ subgroup of
the $E_6$ global symmetry of the Minahan-Nemeschansky theory.

Let $\phi_2$ be a meromorphic quadratic differential with a second-order pole at the puncture,
chosen so that the WKB spectral network is as shown in the leftmost frame of 
\autoref{fig:once-punctured-torus}, and take $\phi_3$ to be a small perturbation,
so that the WKB spectral network is as shown in the second frame.\footnote{Concretely,
 \autoref{fig:once-punctured-torus} was drawn using the following choice.
Let $C$ be the square torus, with modulus $\tau = \I$. 
Let $\vartheta(z)$ be the Jacobi theta function specialized to $\tau = \I$;
this function has a zero at $z=0$.
Then we take $z_1 = \frac14 + \frac{\I}{4}$, $z_2 = -z_1$, $\alpha = \exp(\frac{9 \pi \I}{48})$, $\phi_2 = \alpha^{-2} \frac{\vartheta(z-z_1) \vartheta(z-z_2)}{\vartheta(z)^2} \de z^2$, $\phi_3 = \eps \alpha^{-3} \frac{\vartheta(z-y_1) \vartheta(z-y_2) \vartheta(z-y_3)}{\vartheta(z)^3} \de z^3$ where $y_i$ are the three nontrivial 2-torsion points, and $\eps = \frac{1}{200}$.}
Then we are in the almost-degenerate situation described in \autoref{sec:almost-degenerate-limit},
and we can use the simplified rules there to compute the UV-IR map.

\insfigsvg{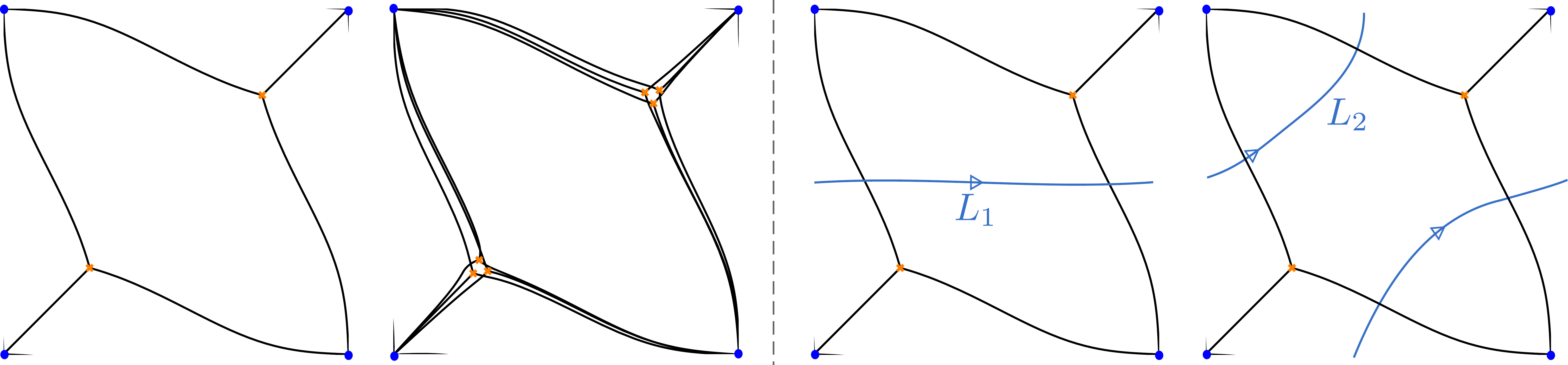}{0.23}{Far left: a spectral network on the 
punctured torus, made up of critical
leaves associated to $\phi_2$ described in the main text, and $\phi_3 = 0$. Middle left: 
a spectral network associated to the same $\phi_2$ and the small perturbation $\phi_3$
described in the main text. Middle right and far right: two loops on the punctured
torus.} 

We consider two line defects in this theory, associated to the loops $L_1$ and $L_2$
shown on the right in \autoref{fig:once-punctured-torus}.
Because these are simple closed curves on 
$C$ they correspond to $\frac12$-BPS line defects (more general links in $C \times \R$
would represent $\frac14$-BPS line defects \cite{Neitzke:2020jik}).

The IR images $F(L_1)$ and $F(L_2)$ are independent of the profile we choose in the 
$x^3$-direction for the links $L_1$ and $L_2$, since any two choices differ by an 
isotopy. 
It is possible, and convenient, to make a choice of profile for 
which there
are no exchanges or higher webs, only detours. These detours occur in clusters 
where $L_1$ or $L_2$ meets the spectral network, as we described in
\autoref{sec:fg-foliations}.
Enumerating all the lifts and summing their contributions is not a task for a human,
at least with the methods we have described here; we implemented it in a Mathematica
program, see \autoref{app:algorithms}.

To describe the result, we recall that the IR skein 
algebra is a quantum torus, with generators $X_\gamma$
as discussed in \autoref{sec:skeinalgebras}. The coefficient of a given $X_\gamma$
in $F(L)$ gives the number and spins of the framed BPS states with charge $\gamma \in \Gamma$
for the line defect $L$. 

\insfigsvg{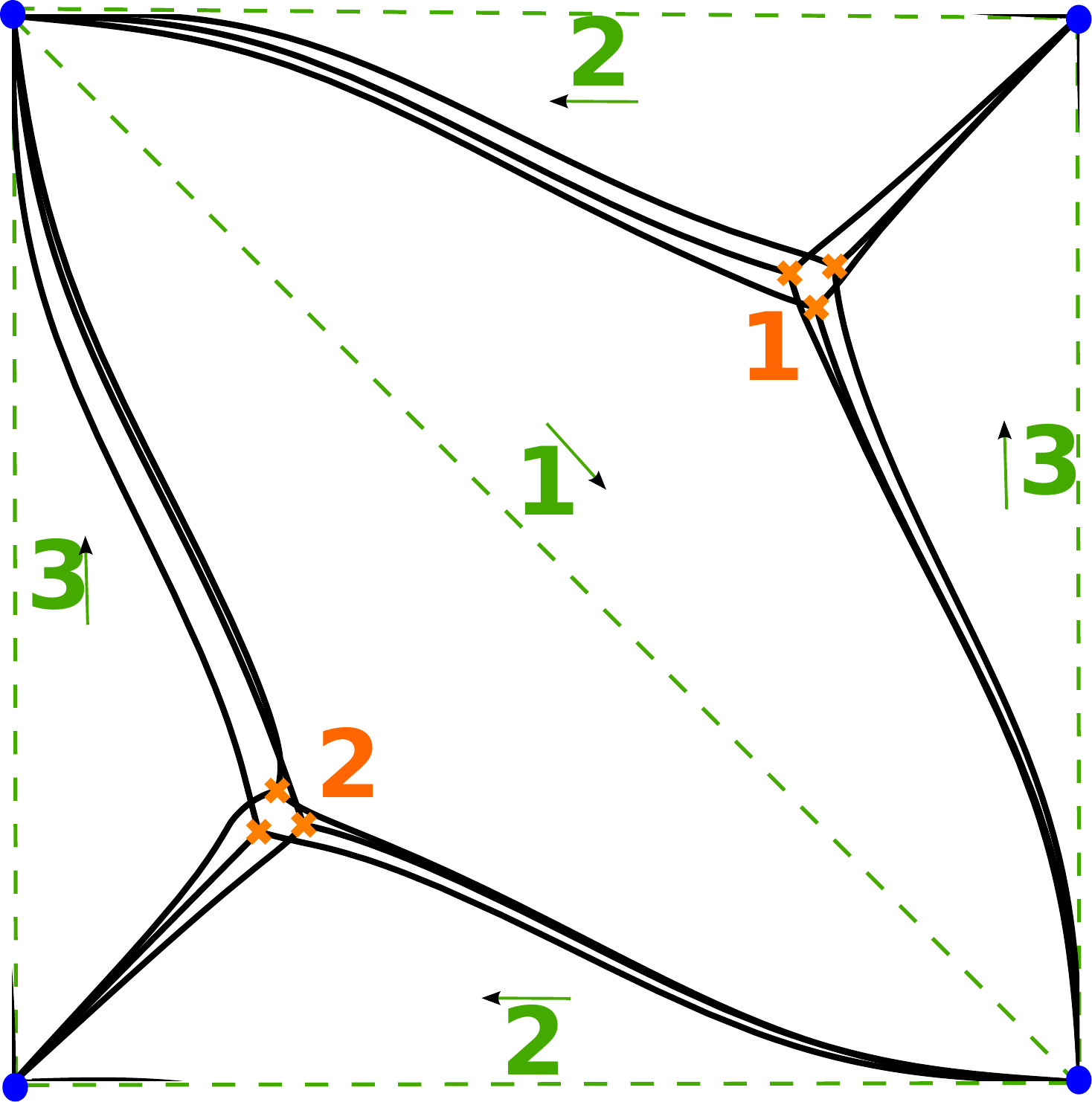}{0.27}{The WKB triangulation associated 
to the spectral networks shown in \autoref{fig:once-punctured-torus}, with the $3$ edges
and $2$ triangles numbered, and an auxiliary choice of orientation fixed on each edge.}

To describe the charge lattice $\Gamma = H_1(\tC,\Z)$,
we use the ideal triangulation shown in \autoref{fig:once-punctured-torus-triangulation}.
Recall from \autoref{sec:fg-foliations} that
there is a canonical 1-cycle $\gamma_{T_a}$ on $\tC$ associated to each 
triangle $T_a$, and two cycles $\gamma_{E_a}^b$
associated to each edge $E_a$. These cycles together 
make up a basis of the charge lattice in the $\fsl(3)$
theory.
To complete this to a basis of the full charge lattice $H_1(\tC,\Z)$ 
for the $\fgl(3)$ theory,
we need to add two more cycles, subject only to the condition that their projections to $C$
generate $H_1(C, \Z)$. We choose the cycles 
$\gamma_A$, $\gamma_B$ shown in \autoref{fig:once-punctured-torus-flavor}.
\insfigsvg{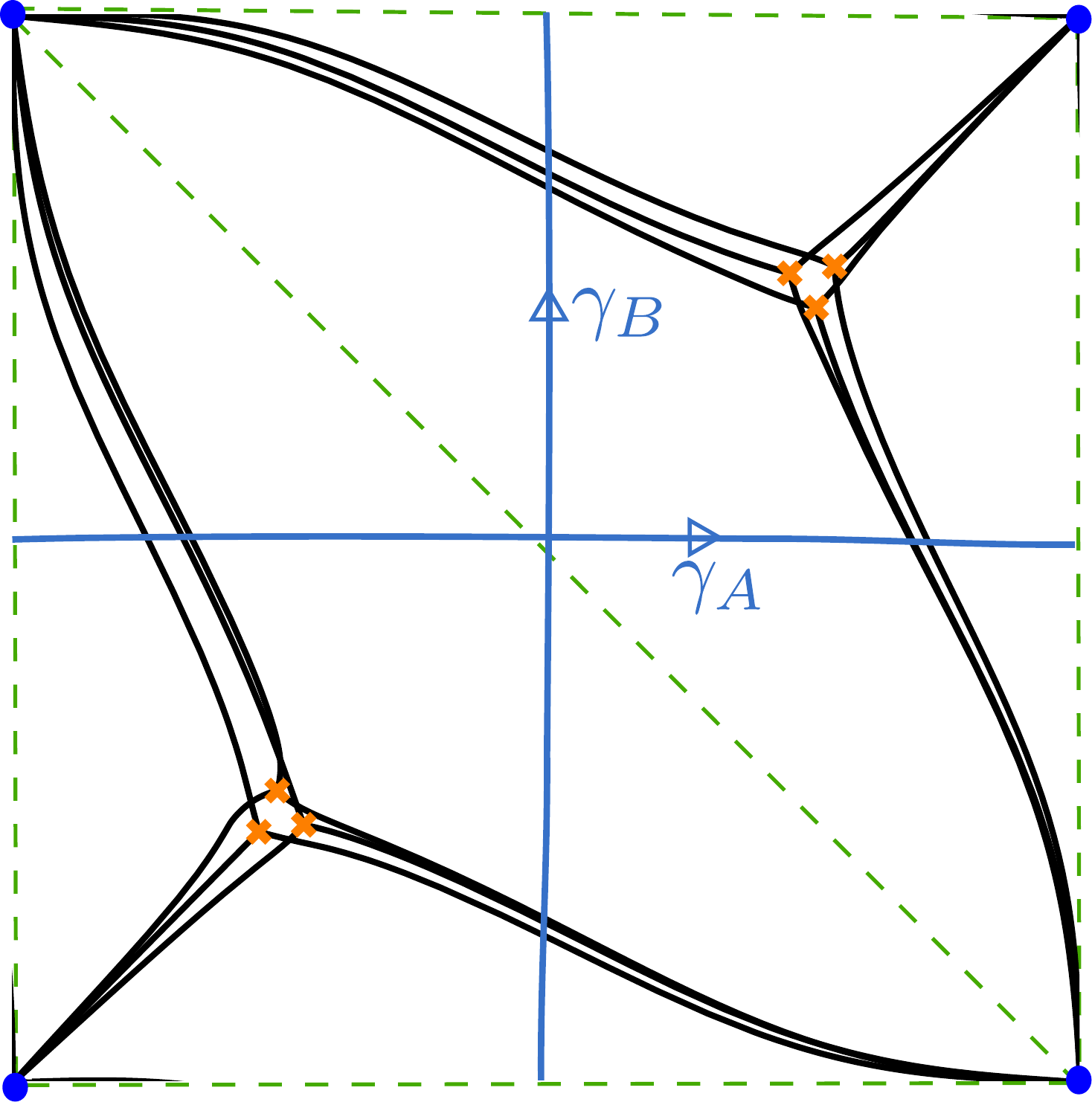}{0.27}{Two $1$-cycles $\gamma_A$, $\gamma_B$ 
on the triple cover $\tC$.}

Then altogether we have a basis for $\Gamma$,
\begin{equation}
	\{ \gamma_{T_1}, \gamma_{T_2}, \gamma^1_{E_1}, \gamma^2_{E_1}, \gamma^1_{E_2}, \gamma^2_{E_2}, \gamma^1_{E_3}, \gamma^2_{E_3}, \gamma_A, \gamma_B \},
\end{equation}
and we represent a general charge $\gamma \in \Gamma$ as a vector in $\Z^{10}$ giving the 
expansion relative to this basis.
Now we can state the results:
\begin{itemize}
\item $F(L_1)$ is a sum of $8$ terms $X_\gamma$, all with coefficient $1$:
\begin{align}
\begin{split}
F(L_1) &= X_{[0,0,0,0,0,0,0,0,1,0]}+X_{[0,0,1,0,0,0,0,0,1,0]}+ 
X_{[1,0,1,0,0,0,0,0,1,0]}+\\
&+ X_{[1,0,1,0,0,0,1,0,1,0]}
   +X_{[1,0,1,1,0,0,0,0,1,0]}+X_{[1,0,1,1,0,0,1,0,1,0]}+ \\
&   + X_{[1,1,1,1,0,0,1,0,1,0]}+X_{[1,1,1,1,0,0,1,1,1,
   0]} \, .
\end{split}
\end{align}
\item $F(L_2)$ is a sum of $19$ terms $X_\gamma$ with coefficient $1$ and $8$ terms
$X_\gamma$ with coefficient $(-q-q^{-1})$:

\begin{align}
\begin{split}
F(L_2) &=X_{[0,0,0,0,0,0,0,0,1,1]}+\\
   &+X_{[0,0,0,0,0,1,0,0,1,1]}+X_{[0,0,1,0,0,1,0,0,1,1]}+X_{[0,1,0,0,0,1,0,0,1,1]}+\\ 
   &+X_{[0,1,0,0,1,1,0,0,1,1]}+X_{[0,1,2,0,0,1,0,0,1,1]}+X_{[0,1,2,0,1,1,0,0,1,1]}+\\
   &+X_{[1,1,1,0,0,1,0,0,1,1]}+X_{[1,1,2,0,0,1,0,0,1,1]}+X_{[1,1,2,0,0,1,1,0,1,1]}+\\
   &+X_{[1,1,2,0,1,1,1,0,1,1]}+X_{[1,1,2,1,1,1,1,0,1,1]}+X_{[2,1,2,0,1,1,0,0,1,1]}+\\
   &+X_{[2,1,2,0,1,1,1,0,1,1]}+X_{[2,1,2,2,1,1,0,0,1,1]}+X_{[2,1,2,2,1,1,1,0,1,1]}+\\
   &+X_{[2,2,2,1,1,1,1,0,1,1]}+X_{[2,2,2,2,1,1,1,0,1,1]}+X_{[2,2,2,2,1,1,1,1,1,1]}+\\
&\left(-q-\frac{1}{q}\right) \left(
   X_{[0,1,1,0,0,1,0,0,1,1]}+
   X_{[0,1,1,0,1,1,0,0,1,1]}+
   X_{[1,1,1,0,1,1,0,0,1,1]}+\right.\\
 &+X_{[1,1,1,1,1,1,0,0,1,1]}+
   X_{[1,1,2,0,1,1,0,0,1,1]}+
   X_{[1,1,2,1,1,1,0,0,1,1]}+\\
 &+\left.X_{[2,1,2,1,1,1,0,0,1,1]}+
   X_{[2,1,2,1,1,1,1,0,1,1]}\right) \, .
\end{split}
\end{align}

\end{itemize}

We remark that these answers have the expected
positivity property when expanded in powers of $-q$, 
and moreover they are characters of representations of $SU(2)_P$; these properties are 
as expected
for the framed BPS spectra of $\frac12$-BPS line defects in 4d $\cN=2$ theories
\cite{Gaiotto:2010be}.

Next we consider the specialization to $q = -1$ and $X_\gamma = 1$:
this gives the total dimension of the space of framed BPS states. We find
$F(L_1) = 8$, $F(L_2) = 35$. On the other hand, in 
\cite{Neitzke:2020jik} we discussed the analogous computation in the 
$\fgl(2)$ theory instead of $\fgl(3)$, with the same surface $C$, the same links $L_1$, $L_2$,
and the same $\phi_2$; 
there the same specialization 
gives $F(L_1) = 3$, $F(L_2) = 6$.
These results are related by $F_{\fgl(3)} = F_{\fgl(2)}^2 - 1$.
This relation is expected on general grounds: it
comes from the fact that if $\rho: GL(2) \to GL(3)$ denotes
the symmetric square representation then $\Tr \rho(A) = (\Tr A)^2 - 1$,
and that the symmetric square map on moduli spaces of flat connections
maps the $SL(2)$-connection with all $X_\gamma = 1$ to 
the $SL(3)$-connection with all $X_\gamma = 1$.\footnote{More generally, for an $SL(3)$-connection 
obtained as the symmetric square of an $SL(2)$-connection, 
the face coordinates are always $1$, and 
both coordinates attached to an edge $E$ are equal to the corresponding edge coordinate 
of the $SL(2)$-connection. 
This fact can be deduced e.g. from comments in the introduction of
\cite{Fock-Goncharov}. We thank Alexander Goncharov for explaining 
this to us.}

Finally we remark that these results for $F(L_1)$ and $F(L_2)$ agree with computations of the $\fsl(3)$ quantum trace of \cite{douglas2021quantum} for the loops $L_1$ and $L_2$,
performed by Daniel Douglas.
This gives support for the hypothesis that the quantum UV-IR map 
on a triangulated surface agrees with the $\fsl(3)$ quantum trace of \cite{douglas2021quantum}.
We thank Daniel Douglas for very helpful related discussions.

\subsection{The \texorpdfstring{$SU(3)$}{SU(3)} gauging of two copies of Minahan-Nemeschansky \texorpdfstring{$E_6$}{E6} theory}
\label{sec:four-punctured-sphere}

Next we consider the class $S$ theory with $C = \C\PP^1$, $\fg = \fgl_3$  and $4$ full punctures,
which is (up to the decoupled $\fgl_1$ part)
a gauging of a diagonal $SU(3)$ flavor 
symmetry in two copies of the Minahan-Nemeschansky $E_6$ theory 
\cite{Gaiotto:2009we}.

To study the UV-IR map
we proceed as we did in \autoref{sec:once-punctured-torus}:
let $\phi_2$ be a meromorphic quadratic differential on $C$ with second-order
poles at the punctures, chosen so that the WKB triangulation is the 
standard tetrahedral triangulation of $C$,
and take $\phi_3$ to be a small perturbation. 
See \autoref{fig:four-punctured-sphere-networks}.

\insfigsvg{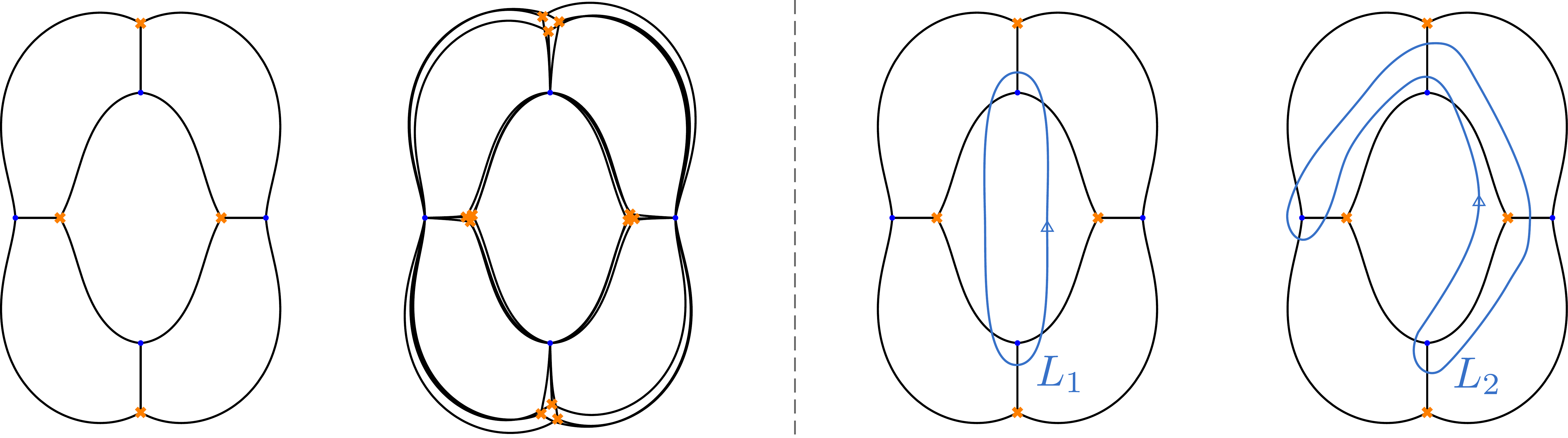}{0.25}{Far left: a spectral network on the four-punctured sphere, made up of critical
leaves associated to $\phi_2$ described in the main text, and $\phi_3 = 0$. Middle left: 
a spectral network associated to the same $\phi_2$ and the small perturbation $\phi_3$
described in the main text. Middle right and far right: two loops on the four-punctured sphere.}

\insfigsvg{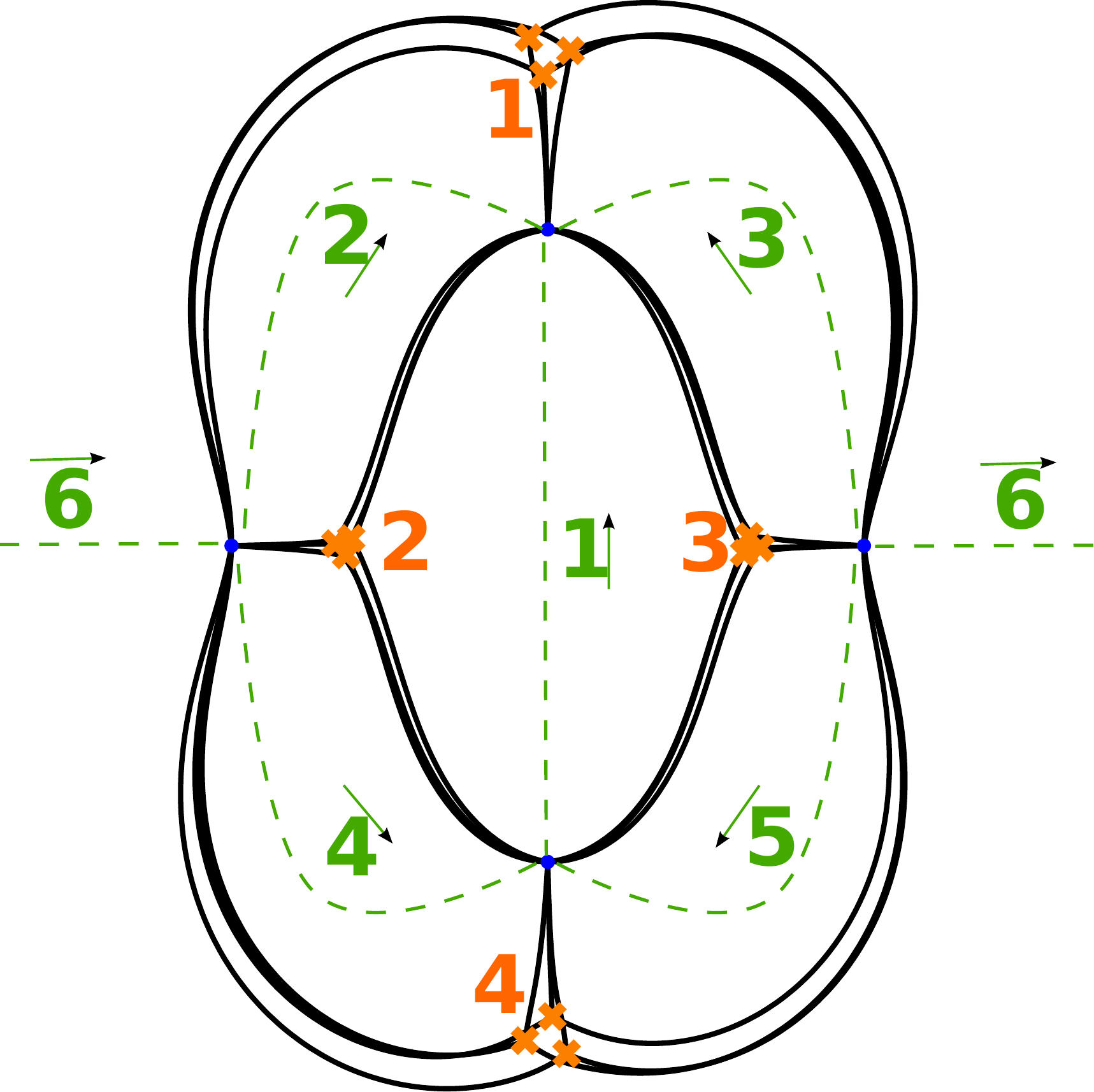}{0.25}{The WKB triangulation associated 
to the spectral networks shown in \autoref{fig:four-punctured-sphere-networks}, 
with the $3$ edges
and $2$ triangles numbered, and an auxiliary choice of orientation fixed on each edge.}

As we did in \autoref{sec:once-punctured-torus} we introduce a basis for
the charge lattice $H_1(\tC,\Z)$: 4 cycles associated to the triangles,
12 associated to the edges, and 3 additional cycles for the extension from
$\fsl(3)$ to $\fgl(3)$, depicted in \autoref{fig:four-punctured-sphere-flavor}.
\insfigsvg{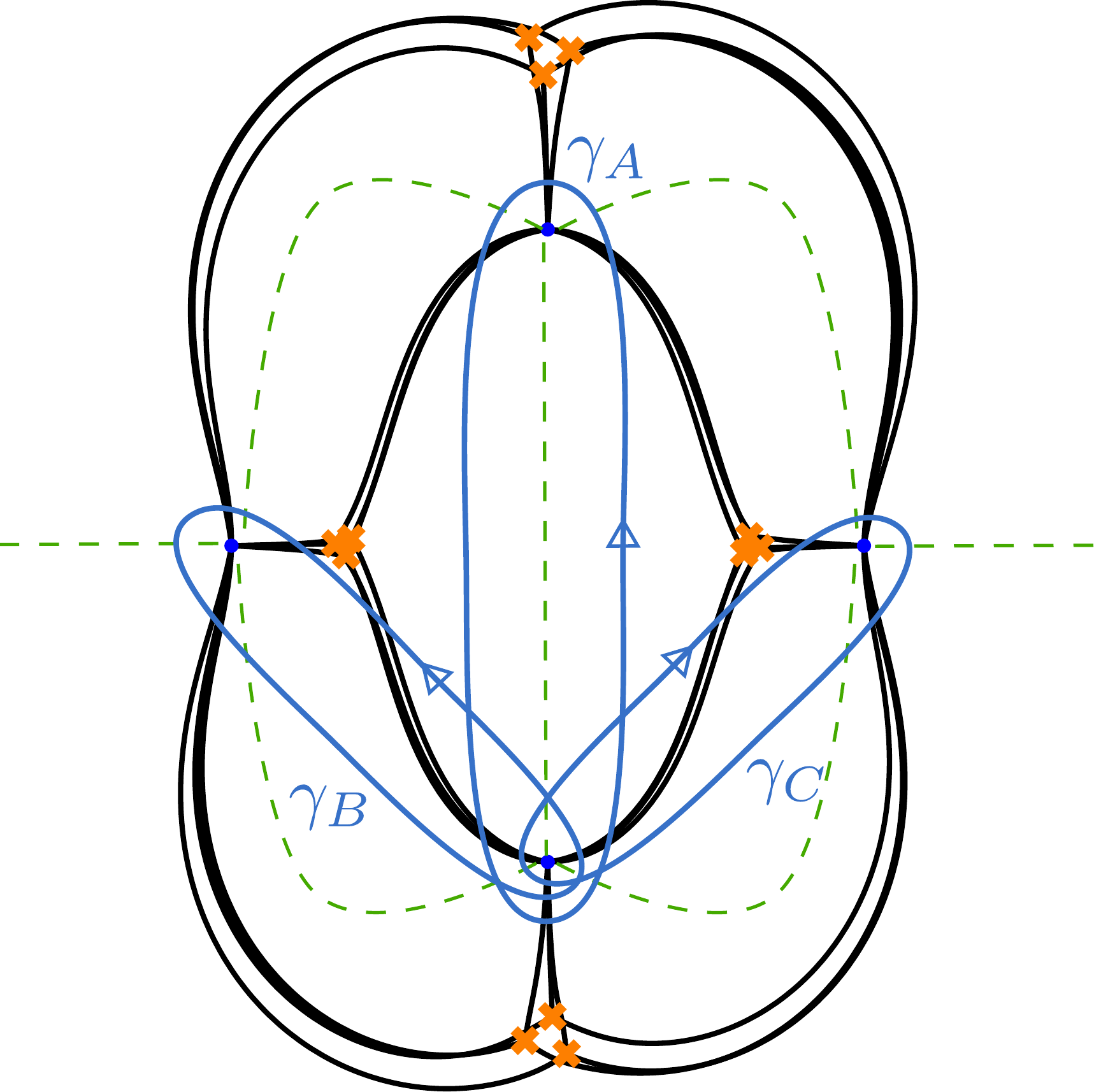}{0.25}{Three $1$-cycles $\gamma_A$, $\gamma_B$, $\gamma_C$ on the triple cover $\tC$.}
Altogether the basis is
\begin{equation}
	\{ \gamma_{T_1}, \gamma_{T_2}, \gamma_{T_3}, \gamma_{T_4}, \gamma^1_{E_1}, \gamma^2_{E_1}, \gamma^1_{E_2}, \gamma^2_{E_2}, \gamma^1_{E_3}, \gamma^2_{E_3}, \gamma^1_{E_4}, \gamma^2_{E_4}, \gamma^1_{E_5}, \gamma^2_{E_5}, \gamma^1_{E_6}, \gamma^2_{E_6}, \gamma_A, \gamma_B, \gamma_C \},
\end{equation}
and we represent the charges $\gamma$ relative to this basis as elements of $\Z^{17}$.
Then, again by computer-aided calculation, we find:
\begin{itemize}
\item $F(L_1)$ is a sum of 48 terms $X_\gamma$, all with coefficient $1$:
\begin{align} 
\begin{split} \label{eq:L1-tetrahedron}
	F(L_1) &= X_{[-1,0,0,0,0,0,-1,-1,-1,-1,0,0,0,0,0,0,1,0,0]} +\\ 
	&+ X_{[-1,0,0,0,0,0,-1,-1,-1,-1,0,0,0,1,0,0,1,0,0]}+ \\ 
	&+ (\text{44 more terms}) \, + \\
	&+ X_{[0,1,1,1,0,0,0,0,0,0,1,0,1,1,0,0,1,0,0]}+\\ 
	&+ X_{[0,1,1,1,0,0,0,0,0,0,1,1,1,1,0,0,1,0,0]}.
\end{split}
\end{align}
Thus the line defect $L_1$ has 48 framed BPS states, all with distinct charges,
and all with spin zero.
\item $F(L_2)$ is more interesting; there are $707$ $X_\gamma$ which appear with coefficient $1$,
$192$ $X_\gamma$ with coefficient $(-q -q^{-1})$, and $16$ $X_\gamma$ with 
coefficient $(q^2 + 2 + q^{-2})$:
\begin{align}
\begin{split} \label{eq:L2-tetrahedron}
	F(L_2) &= X_{[-2,-1,-1,-1,0,0,-1,-1,-1,-1,-1,-1,0,0,-1,-1,0,-1,0]} + (\text{706 more terms}) \, + \\
	&+ \left(-q-q^{-1}\right) \left(X_{[-2,-1,0,0,0,0,0,-1,-1,-1,0,0,1,1,-1,-1,0,-1,0]} + 
	(\text{191 more terms}) \right) \, + \\
    &+ \left(q^2+2+q^{-2}\right) \left(X_{[-1,-1,0,0,0,0,0,-1,-1,0,0,0,1,1,-1,-1,0,-1,0]} + 
	(\text{15 more terms}) \right).
\end{split}
\end{align}
Thus the line defect $L_2$ does have framed BPS states
with spin $\frac12$ and spin $1$; the spin $\frac12$ multiplets appear by themselves,
while every multiplet of spin $1$ is accompanied by a 
multiplet of spin $0$ with the same charge.
\end{itemize}
The full forms of \eqref{eq:L1-tetrahedron} and \eqref{eq:L2-tetrahedron}
are given in a Mathematica notebook included with the arXiv
version of this paper.

Again these answers have the expected
positivity property when expanded in powers of $-q$, 
and are characters of representations of $SU(2)_P$, as expected.
At $q = -1$ and $X_\gamma = 1$, we have again
$F_{\fgl(3)} = F_{\fgl(2)}^2 - 1$ for both defects
(concretely $48 = 7^2 - 1$ and $1155 = 34^2 - 1$).

Moreover, again these results agree with computations of the $\fsl(3)$ quantum trace
of \cite{douglas2021quantum}, performed by Daniel Douglas.

\appendix

\section{Computer computations} \label{app:algorithms}

In \autoref{sec:examples} we reported the results of various computations, most of which
were done with computer assistance. In this section we briefly describe the 
algorithms involved.

\subsection{For computations in \texorpdfstring{$M = \R^3$}{M=R3}}

For the computations in $M = \R^3$ reported in \autoref{sec:examples}, 
we begin by fixing a polygonal 
link $L \subset \R^3$, with vertices either chosen by hand or 
randomly generated.
To compute $F(L)$, the main difficulty is to somehow identify all
possible webs which can attach to $L$.
For this purpose we proceed as follows. (We discuss here only the simple
case where there are no webs containing loops; 
when there are loops, the problem becomes
more complicated.)

For any line segment $S$ in $\R^3$, we let $P_{ij}(S)$ denote a strip in $\R^3$,
with boundary $S$ and foliated by half-leaves
of the foliation $F_{ij}$.
We begin by collecting the strips $P_{ij}(S)$ for all possible $ij$
and all segments $S$ of the link $L$.
We then iteratively add new strips to the collection as follows.
We look at all the strips in the collection and compute their intersections.
If $P_{ij}(S)$ intersects $P_{jk}(S')$, then their intersection is a segment
$S''$, and we add to our collection a new strip $P_{ik}(S'')$.
We then check to see whether this new strip
intersects any of the old ones; if it does, each such intersection 
may generate a new strip, and so on. We continue this iterative process until there
are no new intersections.

Once this process terminates, we look for intersections between
strips $P_{ij}(S)$ and $P_{ji}(S')$ in our collection. 
When such a ``head-on'' intersection exists, 
the intersection segment is a leaf of $F_{ij}$. 
If $S$ and $S'$ are both segments of $L$, then such an intersection is
a leaf of $F_{ij}$ with both ends on $L$, i.e. it is an exchange.
More generally, $S$ and $S'$ may have been generated from intersections of other
strips. In this case the intersection segment is not an exchange but
rather a piece of a larger web. Our algorithm collects a list of 
these intersection segments and afterward pieces them together into webs.
See \autoref{fig:shooting}.

\insfigpng{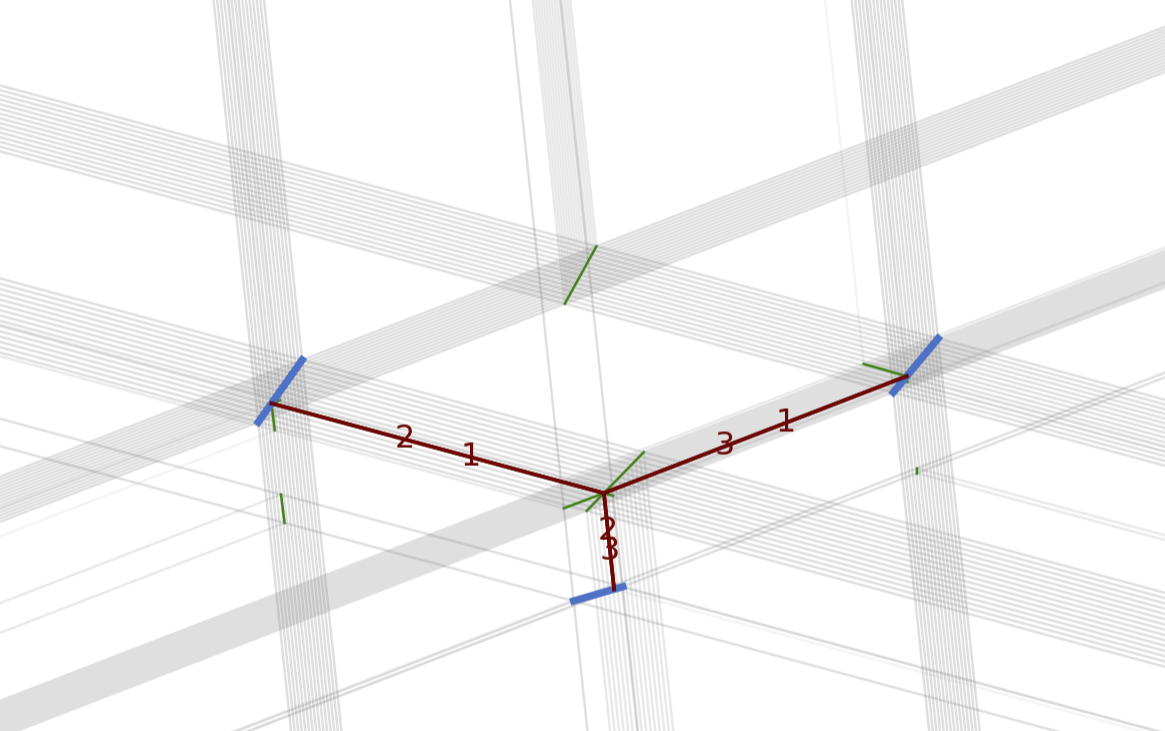}{0.35}{The picture in $\R^3$ generated by applying our
web-finding algorithm to three short link segments. 
Each of the link segments (blue) emits six
strips (gray). When strips of type $ij$ and $jk$ intersect in a segment (green), they can give birth to an additional strip. When a strip of type $ij$ intersects a strip
of type $ji$, the resulting intersection segment is part of a web; in this example
there are just three such segments, which make up a single three-string web (red).}

Once the webs have been identified, there are no major conceptual difficulties
in enumerating all possible lifts, and calculating the factors associated with
each lift according to the rules of \autoref{sec:qab-general}. Having done so, 
since $\tM$ is a
disjoint union of copies of $\R^3$, each lift can be reduced 
using the $GL(1)$ skein relations to a multiple of the empty link. Practically speaking,
this reduction just requires us to compute the writhe of the link diagram with respect
to some projection, which is straightforward. 

We implemented these algorithms in a Python program which we used to make the
various computations in $\R^3$ reported above.
The code (unfortunately very lightly documented),
and a sample Jupyter notebook {\tt sample-computations.ipynb} which can be 
used to reproduce the results reported in \autoref{sec:link-1}-\autoref{sec:link-3}, 
are included as auxiliary 
files with the arXiv version of this preprint.

This algorithm heavily exploits the fact that $M = \R^3$ and all of the
foliations are by straight lines. One could imagine a version of this
algorithm which would work in more general examples, by dividing $M$ up into
domains, choosing coordinates in each domain for which the leaves
are straight lines, and introducing rules for patching
the domains together at their boundaries. This would be in the spirit of a
well-known maneuver in the study of compact Riemann surfaces carrying
holomorphic quadratic differentials: such a surface can always 
be obtained by starting with
a polygon in $\C$ carrying the quadratic differential $\varphi = \de z^2$,
and then gluing the edges together via maps of the form $z \mapsto \pm z + c$.

\subsection{For computations on a Riemann surface \texorpdfstring{$C$}{C}, with nearly degenerate
foliations}

For the computations in $M = C \times \R$ reported in \autoref{sec:examples}, we proceed differently. As explained in \autoref{sec:fg-foliations}, since we 
are working in the limit of nearly degenerate foliations, all the BPS web contributions
come from exchanges, clustered around places where $L$ meets a critical leaf of $\fapprox$ 
or where $L$ meets a generic leaf of $\fapprox$ twice
These are simple enough
to enumerate by hand, given a picture of $L$ (including its profile in the $x^3$ direction)
and the foliation $\fapprox$. 

Once these exchange clusters have been found, the remaining job is to enumerate the lifts $\tL$, compute their weights correctly,
and express them in terms of the basis elements $X_\gamma$ associated to the triangulated 
surface. Implementing this in code requires a lot of tedious bookkeeping, but it is 
conceptually straightforward, 
following the rules we described in \autoref{sec:fg-foliations}. We construct 
a representation of each path as a linear
combination of basic $1$-chains on $\tM$, and then solve a system of linear equations to express
the path as a linear combination of the $X_\gamma$ plus boundaries of $2$-chains.

We implemented these computations, in various examples, in a Mathematica notebook;
in particular this notebook can be used to reproduce the results 
reported in \autoref{sec:once-punctured-torus}-\autoref{sec:four-punctured-sphere}.
This notebook is included as an auxiliary file {\tt state-sums.nb} with
the arXiv version of this preprint.

\newpage

\bibliographystyle{utphys}

\bibliography{q-ab}

\end{document}